\documentclass{article}

\usepackage{PRIMEarxiv}
\usepackage[utf8]{inputenc} % allow utf-8 input
\usepackage[T1]{fontenc}    % use 8-bit T1 fonts
\usepackage{hyperref}       % hyperlinks
\usepackage{url}            % simple URL typesetting
\usepackage{booktabs}       % professional-quality tables
\usepackage{amsfonts, amsmath}       % blackboard math symbols
\usepackage{nicefrac}       % compact symbols for 1/2, etc.
\usepackage{microtype}      % microtypography
\usepackage{lipsum}
\usepackage{fancyhdr}       % header
\usepackage{graphicx}       % graphics
\graphicspath{{media/}}     % organize your images and other figures under media/ folder
\usepackage{epstopdf}% To incorporate .eps illustrations using PDFLaTeX, etc.

\usepackage{caption}
\usepackage{subcaption}

\title{Target-oriented full-waveform inversion based on generalized Rényi entropy using patched Green's function techniques
}

\author{
  Wagner A. Barbosa  \\
  Department of Theoretical and Experimental Physics \\
  Federal University of Rio Grande do Norte \\
  Natal, RN, 59078-970, Brazil.\\
  \texttt{almeidawagner190@gmail.com} \\
   \And
  Sérgio Luiz E. F. da Silva  \\
  Seismic Inversion and Imaging Group \\
  Fluminense Federal University \\
  Niterói, RJ, 24210-346, Brazil\\
  \texttt{sergioluizsilva@id.uff.br} \\
   \And
  Erick de la Barra \\
  Department of Physics\\
  University of La Serena \\
  La Serena, CO, Chile.\\
  \texttt{erick.delabarra@userena.cl} \\
    \And
  João M. de Araújo  \\
  Department of Theoretical and Experimental Physics \\
  Federal University of Rio Grande do Norte \\
  Natal, RN, 59078-970, Brazil.\\
  \texttt{joaomedeiros@fisica.ufrn.br} \\
}

\begin{document}
\maketitle

\begin{abstract}
The estimation of physical parameters from data analysis is a crucial point for the description and modeling of many complex systems. Based on Rényi $\alpha$-Gaussian distribution and patched Green's function (PGF) techniques, we propose a robust framework for data inversion using a wave-equation based methodology named full-waveform inversion (FWI). We show the effectiveness of our proposal by considering two distinct realistic P-wave velocity models, in which the first one is inspired in the Kwanza Basin in Angola and the second in a region of great economic interest in the Brazilian pre-salt field. We call our proposal by the abbreviation $\alpha$-PGF-FWI. The results reveal that the $\alpha$-PGF-FWI is robust against additive Gaussian noise and non-Gaussian noise with outliers in the limit $\alpha \rightarrow 2/3$, being $\alpha$ the Rényi entropic index.
\end{abstract}

% keywords can be removed

\section{Introduction}\label{sec1}
\label{sec:introduction}
Full-waveform inversion (FWI) is a powerful methodology to estimate subsurface physical parameters by exploring the complete waveforms recorded in a seismic survey ~\cite{VirieuxOperto_2009_overview,FichtnerBook}. From a practical point of view, FWI is formulated as a local optimization problem, in which the misfit function to be minimized is based on the least-squares distance between the modeled data and the observed data ~\cite{Tarantola_1984_FWI_Origem}. In this regard, the modeled data are obtained by solving the wave equation at specific positions of the subsurface model, which are associated with the seismic acquisition geometry in order to compare with the observed data. Indeed, the construction of quantitative models from the ample physics provided by the wave equation solution is very useful for describing and modeling complex systems. For this reason, this technique has been employed in a wide variety of applications from geophysics~\cite{RealDataApplicationsFWI2018,OptimalTransportRealData2021} to other scientific fields such as biomedical imaging~\cite{Bernard_FWIultrasonic_2017,Guasch_BrainAWI_2020} and astrophysics~\cite{Hanasoge_FWIAstrofisica0_2014,Hanasoge_FWIAstrofisica_2014}.

Despite the FWI potentials, it is inherently an ill-posed problem in the sense of Hadamard, which means that at least one of the following features is violated: the solution (i) exists, (ii) is unique; and (iii) depends continuously on the observed data. The characteristics (i) and (ii) are commonly violated in FWI because in a typical geophysical survey there is information only on the positions associated with the seismic receivers that cover a small area of the region of interest, which leads the FWI to solve an inconsistent and overdetermined system of equations. Regarding the characteristic (iii), the FWI solution is unstable, as a small variation in the noise level of the observed data leads to discontinuous changes in the reconstructed subsurface model \cite{FichtnerBook}. Furthermore, the least-squares FWI (hereinafter classical FWI) assumes that the errors obeys Gaussian statistics \cite{MENKE198479}, which is not always true, for instance, in geophysical problems \cite{tarantola2005book,Constable_1988_geophysicalJI}.

In the FWI case, the errors come from the computation of differences between the modeled data and the observed data, and therefore, it includes uncertainties associated with seismic noise and incomplete modeling of wave physical phenomena. Indeed, errors are seldom Gaussian in FWI applications. Thus, a wide variety of criteria has been proposed in the literature to mitigate the effects of non-Gaussian errors in the data inversion process \cite{residualnormbrossier}. A very common robust criterion to non-Gaussian errors, especially to erratic data (outliers), is the misfit function based on the $l_1$-norm of the error. Such criterion is based on the assumption that the errors obey the Laplace distribution. Its success is associated with the long tails of the Laplace distribution \cite{tarantola2005book}. For this reason, inverse problems based on the Laplace distribution have been extended in the context of generalized statistical mechanics in order to control the weighting performed by the Laplace distribution's tails \cite{EPJ_PLUS_PSI_Laplace_2021}. However, misfit functions based on Laplace distributions suffer from a singularity issue whenever the residual data is very close to zero \cite{tarantola2005book}.

In an attempt to obtain robust and non-singular misfit functions, the geophysical data inversion has been formulated in the context of generalization of Gauss' law of error.  For instance, Ref.~\cite{q_GaussianPhysicaA} formulated the FWI in the context of Tsallis statistics (also known as \textit{q}-statistics) based on the \textit{q}-generalization of Gauss' error law \cite{Suyari_2005_IEEE_TsallisErrorLaw}. In this regard, the classical and Cauchy distribution based misfit functions are particular cases in the $q \rightarrow 1$ and $q = 2$ limits, respectively. Indeed,  generalizations of Gauss' law of error based on the foundations of statistical physics have been successfully applied to perform robust physical parameters' estimation in non-linear geophysical problems, such as misfit functions based on Student’s t distribution \cite{studentT_1_1989,studentT_tristan_2012,Ubaidillah_2017_studentT}, deformed Gaussian distributions \cite{GeneralizedGaussian_SEG2015,kappa_GaussianPRE,Sacchi_SEG_generalized_2020,PSI_Jackson_2021,daSilva_et_al_NewKappaGaussian}, generalized maximum likelihood approaches \cite{HASEGAWA20093399,Ferrari_2010_AnnalsStat_lqLikelihood,PhysRevE.104.024107}, non-parametric methods \cite{Carvalho_2021_Geophys_J_Int}, as well as in the Rényi framework \cite{extensiveANDnonextensive_physicaA}.

In this work, we formulate the FWI based on the generalization of Gauss' error law linked to the Rényi entropy (or $\alpha$-entropy). The Rényi $\alpha$-entropy \cite{Renyi1965informationTheory,renyi1961measures} was proposed in the context of information theory as a generalization of the  Boltzmann - Gibbs - Shannon (BGS) entropy \cite{lenzi_et_al_2000_physicaA_renyi}, which is very useful to modeling and describing several complex systems in ecology \cite{RenyiApp_Ecology_2015,XU2021107668}, machine learning \cite{MachineLearningAppRenyi_2019,e22020186}, as well as in quantum entanglement \cite{QuantumRenyiapp_2015,science.aau4963_2019}. By considering that the errors are independent and identically distributed by a $\alpha$-generalized Gaussian distribution, which arises from the maximization of the $\alpha$-entropy, we place a robust misfit function in the broad context of the FWI based on the Gauss' law of error in the Rényi framework. This allows us to perform estimates of physical parameters in an unbiased way.

In addition to mitigating the effects of non-Gaussian errors, another very important issue of FWI is the high computational cost. From a computational point of view, the FWI applications in the mapping of subsurface models with a large extension is limited due to the computational cost of the procedure for solving the wave equation, which is performed several times during the data inversion process. In this way, to avoid solving the wave equation in the entire physical domain, we propose a new misfit function based on Rényi statistics along with a technique for solving the wave equation only in a target region. Such a methodology inspired from the condensed matter physics named Patched Green's Function (PGF) \cite{PhysRevB.90.115408,PhysRevB.91.125408,FERREIRA2002355}, in which some of us have recently  generalized the PGF technique for to apply it in problems of target-oriented modeling \cite{Moura2020,almeida2020,FWIPGF2021}. The PGF is a powerful methodology to reduce the computational cost in comparison to the classical modeling techniques. This gain in computational time is directly linked to the fact that the PGF computes the wave field just at a target area and receivers positions. 

This paper is organized as follows: First, in Section~\nameref{sec:methodology}, after presenting the FWI theoretical foundations in its classical approach, we introduce FWI in the context of Rényi statistics and the PGF technique employing the Lagrangian formalism.  Then, in Section~\nameref{sec:numericalexample}, we illustrate how our proposal deals with non-Gaussian errors by presenting two numerical examples. In the first one, we consider a Marmousi case study, in which the main goal consists of investigating the robustness of the FWI based on the Rényi $\alpha$-Gaussian distribution regarding erratic data, and, in the second one, we present a realistic application of target-oriented waveform inversion to estimates model parameters in the context of typical Brazilian pre-salt reservoirs. To conclude, in Section~\nameref{sec:conclusion}, we present our final remarks and perspectives.

%%%%%%%%%%%%%%%%%%%%%%%%%%%%%%%%%%%%%%%%%%%%%
\section{Methodology \label{sec:methodology}} %%
%%%%%%%%%%%%%%%%%%%%%%%%%%%%%%%%%%%%%%%%%%%%%

FWI is a non-linear inverse problem, in which the forward problem consists of modeling the wave propagation through the numerical solution of a wave equation \cite{VirieuxOperto_2009_overview}. In this work, we consider the acoustic approximation that satisfies the following equation:
\begin{equation}
    \label{eq:waveEquationTimeDomain}
    \nabla^2 \Psi_s(\mathbf{x},t) - \frac{1}{v_P^2(\mathbf{x})}\frac{\partial^2 \Psi_s(\mathbf{x},t)}{\partial t^2} = f_s(t) \, \delta\left(\mathbf{x}-\mathbf{x}_s\right),
\end{equation}
where $\Psi_s$ is the pressure wavefield generated by the seismic source $f_s$, $v_P$ is the P-wave velocity model, $\textbf{x} \in \mathbb{R}^2$ and $t \in \mathbb{R}^+$ represent, respectively, the spatial coordinates and the time, and $f_s(t) \, \delta\left(\mathbf{x}-\mathbf{x}_s\right)$ denotes the source term at the position $\mathbf{x} = \textbf{x}_s$. The forward problem can also be solved in the frequency domain. Applying the Fourier transform to Eq.~\eqref{eq:waveEquationTimeDomain}, we obtain the acoustic wave equation in the frequency domain (also known as Helmholtz equation):
\begin{equation}
    \label{eq:wave}
    \nabla^2 \psi_s(\mathbf{x},\omega) + \frac{\omega^2}{v_P^2(\mathbf{x})}\psi_s(\mathbf{x},\omega) = F_s(\omega) \, \delta\left(\mathbf{x}-\mathbf{x}_s\right),
\end{equation}
where $\omega$ is the angular frequency, $\psi_s$ and $F_s$ are the Fourier transform of $\Psi_s$ and $f_s$, respectively. 

The inverse problem consists of inferring the subsurface physical parameters (in our case, the P-wave velocities of the medium) from indirect observations of seismic waveforms (observed data). In the classical approach, the FWI is formulated as a constrained least-squares optimization task as follows:
\begin{equation}
	\underset{m, \psi }{\min } \hspace{0.2cm}
\frac{1}{2} \sum _{\omega} \sum _{s,r} 
\Bigg( \Upsilon_{s,r} \psi_s(\omega) - d_{s,r}(\omega) \Bigg)^\dagger \Bigg( \Upsilon \psi_s(\omega) - d_{s,r}(\omega)  \Bigg),
\label{eq:FWIformulation_classical}    
\end{equation}
subject to,
\begin{equation}
\textbf{A}(m,\omega) \psi_s(\omega) = \textbf{S}_s(\omega),
\label{eq:FWIformulation_classical_constraint}
\end{equation}
where the constraint in the latter equation represents the frequency-domain wave equation \eqref{eq:wave} in a compact form with $\textbf{A}(m,\omega) = \nabla^2  + m \omega^2$ and $\textbf{S}_s(\omega) =  F_s(\omega) \, \delta\left(\mathbf{x}-\mathbf{x}_s\right)$, in which $m = m(\textbf{x}) = 1/v_P^2(\textbf{x})$ denotes the model parameters (in this case, the squared slowness) and the operator $\textbf{A}$ is known as impedance matrix (or Helmholtz matrix) \cite{Marfurt_1984,Pratt_1999}.  It is worth emphasizing that the spatial coordinate (\textbf{x}) is implicit in Eqs.~\eqref{eq:FWIformulation_classical} and \eqref{eq:FWIformulation_classical_constraint} and henceforth for a simplified notation. $\Upsilon u_s$ and $d_s$ represent modeled data and observed data, respectively, where $\Upsilon$ is a sampling operator (onto the receiver \textit{r} of the source \textit{s}), $\psi_s$ is the solution of Eq.~\eqref{eq:wave}, and $d_{s,r}$ is the observed data. The superscript $\dagger$ refers to the adjoint operation (i.e., transpose conjugate).

In the minimization problem \eqref{eq:FWIformulation_classical}, quasi-Newton methods are widely employed for finding an informative local minimum. In this framework, model parameters are iteratively updated along a descent direction, which can be expressed as:
\begin{equation}
     \mathbf{m}_{j+1}=\mathbf{m}_{j}-\beta_{j} \mathbf{H}_j^{-1} \nabla_\textbf{m} \mathbf{\phi }(\mathbf{m}_j), \quad \text{with} \quad j = 0, 1, 2, ..., N_{iter},
     \label{eq:methodquasiNewton}
\end{equation}
where $\mathbf{m} = 1/v_P^2(\textbf{x})$ is the model parameter, $\beta_j > 0$ is a step-length computed in the $j$th iteration being $N_{iter}$ the maximum number of iterations. In large-scales problems, $\mathbf{H}^{-1}$ is the \textit{l}-BFGS approximation of the inverse Hessian matrix calculated from previous gradients of a misfit function, $\nabla_\textbf{m}\phi(\textbf{m})$ \cite{SEISCOPE_optimization_2016}. 

In this way, it is remarkable that the gradient of the misfit function is essential in data inversion via FWI. In the classical approach, the misfit function is given by:
\begin{equation}
\underset{m}{\min } \hspace{0.2cm} \phi_1(m) =
\frac{1}{2} \sum_{\omega} \sum_{s,r} \Bigg( \Upsilon \psi_s(m,\omega) - d_{s,r}(\omega) \Bigg)^\dagger \Bigg( \Upsilon \psi_s(m,\omega) - d_{s,r}(\omega)  \Bigg)
\label{eq:inverseproblemformulation_2}
\end{equation}
and the misfit function gradient is:
\begin{equation}
\nabla_m \phi_1(m) := \frac{\partial \phi_1(m) }{\partial m} =
 \sum_{\omega} \sum_{s,r} \Re\Bigg\{\Bigg(\Upsilon \frac{\partial \psi_s(m,\omega)}{\partial m} \Bigg)^\dagger \Bigg( \Upsilon \psi_s(m,\omega) - d_{s,r}(\omega) \Bigg) \Bigg\}
\label{eq:inverseproblemformulation_2_gradient}
\end{equation}
where $\psi_s(m,\omega)$ is a solution of the wave equation. However, note that at each iteration of the FWI, the derivative of the modeled data in relation to each model parameter must be computed, which is computationally very expensive. For this reason, we now present an efficient way to compute $\nabla_m \phi_1(m)$.

In this way, note that the solution of the problem formulated in \eqref{eq:FWIformulation_classical} can be alternatively done by minimizing the following augmented Lagrangian functional \cite{Haber_2000}:
\begin{eqnarray}
    \mathcal{L}(m, \psi, \lambda)  = \frac{1}{2} \sum_{\omega} \sum_{s,r} \Bigg( \Upsilon \psi_s(\omega) - d_{s,r}(\omega) \Bigg)^\dagger \Bigg( \Upsilon \psi_s(\omega) - d_{s,r}(\omega)  \Bigg)  \nonumber \\ + \sum_{\omega} \sum _{s} \Big\langle \lambda_s (\omega) , \textbf{A}(m,\omega)\psi_s(\omega)-\textbf{S}_s (\omega) \Big\rangle _{\mathbf{x}},
    \label{eq:lagrangianclassicalFWI}
\end{eqnarray}
in which $\lambda$ is the Lagrange multiplier and $\langle .  \rangle_{\mathbf{x}}$ denotes the dot product on spatial coordinates $\mathbf{x}$.

The minimization of Eq.~\eqref{eq:lagrangianclassicalFWI} consists of computing the Lagrangian stationary point. Thus, taking the derivative of $\mathcal{L}(m, \psi_s , \lambda_s)$ with respect to model parameters, pressure wavefield and the Lagrange multiplier, we have:
\begin{equation}
\frac{\partial \mathcal{L}(m, \psi, \lambda)}{\partial m} =
\sum_\omega \sum_{s} \left(\frac{\partial \textbf{A}(m, \omega) \psi_s (\omega) }{\partial m}\right)^\dagger \lambda_s (\omega),
\label{eq:derivada_lagrangeano_m}
\end{equation}
\begin{equation}
\frac{\partial \mathcal{L}(m, \psi, \lambda)}{\partial \psi} =
\sum_\omega  \sum _{s,r} \Upsilon^\dagger\Big(\Upsilon \psi_s (\omega) - d_{s,r}(\omega)\Big) +
\sum_\omega \sum _{s} \textbf{A}^\dagger(m,\omega) \lambda_s(\omega),
\label{eq:derivada_lagrangeano_u}
\end{equation}
and
\begin{equation}
\frac{\partial \mathcal{L}(m, \psi, \lambda)}{\partial \lambda} =
\sum_\omega \sum_{s} \Big( \textbf{A}(m,\omega) \lambda_s (\omega) - \textbf{S}_s(\omega)\Big).
\label{eq:derivada_lagrangeano_v}
\end{equation}
If  $(\hat{m}, \hat{\psi}, \hat{\lambda})$ is at the stationary point of the Lagrangian \eqref{eq:lagrangianclassicalFWI}, the constraint in Eq.~\eqref{eq:FWIformulation_classical} is always satisfied after analyzing the latter equation \eqref{eq:derivada_lagrangeano_v}:
\begin{equation}
\frac{\partial \mathcal{L}(m, \psi, \lambda)}{\partial \lambda}\Biggr|_{\substack{m=\hat{m} \\\psi=\hat{\psi}\\\lambda=\hat{\lambda}}} = 0
\quad \implies \quad \textbf{A}(\hat{m},\omega) \hat{\psi}_s (\omega) = \textbf{S}_s (\omega),
\label{eq:derivada_lagrangeano_v2}
\end{equation}
which means the wave equation is solved in each FWI iteration.

Now, analyzing Eq.~\eqref{eq:derivada_lagrangeano_u} at the stationary point of the Lagrangian \eqref{eq:lagrangianclassicalFWI}, we obtain:
\begin{equation}
\frac{\partial \mathcal{L}(m, \psi, \lambda)}{\partial \psi}\Biggr|_{\substack{m=\hat{m} \\\psi=\hat{\psi}\\\lambda=\hat{\lambda}}} = 0 \quad
\implies \quad \textbf{A}^\dagger(\hat{m},\omega) \hat{\lambda}_s (\omega) =
- \sum _{r} \Upsilon^\dagger\Big(\Upsilon \hat{\psi}_s (\omega) - d_{s,r}(\omega)\Big),
\label{eq:derivada_lagrangeano_u2}
\end{equation}
which is a wave equation similar to Eq.~\eqref{eq:derivada_lagrangeano_v2}, but with wavefield $\lambda_s$ and the source term of the form $- \sum _{r} \Upsilon^\dagger\Big(\Upsilon \hat{\psi}_s (\omega) - d_{s,r}(\omega)\Big)$. 
Thus, the misfit function gradient is efficiently computed by solving the wave equation in only two moments (in the forward problem and in obtaining the Lagrange multiplier $\lambda$), while in the traditional approach the wave equation is calculated several times (for each element of the model space and according to the order of approximation of the derivative via finite differences).

In summary, to obtain the gradient of the misfit function, it is enough to compute the modeled data by solving the wave equation and correlating it (see Eq.~\eqref{eq:derivada_lagrangeano_m}) with the solution $\lambda$ of the wave equation in Eq.~\eqref{eq:derivada_lagrangeano_u2}:
\begin{equation}
\nabla_m \phi_1(m) := \frac{\partial \mathcal{L}(m, \psi, \lambda)}{\partial m} =
\sum_\omega \sum_{s} \omega^2 \psi_s^\dagger(m,\omega) \lambda_s (m,\omega),
\end{equation}
with
\begin{equation}
\textbf{A}^\dagger(m,\omega) \lambda_s (\omega) =
- \sum _{r} \Upsilon_{s,r}^\dagger\Big(\Upsilon \psi_s (m,\omega) - d_{s,r}(\omega)\Big).
\label{eq:adjoinwaveequation_classical}
\end{equation}
The latter equation is known as adjoint wave equation, in which the so-called adjoint wavefield $\lambda$ is computed by back propagating the errors (see Ref.~\cite{Plessix_AdjointReview_GJI_2006} for more details).

Although the classical approach is quite popular, it is doomed to fail if the errors are non-Gaussian. Indeed, if there are a handful of spurious measurements (outliers) in the dataset, the classical approach estimates biased parameters \cite{ClaerboutRobustErraticData1973,Constable_1988_geophysicalJI}. It is possible to see such behavior through the analysis of the adjoint-source (right-hand term in Eq.~\eqref{eq:adjoinwaveequation_classical}). In fact, if there is an outlier into the observed data ($d_{s,r} \rightarrow \infty$), we notice that the adjoint wavefield diverges ($\lambda_{s} \rightarrow \infty$) since infinite energy is inserted into the reverse wave propagation, and therefore $\nabla_m \phi_1(m) \rightarrow \infty$.

%%%%%%%%%%%%%%%%%%%%%%%%%%%%%%%%%%%%%%%%%%%%%
\subsection*{FWI based on Rényi $\alpha$-Gaussian distribution} %%
%%%%%%%%%%%%%%%%%%%%%%%%%%%%%%%%%%%%%%%%%%%%%

The Rényi entropy (or $\alpha$-entropy) \cite{Renyi1965informationTheory,renyi1961measures} is a one-parameter generalization of the classic BGS entropy, which is useful in information theory. For a continuous random variable $X$ with a probability density function $p(x)$, the $\alpha$-entropy is defined as:
\begin{equation}
        \mathcal{H}_\alpha\Big[p(x)\Big] = \frac{1}{1-\alpha} \ln\Bigg( \int p^\alpha(x) \hspace{.1cm}  dx\Bigg),
        \label{eq:RenyiEntropy}
\end{equation}
where $\alpha > 0$ and $\alpha \neq 1$. It is worth emphasizing that the BGS entropy is recovered at the limit $\alpha \rightarrow 1$:
\begin{equation}
        \lim_{\alpha\to1} \mathcal{H}_\alpha \overset{\Delta}{=} \lim_{\alpha\to1}  \frac{ -\int p^\alpha(x) \hspace{.1cm} \ln\big(p(x)\big) dx}{ \int p^\alpha(x) \hspace{.1cm}  dx} = -\int p(x) \hspace{.1cm} \ln\big(p(x)\big) dx = S_{BGS},
    \end{equation}
in which we have employed the L'Hôpital rule ($\overset{\Delta}{=}$).

From the maximum entropy principle (MEP) for the Rényi $\alpha$-entropy, several statistical distributions have emerged to model and describe a wide variety of complex systems, such as power-law decay in Hamiltonian systems \cite{PhysRevLett.93.130601} and statistical inference \cite{LEONENKO20101981}. In this work, we consider an optimal probability function which is derived from the maximization of the $\alpha$-entropy subject to the normalization condition:
\begin{equation}
    \int p(x) dx  = 1,
    \label{eq:normalizationcondition}
\end{equation}
and the unity variance
    \begin{equation}
        \int x^2 p(x) dx = 1.
        \label{eq:SecondMoment}
    \end{equation}
In this regard, the $\alpha$-generalized Gaussian probability distribution (or $\alpha$-Gaussian distribution) is a distribution function resulting from the MEP for the Rényi $\alpha$-entropy
(Eq.~\eqref{eq:RenyiEntropy}) subject to normalization condition (Eq.~\eqref{eq:normalizationcondition}) and the unity variance (Eq.~\eqref{eq:SecondMoment}) \cite{t_Student2003,t_Student_JOHNSON2007,RenyMaximum2019}, which is given by:
\begin{equation}
        p_\alpha(x) = Z_\alpha \Bigg[1-\Bigg(\frac{\alpha-1}{3\alpha-1}\Bigg) x^2\Bigg]^\frac{1}{\alpha-1}_+,
        \label{eq:alphaDistribution}
    \end{equation}
    where $[y]_+ = \max\{0,y\}$ and $Z_\alpha$ is the normalizing constant given by \cite{t_Student_JOHNSON2007}: 
    \begin{equation}
        Z_\alpha = \sqrt{\frac{1-\alpha}{\pi(3\alpha-1)}} \hspace{.1cm} \frac{\Gamma\big(\frac{1}{1-\alpha}\big)}{\Gamma\big(\frac{1+\alpha}{2(1-\alpha)}\big)}
\end{equation}
for $\frac{1}{3} <\alpha < 1$, in which $\Gamma$ represents the Gamma Function.

By considering that the errors $x$ are independent and identically distributed by the $\alpha$-Gaussian distribution (Eq.~\eqref{eq:alphaDistribution}), we obtain the $\alpha$-misfit function using the probabilistic maximum log-likelihood function: 
    \begin{equation}
        \Theta_\alpha(x)=-\ln\Bigg(\prod_{i=1}^N p_{\alpha}(x_i)\Bigg) = -\ln\Bigg\{\prod_{i=1}^N A_\alpha\Bigg[1-\Bigg(\frac{\alpha-1}{3\alpha-1}\Bigg)x^2_i\Bigg]^\frac{1}{\alpha-1}_+\Bigg\},
        \label{eq:alpha_likelihood}
    \end{equation}
which can be written as:
    \begin{equation}
        \Theta_\alpha(x)+ \ln[A_\alpha] = \frac{1}{1-\alpha} \sum_{i=1}^N \ln\Bigg[ 1-\Bigg(\frac{\alpha-1}{3\alpha-1}\Bigg)x^2_i\Bigg]_+,
        \label{eq:alpha_likelihood2}
    \end{equation}
in which $x = \{ x_1, x_2, x_3, \cdot\cdot\cdot, x_N\}$. We notice that maximizing the latter equation \eqref{eq:alpha_likelihood2} is equivalent to minimizing the following function:
    \begin{equation}
        \phi_\alpha(x) = \frac{1}{1-\alpha} \sum_{i=1}^N \ln\Bigg[ 1-\Bigg(\frac{\alpha-1}{3\alpha-1}\Bigg)x^2_i\Bigg]_+.
        \label{eq:alpha_likelihood3}
    \end{equation}

In this way, the FWI based on $\alpha$-Gaussian distribution (hereafter $\alpha$-FWI) is formulated as the following minimization task:
\begin{equation}
\underset{m, \psi }{\min } \hspace{0.2cm} \frac{1}{1-\alpha} \sum_{\omega} \sum_{s,r} \ln\Bigg[ 1-\Bigg(\frac{\alpha-1}{3\alpha-1}\Bigg)\Bigg( \Upsilon \psi_s(\omega) - d_{s,r}(\omega) \Bigg)^\dagger \Bigg( \Upsilon \psi_s(\omega) - d_{s,r}(\omega)  \Bigg)\Bigg]_+
\label{eq:alphamisfitfWI}
\end{equation}
subject to the wave equation, $\textbf{A}(m,\omega) \psi_s(\omega) = \textbf{S}_s(\omega)$. Thus, the solution of the optimization problem formulated in \eqref{eq:alphamisfitfWI} can be done by minimizing the following augmented Lagrangian $\alpha$-functional:
\begin{align}
    \label{eq:lagrangianalphaFWI}
    \mathcal{L}_\alpha(m, \psi, \Lambda)  &= \frac{1}{1-\alpha} \sum_{\omega} \sum_{s,r} \ln\Bigg[ 1-\Bigg(\frac{\alpha-1}{3\alpha-1}\Bigg)\Bigg( \Upsilon \psi_s(\omega) - d_{s,r}(\omega) \Bigg)^\dagger \Bigg( \Upsilon \psi_s(\omega) - d_{s,r}(\omega)  \Bigg)\Bigg]_+ \\ 
    &+ \sum_{\omega} \sum _{s} \Big\langle \Lambda_s (\omega) , \textbf{A}(m,\omega)\psi_s(\omega)-\textbf{S}_s (\omega) \Big\rangle _{\mathbf{x}},& \nonumber%\qedhere
\end{align}
where $\Lambda$ is a Lagrangian multiplier.

Computing the stationary point of  Eq.~\eqref{eq:lagrangianalphaFWI}, we have:
\begin{equation}
\frac{\partial \mathcal{L}_\alpha (m, \psi, \Lambda)}{\partial m} =
\sum_\omega \sum_{s} \left(\frac{\partial \textbf{A}(m, \omega) \psi_s (\omega) }{\partial m}\right)^\dagger \Lambda_s (\omega),
\label{eq:derivada_lagrangeano_m_alpha}
\end{equation}
\begin{eqnarray}
\frac{\partial \mathcal{L}_\alpha(m, \psi, \Lambda)}{\partial \psi} =
\sum_\omega  \sum _{s,r} \frac{2 \Upsilon^\dagger\Big(\Upsilon \psi_s (\omega) - d_{s,r}(\omega)\Big)}{3\alpha - 1 -(\alpha-1)\Big(\Upsilon \psi_s (\omega) - d_{s,r}(\omega)\Big)^\dagger \Big(\Upsilon \psi_s (\omega) - d_{s,r}(\omega)\Big)} \nonumber \\ +
\sum_\omega \sum _{s} \textbf{A}^\dagger(m,\omega) \Lambda_s(\omega),
\label{eq:derivada_lagrangeano_u_alpha}
\end{eqnarray}
and
\begin{equation}
\frac{\partial \mathcal{L}_\alpha(m, \psi, \Lambda)}{\partial \Lambda} =
\sum_\omega \sum_{s} \Big( \textbf{A}(m,\omega) \Lambda_s (\omega) - \textbf{S}_s(\omega)\Big).
\label{eq:derivada_lagrangeano_v_alpha}
\end{equation}
We notice that Eqs.~\eqref{eq:derivada_lagrangeano_m_alpha} and \eqref{eq:derivada_lagrangeano_v_alpha} are exactly Eqs.~\eqref{eq:derivada_lagrangeano_m} and \eqref{eq:derivada_lagrangeano_v}, respectively. This result was already expected, since the wave equation constraint  is intrinsic to FWI. However, we notice that Eq.~\eqref{eq:derivada_lagrangeano_u_alpha} associated with the pressure wavefield is quite different from the one resulting from the classical approach (Eq.~\eqref{eq:derivada_lagrangeano_u}). In this way, if  $(\hat{m}, \hat{\psi}, \hat{\Lambda})$ is at the stationary point of the Lagrangian \eqref{eq:lagrangianalphaFWI}, we obtain:
\begin{equation}
\frac{\partial \mathcal{L}_\alpha(m, \psi, \Lambda)}{\partial \psi}\Biggr|_{\substack{m=\hat{m} \\\psi=\hat{\psi}\\\Lambda=\hat{\Lambda}}} = 0,
\end{equation}
which yields the following $\alpha$-adjoint equation,
\begin{equation}
\textbf{A}^\dagger(\hat{m},\omega) \hat{\Lambda}_s (\omega) =
- \sum _{r} \frac{2 \Upsilon^\dagger\Big(\Upsilon \psi_s (\omega) - d_{s,r}(\omega)\Big)}{3\alpha - 1 -(\alpha-1)\Big(\Upsilon \psi_s (\omega) - d_{s,r}(\omega)\Big)^\dagger \Big(\Upsilon \psi_s (\omega) - d_{s,r}(\omega)\Big)}.
\label{eq:derivada_lagrangeano_u2_alpha}
\end{equation}
Note that Eq.~\eqref{eq:derivada_lagrangeano_u2_alpha} becomes Eq.~\eqref{eq:derivada_lagrangeano_u2} in the limit $\alpha \rightarrow 1$.

Thus, the $\alpha$-misfit function is given by:
\begin{equation}
\underset{m}{\min } \hspace{0.2cm} \phi_\alpha(m) =
\frac{1}{1-\alpha} \sum_{\omega} \sum_{s,r} \ln\Bigg[ 1-\Bigg(\frac{\alpha-1}{3\alpha-1}\Bigg)\Bigg( \Upsilon \psi_s(m,\omega) - d_{s,r}(\omega) \Bigg)^\dagger \Bigg( \Upsilon \psi_s(m,\omega) - d_{s,r}(\omega)  \Bigg)\Bigg]_+,
\label{eq:final_misfitfunction}
\end{equation}
valid for $\frac{1}{3} < \alpha < 1$, in which the gradient of the $\alpha$-misfit function is computed by solving the wave equation and correlating it with the solution $\Lambda$ of the wave equation in Eq.~\eqref{eq:derivada_lagrangeano_u2_alpha}:
\begin{equation}
\nabla_m \phi_\alpha(m) := \frac{\partial \mathcal{L}_\alpha(m, \psi, \Lambda)}{\partial m} =
\sum_\omega \sum_{s} \omega^2 \psi_s^\dagger(m,\omega) \Lambda_s (m,\omega),
\end{equation}
with
\begin{equation}
\textbf{A}^\dagger(\hat{m},\omega) \hat{\Lambda}_s (\omega) =
- \sum _{r} \frac{2 \Upsilon^\dagger\Big(\Upsilon \psi_s (m,\omega) - d_{s,r}(\omega)\Big)}{3\alpha - 1 -(\alpha-1)\Big(\Upsilon \psi_s (m,\omega) - d_{s,r}(\omega)\Big)^\dagger \Big(\Upsilon \psi_s (m,\omega) - d_{s,r}(\omega)\Big)},
\label{eq:adjoinwaveequation_classical_alpha}
\end{equation}
where the classical misfit function \eqref{eq:inverseproblemformulation_2} is a particular case in the $\alpha \rightarrow 1$ limit-case.

In contrast to the classical approach, the $\alpha$-misfit function mitigates the effects of non-Gaussian errors, specially outliers. It is possible to see such behavior through the analysis of the $\alpha$-adjoint-source (right-hand term in Eq.~\eqref{eq:adjoinwaveequation_classical_alpha}). In fact, if there is an outlier into the observed data ($d_{s,r} \rightarrow \infty$), we notice that the adjoint wavefield tends to zero ($\Lambda_{s} \rightarrow 0$), and therefore $\nabla_m \phi_\alpha(m) \rightarrow 0$ in that case.  Indeed, the $\alpha$-misfit function magnifies small errors and suppresses large one, since the $\alpha$-adjoint-source  \eqref{eq:adjoinwaveequation_classical_alpha} is proportional to the errors $\Lambda_{s} \propto \Delta d_{s,r} = \Upsilon\psi_{s}(m) - d_{s,r}$ for small errors and inverse of the errors $\Delta d_{s,r}$, $\Lambda_{s} \propto 1/\Delta d_{s,r}$ for large ones. The classical approach, on the other hand, magnificent $\alpha \rightarrow 1$) linearly suppresses small errors and magnifies large errors ($\lambda_{s} \propto \Delta d_{s,r}$), which explain the sensitivity of this approach to non-Gaussian errors  \cite{ClaerboutRobustErraticData1973}.

%%%%%%%%%%%%%%%%%%%%%%%%%%%%%%%%%%%%%%%%%%%%%%%%%%%%%%%%%%%%%%%
%%%%%%%%%%%%%%%%%%%%%%%%%%%%%%%%%%%%%%%%%%%%%%%%%%%%%%%%%%%%%%%
\subsection*{Target-oriented FWI using PGF techniques}
%%%%%%%%%%%%%%%%%%%%%%%%%%%%%%%%%%%%%%%%%%%%%%%%%%%%%%%%%%%%%%%
%%%%%%%%%%%%%%%%%%%%%%%%%%%%%%%%%%%%%%%%%%%%%%%%%%%%%%%%%%%%%%%

As discussed earlier, solving the wave equation is a critical issue of the FWI technique. In this way, PGF technique appears as a powerful alternative to reduce the computational cost of imaging problems based on wave equation. In the PGF framework, the solution of the frequency-domain wave equation (see, for instance, the constraint in Eq.~\eqref{eq:FWIformulation_classical}) is equivalent to solving the following linear system of equations \cite{Moura2020,almeida2020}:
\begin{equation}\label{impedance}
    \mathbf{G}^{-1}\boldsymbol{\psi}_s = \frac{F_\omega}{h^2}\mathbf{U}_s,
\end{equation}
where the impedance matrix \textbf{A} is equivalent to the inverse of the Green function $\mathbf{G}^{-1}$, $\boldsymbol{\psi}_s$ is the discretized pressure wavefield, and $\frac{F_\omega}{h^2}\mathbf{U}_s = F_s(\omega) \, \delta\left(\mathbf{x}-\mathbf{x}_s\right)$ is the source term in which $h$ is the grid spacing (in meters), $F_\omega$ is the amplitude of the Fourier coefficients of the seismic source and $\mathbf{U}_{s}$ is a column vector with zero elements everywhere except on the source position $\textbf{x}_s$ representing $\delta\left(\mathbf{x}-\mathbf{x}_s\right)$. 

%By considering that the Green function in Eq.~\eqref{impedance} is invertible, we may apply the so-called Dyson equation to compute $\mathbf{G}^{-1}$ \cite{doniach1974green,economou2006,sheng2006introduction}. In this regard, the total Green function $\mathbf{G}$ is associated to a self-energy $\mathbf{g}$ by employing a connection potential term $\mathbf{V}$. We highlight that the connection potential is a very important term in the Dyson equation, as it is responsible for connecting the elements necessary for the calculation of the recursive process, which will be described later. In this way, the impedance matrix $\mathbf{G}^{-1}$ can be described for the entire computational domain \cite{Moura2020,FERREIRA2002355}:
%\begin{equation}
  %  \label{eq:Dyson1}
 %   \mathbf{G}^{-1} = \mathbf{g}^{-1} - \mathbf{V}. 
%\end{equation} 
%\textcolor{red} {A principal motivação  para construirmos um propagador de onda usando o método (PGF)  tem como objetivo principal, melhorar o tempo  computacional de problemas relacionados ao FWI orientado a uma região alvo. O FWI convencional, mesmo sendo direcionado a uma região alvo, ele gera uma atualização do campo de onda  em todo o modelo computacional, isso leva a um alto custo computacional, pois sabemos que para reconstruir um modelo de velocidade é necessário realizar várias atualização}

It is worth noting that the PGF technique is very useful and powerful since the wave equation solution computational cost is drastically reduced and, consequently, the target-oriented FWI runtime is reduced without losing quality in the seismic data inversion process. In contrast, the conventional FWI formulation is very costly from a computational point of view, since it requires the computation of the wavefield for the entire physical domain.

%\textcolor{red}{O método PGF é  bem utilizada em física da matéria condensada para o cálculo do transporte de ondas eletrônicas [ \cite{PhysRevB.90.115408, PhysRevB.91.125408}], essa metodologia nos permiti  calcular a função de Green total de um  sistema, através do uso da Equação de Dyson } 
The PGF method is widely used in condensed matter physics for modeling transport of electronic waves \cite{PhysRevB.90.115408, PhysRevB.91.125408} through the calculation of the complete Green's function $\mathbf{G}$ of a system using the so-called Dyson equation. By considering that the Green function in Eq.~\eqref{impedance} is invertible, we may apply the Dyson equation to compute $\mathbf{G}^{-1}$ \cite{doniach1974green,economou2006,sheng2006introduction}. In this regard, the total Green's function $\mathbf{G}$ is associated to a self-energy $\mathbf{g}$ by employing a connection potential term $\mathbf{V}$. We highlight that the connection potential is a very important term in the Dyson equation, as it is responsible for connecting the elements necessary for the calculation of the recursive process, which will be described later. In this way, the impedance matrix $\mathbf{G}^{-1}$ can be described for the entire computational domain \cite{Moura2020,FERREIRA2002355}:
\begin{equation}
    \label{eq:Dyson}
     \mathbf{G} = \mathbf{g} + \mathbf{gVG}.
\end{equation} 
The idea behind Dyson equation is, to make the connection between the main elements of the matrix referring to each Green's function, in which this connection process is carried out through the connection potential \textbf{V}.

In its original formulation, the PGF technique is only valid for wave propagation in a homogeneous medium. However, Ref.~\cite{Moura2020} reformulated this technique for propagating acoustic waves in disordered (non-homogeneous) media. In this regard, the solution of the wave equation in \eqref{impedance} is efficiently solved by computing the following Green functions \cite{almeida2020}:
\begin{equation}
    \mathbf{G}_{es} = \left(\textbf{1}-\mathbf{g}_{ee}\mathbf{V}_{ei} \mathbf{g}_{ii}\mathbf{V}_{ie}\right)^{-1}\mathbf{g}_{es},   \label{G-g1}
\end{equation}
\begin{equation}
    \mathbf{G}_{ts} = \mathbf{g}_{ti}\mathbf{V}_{ie}\mathbf{G}_{es},\label{G-t}
    \end{equation}
\begin{equation}    
    \mathbf{G}_{rs} = \mathbf{g}_{rs}+  \mathbf{g}_{re}\mathbf{V}_{ei}\mathbf{g}_{ii}\mathbf{V}_{ie}\mathbf{G}_{es}.
    \label{G-g}
\end{equation}
where the Green function is computed just in the target region (represented by index \textit{t}) and at the position of sources (\textit{s}) and receivers (\textit{r}). The subscripts $e$ and $i$ refer, respectively, to the elements of the outer edge and the inner edge of the target region, connected through the employment of the potential $\textbf{V}_{i,e}$.  Equations~\eqref{G-g1} and \eqref{G-g} represent the Green function computed from the position of sources to receivers. Equation~\eqref{G-t} is the Green function.

In summary, the PGF method is performed in two steps: First, we find the element that connect the edges of the target  $e$ and $i$. In this regard, we calculate the matrices $\mathbf{g}_{rs}$ $\mathbf{g}_{es}$ and $\mathbf{g}_{ee}$ at outside the target area. In the second step called "fill-in", we compute speed terms on the diagonal of the impedance matrix $\mathbf{G}^{-1}$ through the connection potential V in all seismic inversion process. Then, we compute the elements of the Green function relating the inner edge $\mathbf{g}_{ii}$ of the target region with the points within the target area $\mathbf{g}_{ti}$. The split-in step is performed throughout the inversion process (see Ref.~\cite{FWIPGF2021} for more details). 

%%%%%%%%%%%%%%%%%%%%%%%%%%%%%%%%%%%%%%%%%%%%%%%
%%%%%%%%%%%%%%%%%%%%%%%%%%%%%%%%%%%%%%%%%%%%%%%
\section{Numerical Experiments \label{sec:numericalexample}}
%%%%%%%%%%%%%%%%%%%%%%%%%%%%%%%%%%%%%%%%%%%%%%%
%%%%%%%%%%%%%%%%%%%%%%%%%%%%%%%%%%%%%%%%%%%%%%%

To illustrate how the $\alpha$-PGF-FWI deals with non-Gaussian errors, we present two numerical examples. In the first one, we consider a Marmousi case study, in which the main goal consists of investigating the robustness of the $\alpha$-FWI regarding outliers in the dataset. Then, in the second one, we present an application of the  $\alpha$-PGF-FWI to estimates the model parameters in the context of typical Brazilian pre-salt reservoirs. In all numerical experiments, we consider the \textit{limited}-\textit{memory} Broyden-Fletcher-Goldfarb-Shanno (\textit{l}-BFGS) algorithm \cite{ByrdNocedalDetails} to solve the minimization problems (Eq.~\eqref{eq:methodquasiNewton}), which is a very efficient quasi-Newton method for deal with large-scale optimization tasks \cite{ByrdNocedalDetails}. The \textit{l}-BFGS fetches an informative (local) minimum of the misfit function using its gradient $\nabla_m \phi(m)$. In this way, the \textit{l}-BFGS computes an approximation of the inverse Hessian matrix $\textbf{H}^{-1}$ from previous gradients by imposing a secant condition \cite{ByrdNocedalDetails}. Moreover, we compute the step-length $\beta$ along the descent-direction search of the gradient which satisfies the Wolfe conditions \cite{wolfeOriginal}.  In addition, we consider a Ricker wavelet as the seismic source \cite{rickerSource}, which is defined as: $f_s(t)=(1-2\pi^2\mu_p^2t^2) \exp(-\pi^2\mu_p^2t^2)$, where $\mu_p$ is the peak frequency.

%%%%%%%%%%%%%%%%%%%%%%%%%%%%%%%%%%%%%%%%%%%%%%%%%%%%%%%%%%%%%%%%%%%%%%%%%
%%%%%%%%%%%%%%%%%%%%%%%%%%%%%%%%%%%%%%%%%%%%%%%%%%%%%%%%%%%%%%%%%%%%%%%%%
\subsection*{Marmousi case study: robustness to erratic data}
%%%%%%%%%%%%%%%%%%%%%%%%%%%%%%%%%%%%%%%%%%%%%%%%%%%%%%%%%%%%%%%%%%%%%%%%%
%%%%%%%%%%%%%%%%%%%%%%%%%%%%%%%%%%%%%%%%%%%%%%%%%%%%%%%%%%%%%%%%%%%%%%%%%

\begin{figure}[]
\begin{subfigure}{.5\textwidth}
  \centering
  \caption{}
  \includegraphics[width=\linewidth]{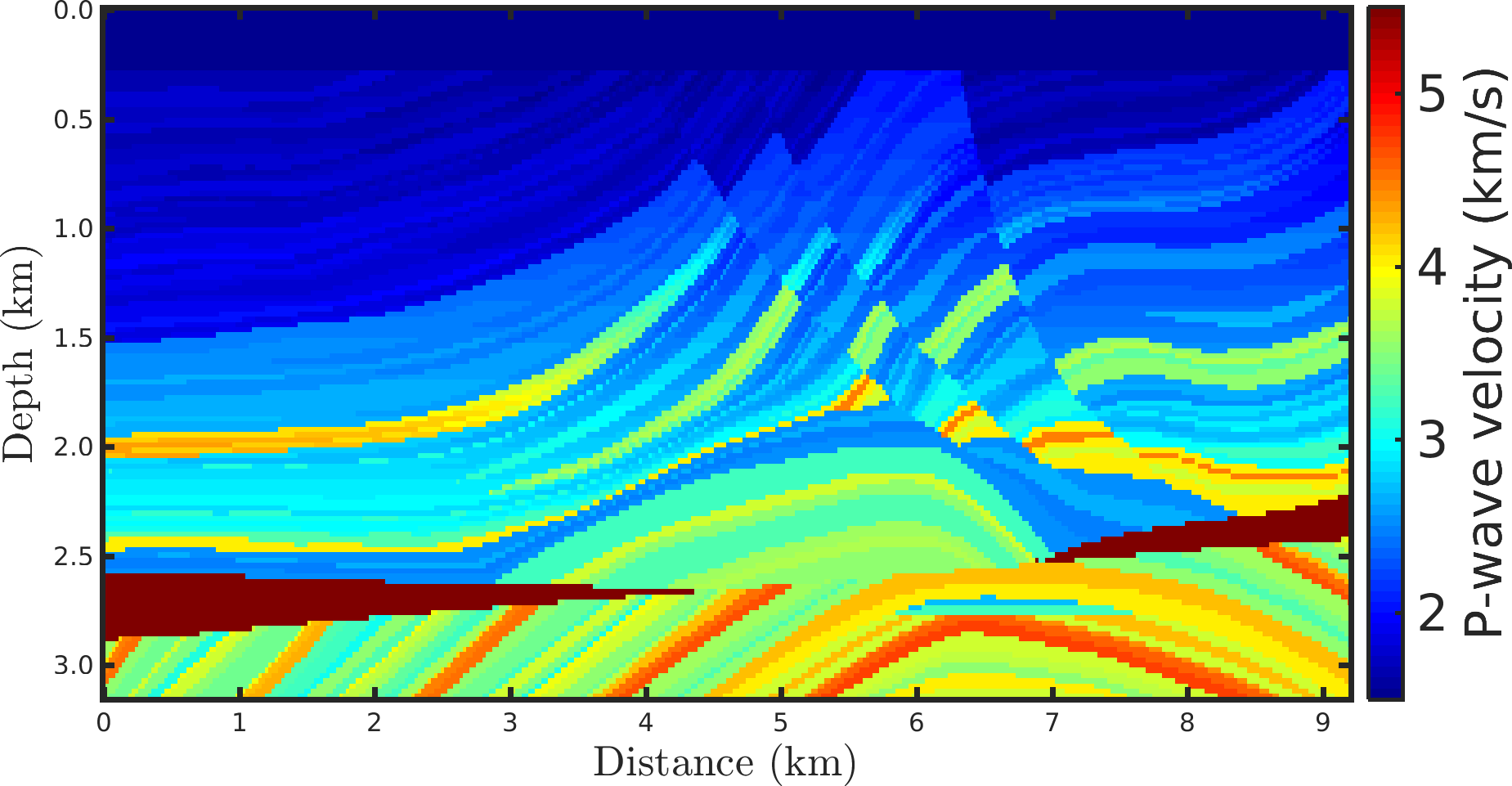}
  \label{fig:true_model_marmousi_test}
\end{subfigure}
\begin{subfigure}{.5\textwidth}
  \centering
  \caption{}
  \includegraphics[width=\linewidth]{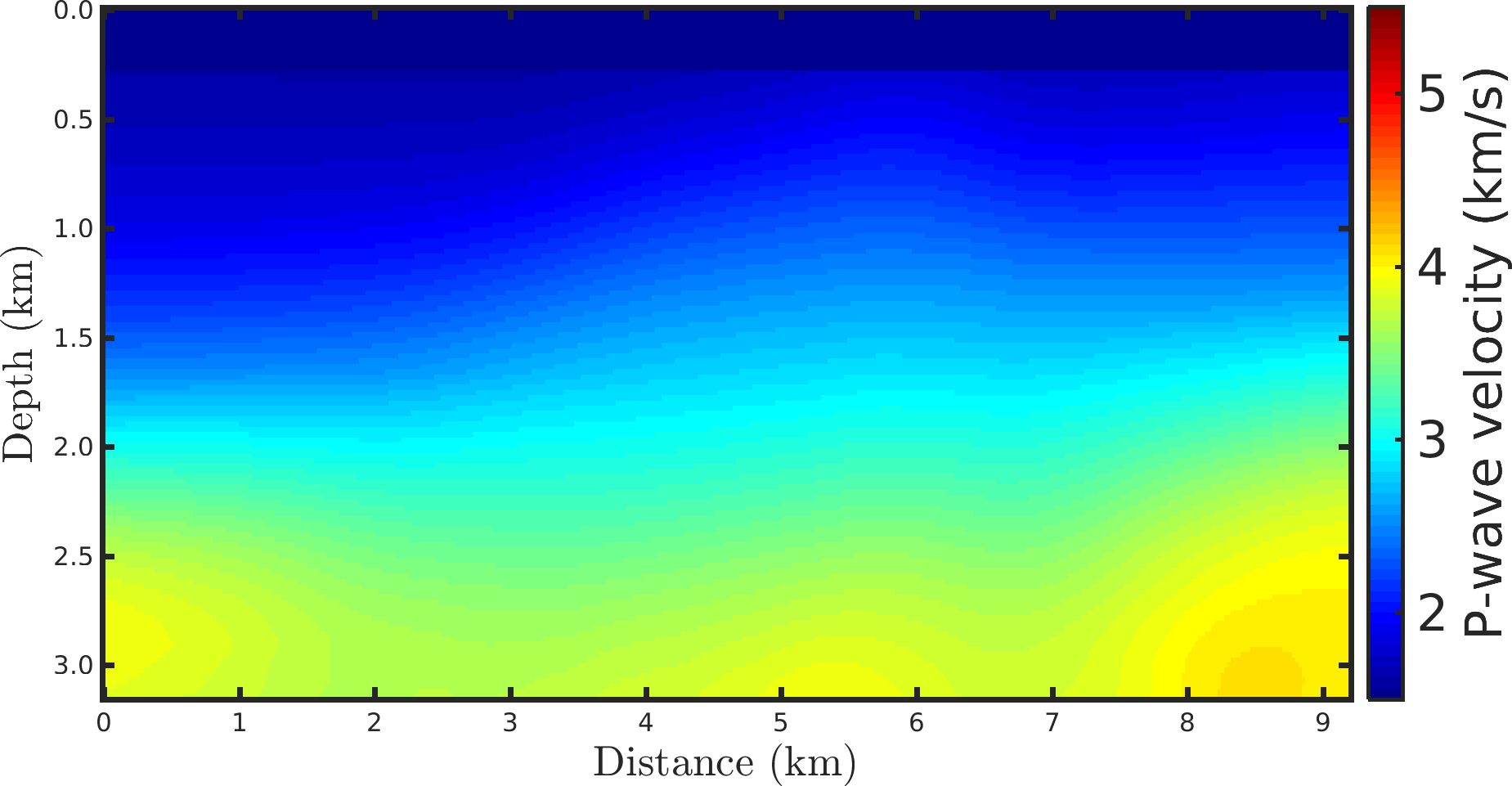}
  \label{fig:initial_model_marmousi_test}
\end{subfigure}
\caption{(a) Realistic P-wave velocity model used as  the ground truth (true model) representing the Kwanza Basin in Angola, which is known as Marmousi model. (b) Initial model employed in the $\alpha$-FWI tests.}
\label{fig:used_models_marmousi_test}
\end{figure}

\begin{figure}[]
\begin{subfigure}{.24\textwidth}
  \centering
  \caption{}
  \includegraphics[width=\linewidth]{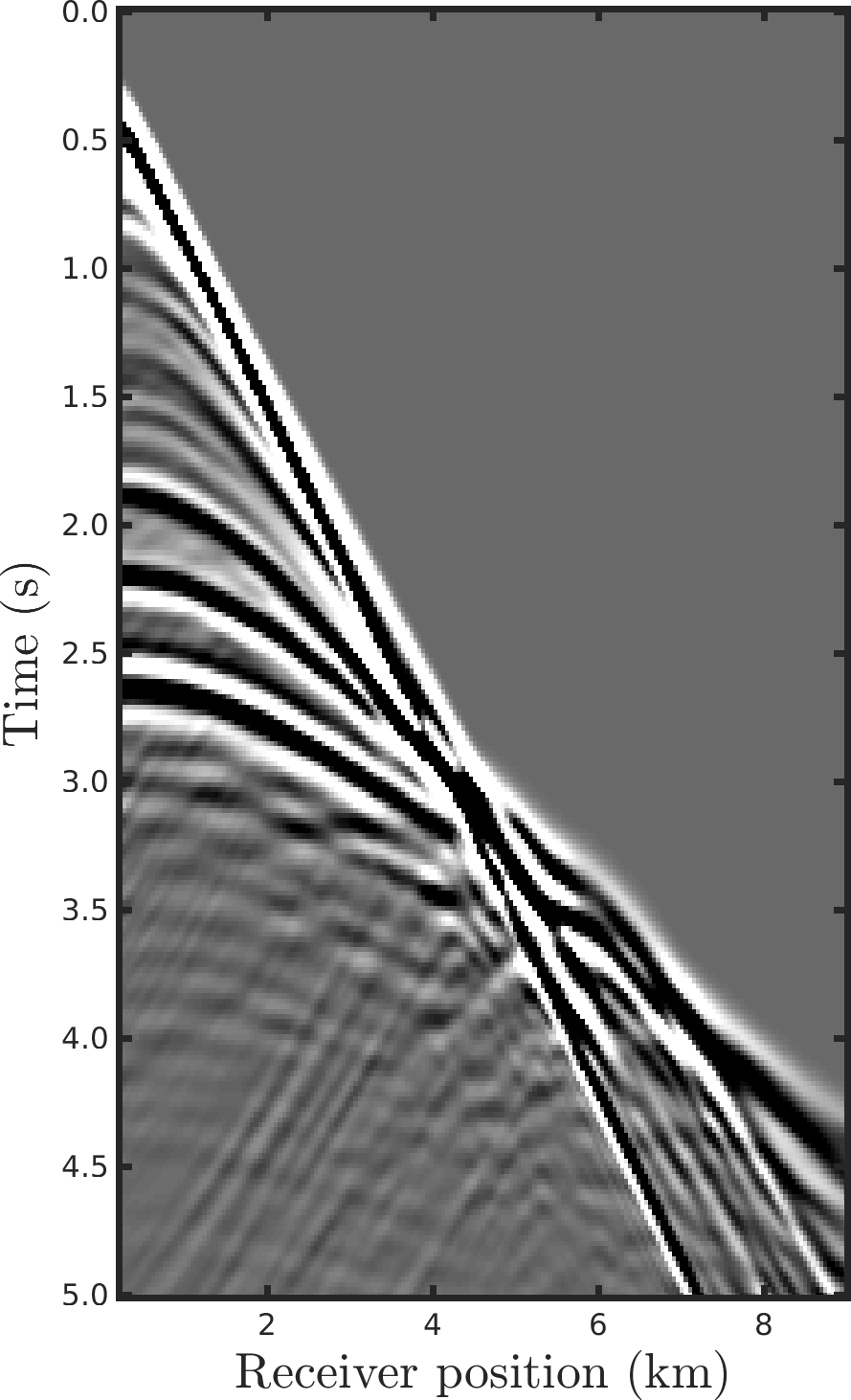}
  \label{fig:noiseless_seismograms_marmousi_test}
\end{subfigure}
\begin{subfigure}{.24\textwidth}
  \centering
  \caption{}
  \includegraphics[width=\linewidth]{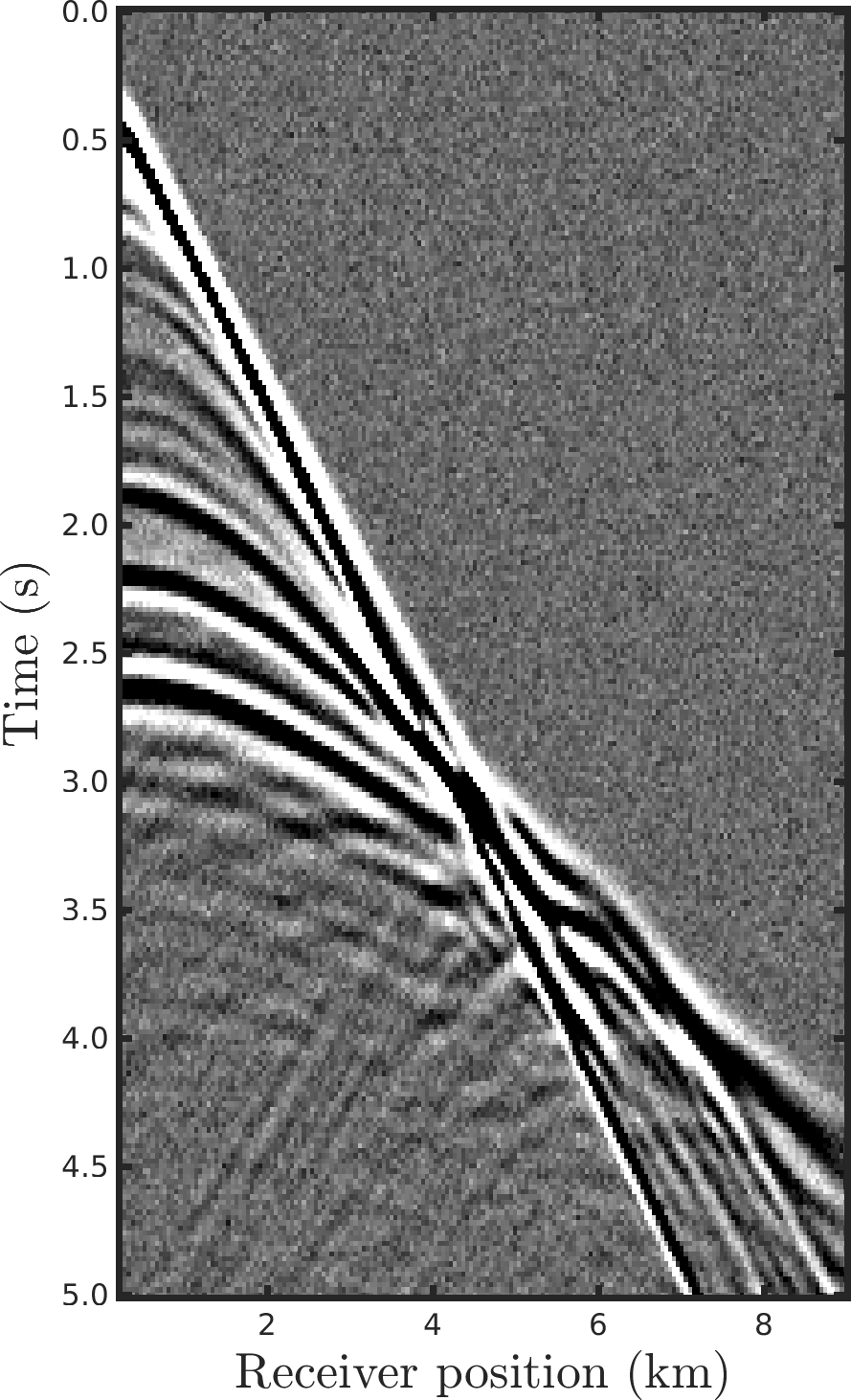}
  \label{fig:snr70_seismograms_marmousi_test}
\end{subfigure}
\begin{subfigure}{.24\textwidth}
  \centering
  \caption{}
  \includegraphics[width=\linewidth]{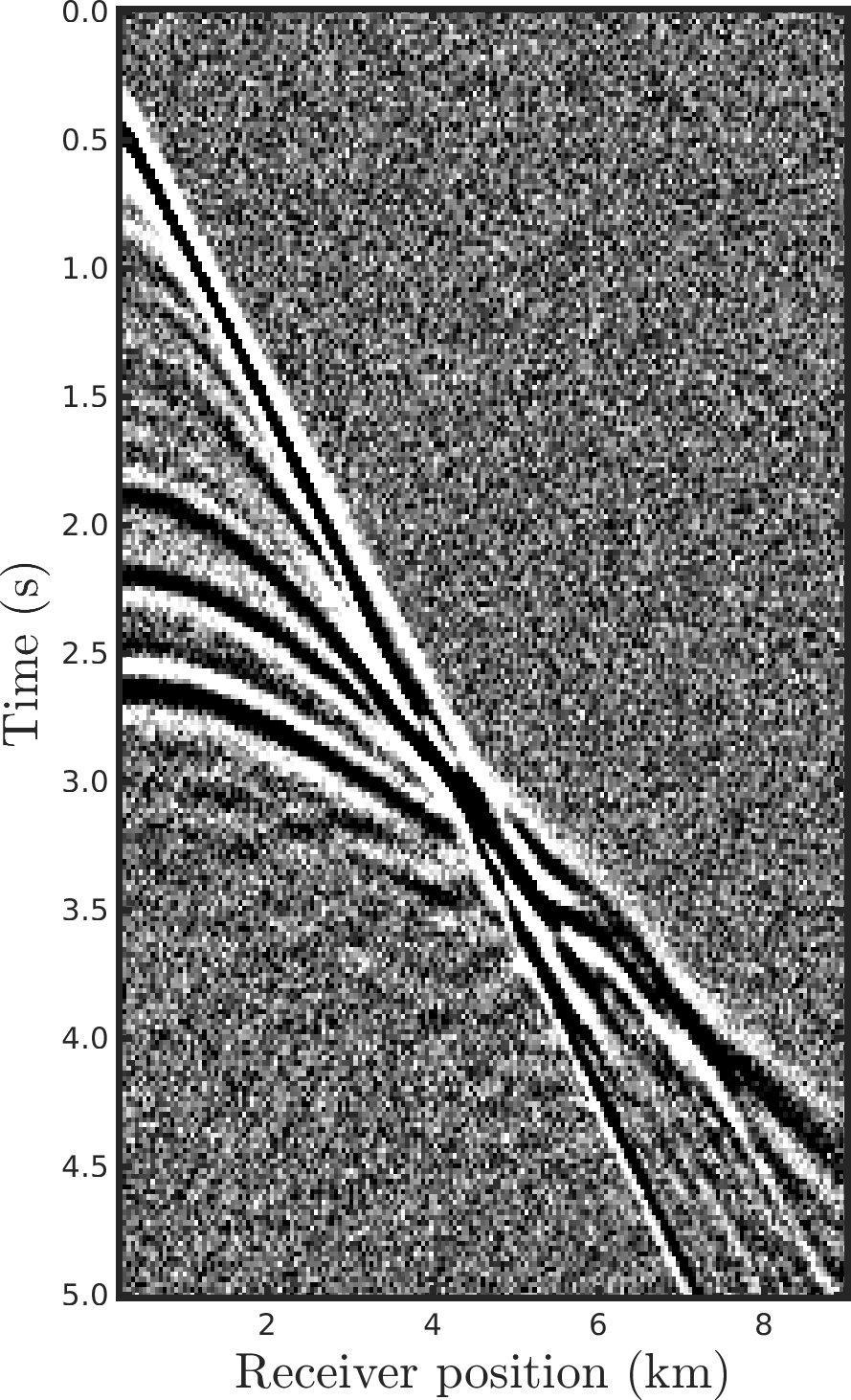}
  \label{fig:snr60_seismograms_marmousi_test}
\end{subfigure}
\begin{subfigure}{.24\textwidth}
  \centering
  \caption{}
  \includegraphics[width=\linewidth]{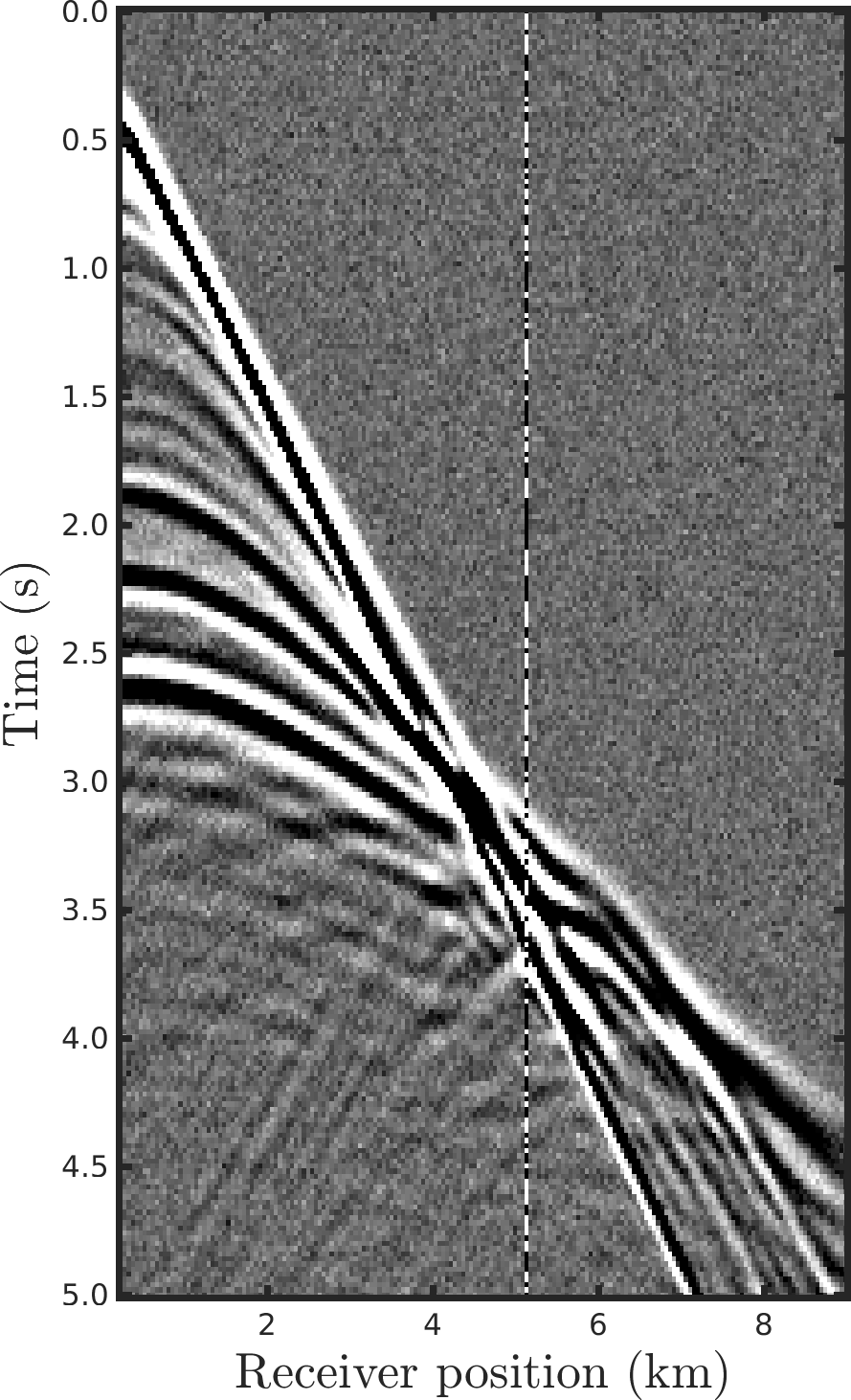}
  \label{fig:snr70_spike_seismograms_marmousi_test}
\end{subfigure}
\caption{Seismograms (shot-gathers) generated through a seismic survey with the first seismic source at lateral distance $216m$, for the (a) noiseless data case; the Gaussian noise cases with (b) $SNR = 70dB$ and (c) $SNR = 60dB$; and (d) the Gaussian noise case $SNR = 70dB$ and erratic data (outliers).}
\label{fig:used_seismograms_marmousi_test}
\end{figure}

To analyze the robustness of the $\alpha$-FWI regarding erratic data (outliers), we consider a 2D acoustic velocity model which is widely used in geophysical imaging tests, named the Marmousi model. Such a model presents a complex velocity geometry, as depicted in Fig.~\ref{fig:true_model_marmousi_test}, and it is based on a realistic region of the Kwanza Basin in Angola \cite{marmousi}. By using the Marmousi model (true model), Fig.~\ref{fig:true_model_marmousi_test}, we generate a seismic dataset considering a fixed-spread acquisition at $24m$ in-depth with $183$ equally spaced seismic sources each $48m$, from $216m$ to $8952m$, in which we employ a Ricker wavelet with $\mu_p = 5Hz$  as the seismic source. Furthermore, we consider $23$ equally spaced receivers located every $408m$, from $96m$ to $9072m$, deployed at the ocean floor ($240m$ in-depth) to simulate an ocean bottom nodes (OBN) acquisition, which is a marine seismic acquisition very employed in the last years. The acquisition time was $5s$. Figure~\ref{fig:used_seismograms_marmousi_test} shows some examples of shot-gathers generated by the first seismic sources.

We performed FWI taking into account a noiseless data scenario, to validated our algorithms, and three noisy data scenarios: in the two first ones, we consider a dataset polluted by Gaussian noise, and them, in the last one, a seismic dataset contaminated by spiky-noise (outliers) and Gaussian noise to simulate a realistic circumstance like to Ref.~\cite{Elboth_Geophysics_74_SweelNoise_2009}. In the case of the first two scenarios, we consider seismic data contaminated by white Gaussian background noise with signal-to-noise ratio (SNR) of $70dB$ and $60dB$, respectively, as depicted in Figs.~\ref{fig:snr70_seismograms_marmousi_test} and \ref{fig:snr60_seismograms_marmousi_test}. We notice that the SNR is computed by the ratio between the noiseless observed data power and the amplitude noise power. In the fourth scenario, we consider a data set contaminated by Gaussian noise with $SNR = 70dB$ and that only one of the seismic traces from each receiver is contaminated by outliers (see vertical line in Fig.~\ref{fig:snr70_spike_seismograms_marmousi_test}).

For each noisy-scenario, we carried out eight data inversions, in which the first one refers to the classical approach ($\alpha \rightarrow 1$), and the last seven ones are based on the $\alpha$-misfit function with $\alpha = 0.35$, $0.45$, $0.55$, $0.65$, $0.75$, $0.85$ and , $0.95$. In all numerical experiment, we computed $200$ \textit{l}-BFGS iterations from the above-described geometry seismic acquisition and the initial model depicted in Fig.~\ref{fig:initial_model_marmousi_test}.

\begin{figure}[]
\begin{subfigure}{.5\textwidth}
  \centering
  \caption{classical FWI ($\alpha \rightarrow 1$)}
  \includegraphics[width=\linewidth]{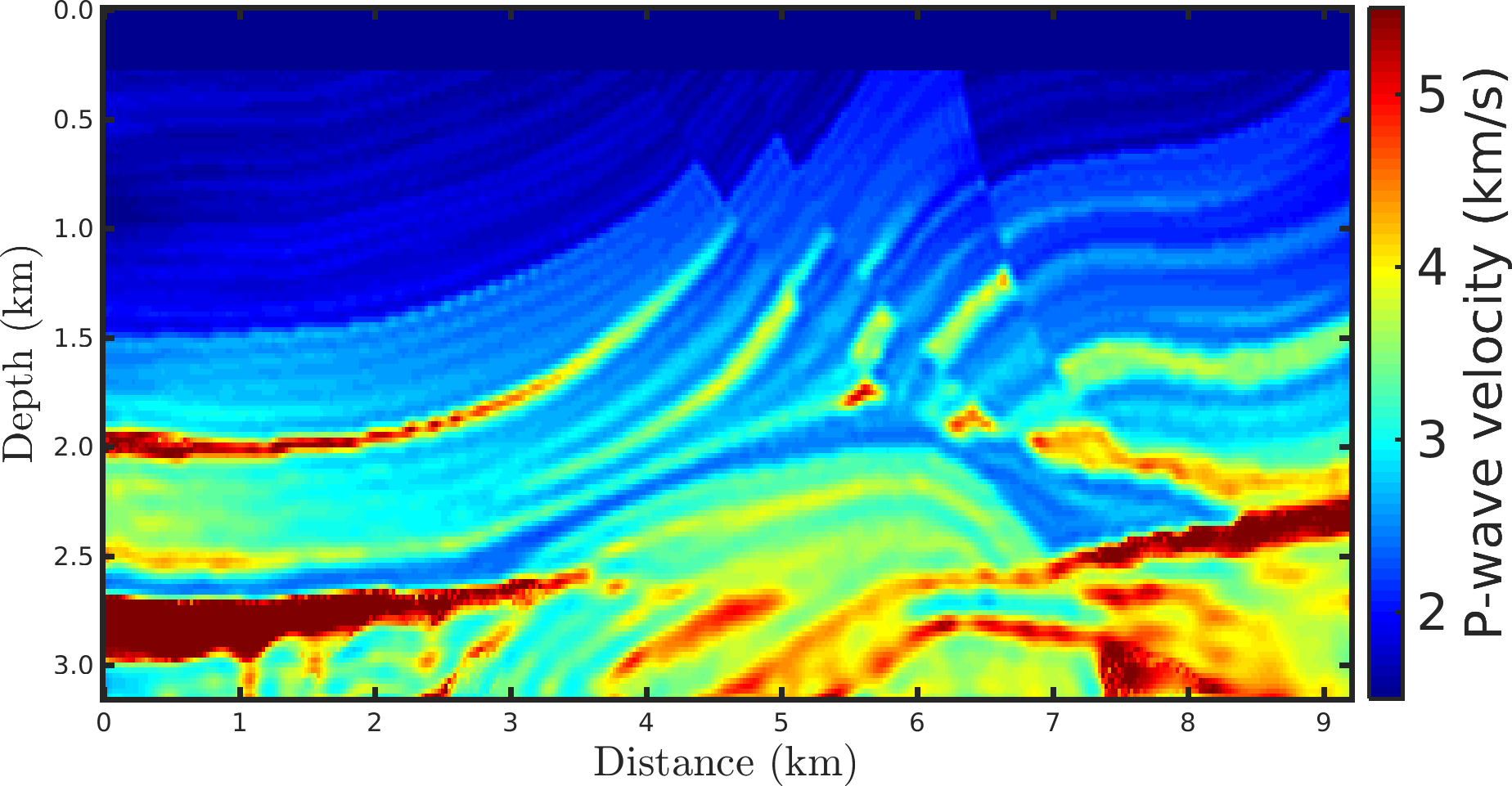}
  \label{fig:noiseless_data_classicalFWI_timedomain}
\end{subfigure}
\begin{subfigure}{.5\textwidth}
  \centering
  \caption{$\alpha$-FWI  with $\alpha = 0.95$}
  \includegraphics[width=\linewidth]{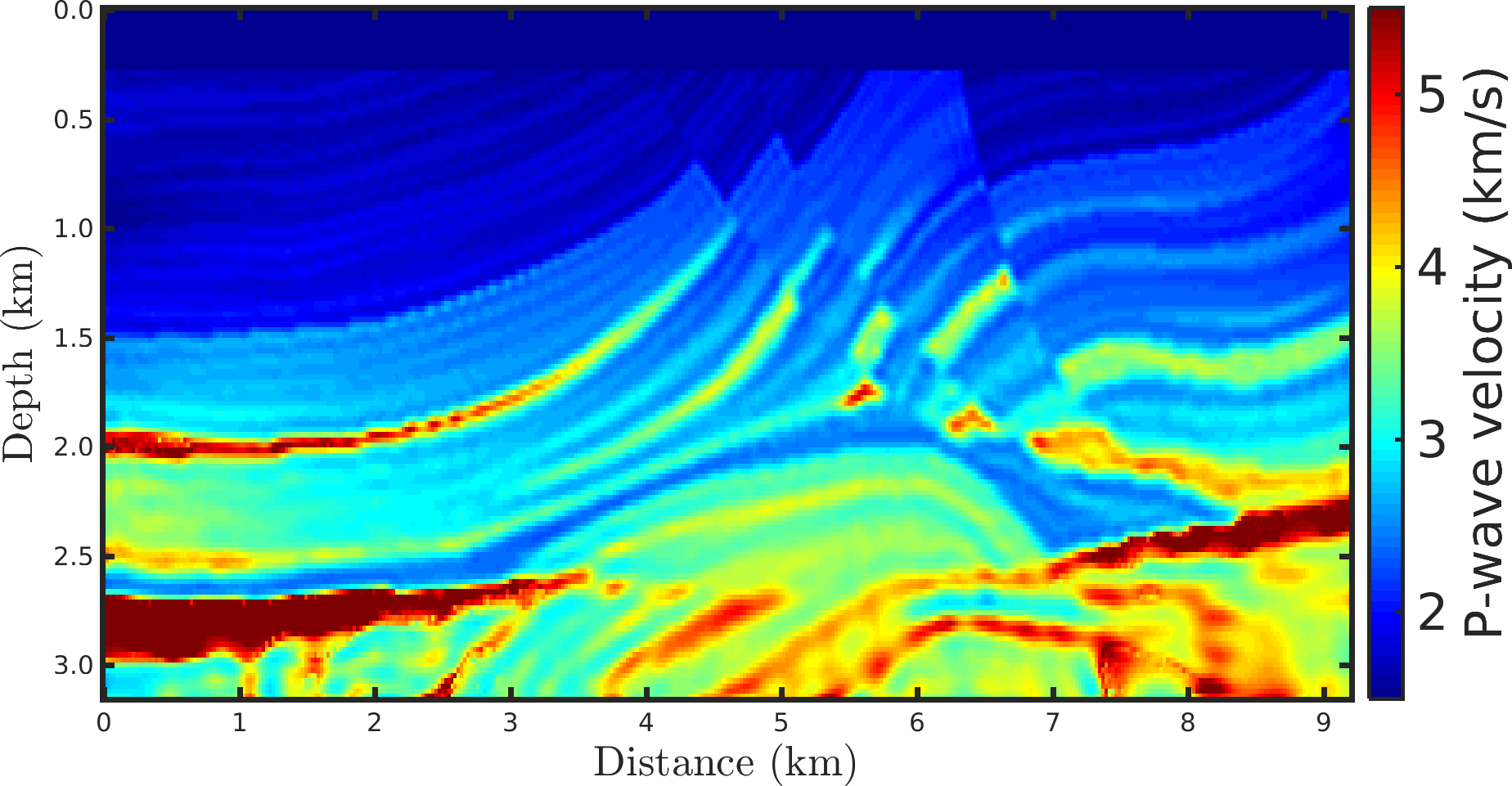}  
  \label{fig:noiseless_data_aFWI_timedomain_a_0_95}
\end{subfigure}
\begin{subfigure}{.5\textwidth}
  \centering
  \caption{$\alpha$-FWI  with $\alpha = 0.85$}
  \includegraphics[width=\linewidth]{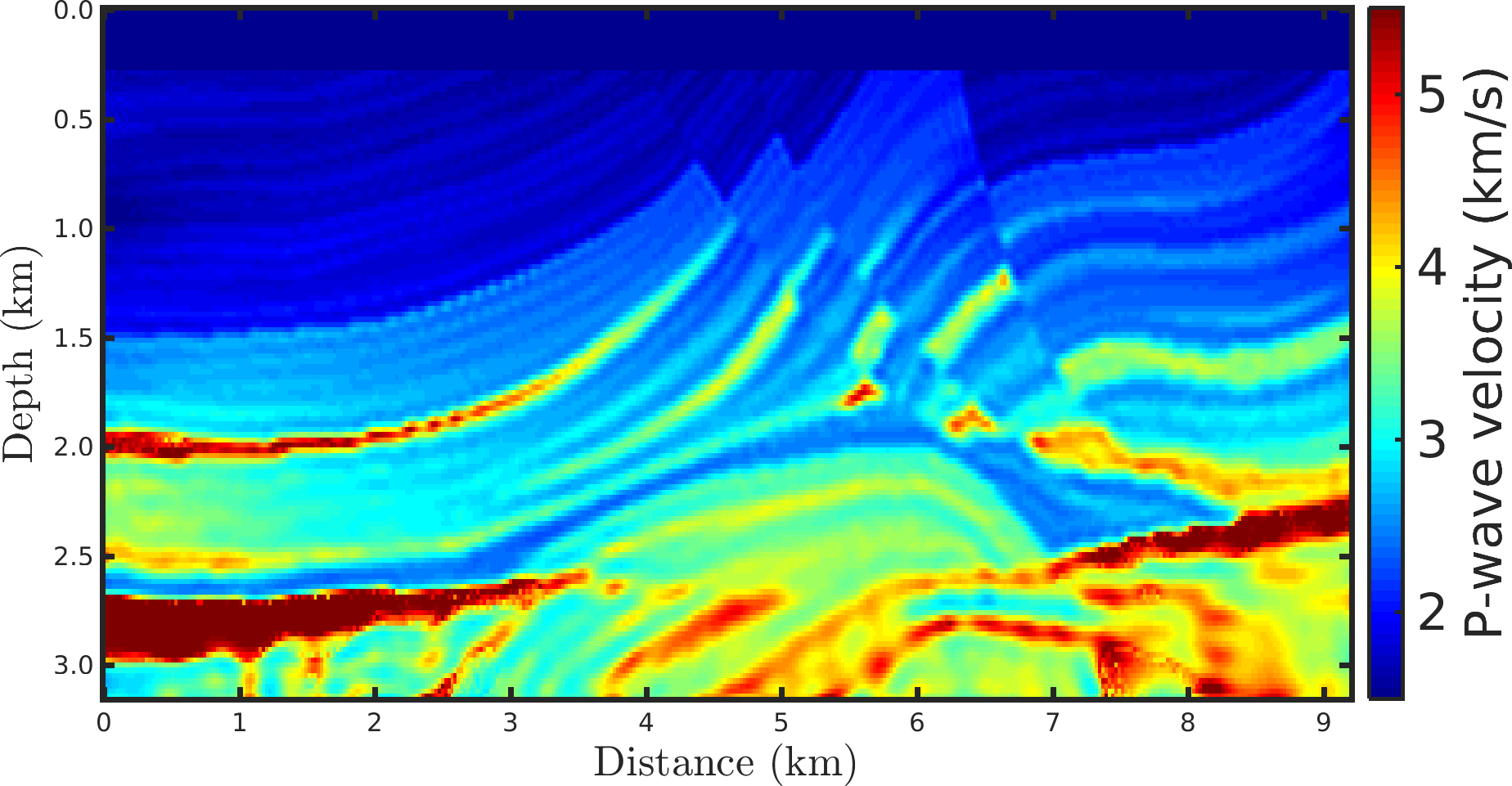}
  \label{fig:noiseless_data_aFWI_timedomain_a_0_85}
\end{subfigure}
\begin{subfigure}{.5\textwidth}
  \centering
  \caption{$\alpha$-FWI  with $\alpha = 0.75$}
  \includegraphics[width=\linewidth]{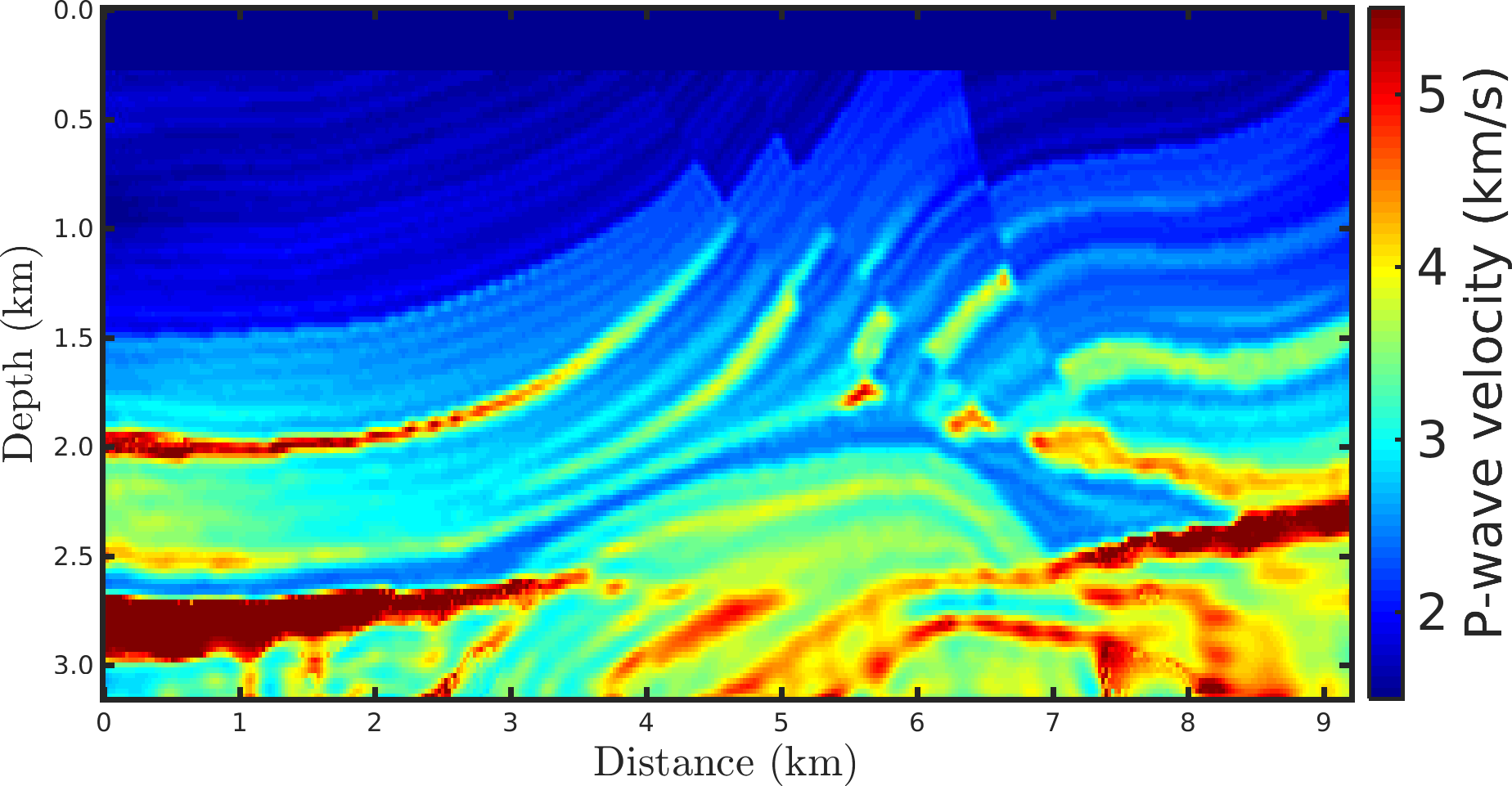}  
  \label{fig:noiseless_data_aFWI_timedomain_a_0_75}
\end{subfigure}
\begin{subfigure}{.5\textwidth}
  \centering
  \caption{$\alpha$-FWI  with $\alpha = 0.65$}
  \includegraphics[width=\linewidth]{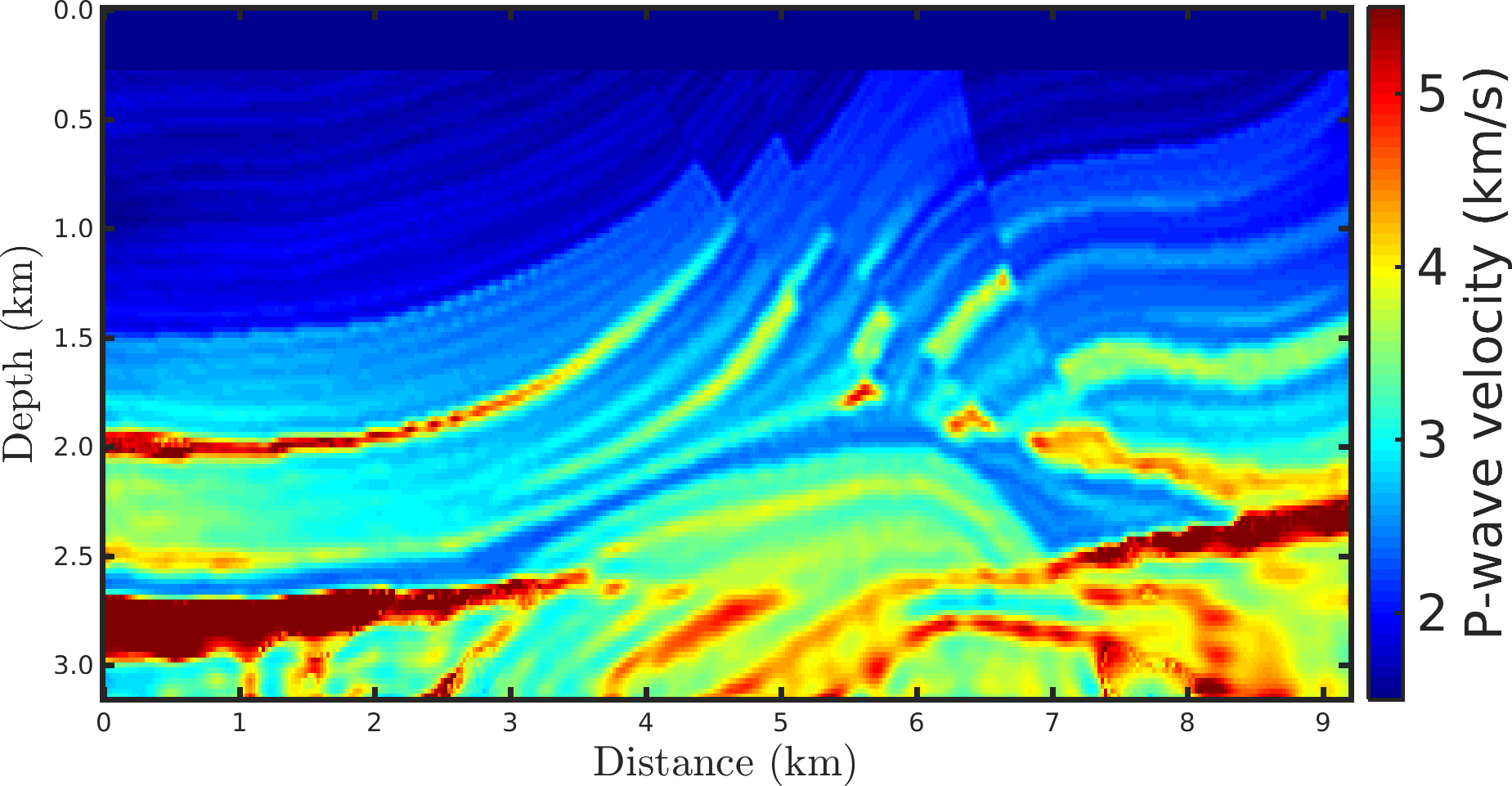}
  \label{fig:noiseless_data_aFWI_timedomain_a_0_65}
\end{subfigure}
\begin{subfigure}{.5\textwidth}
  \centering
  \caption{$\alpha$-FWI  with $\alpha = 0.55$}
  \includegraphics[width=\linewidth]{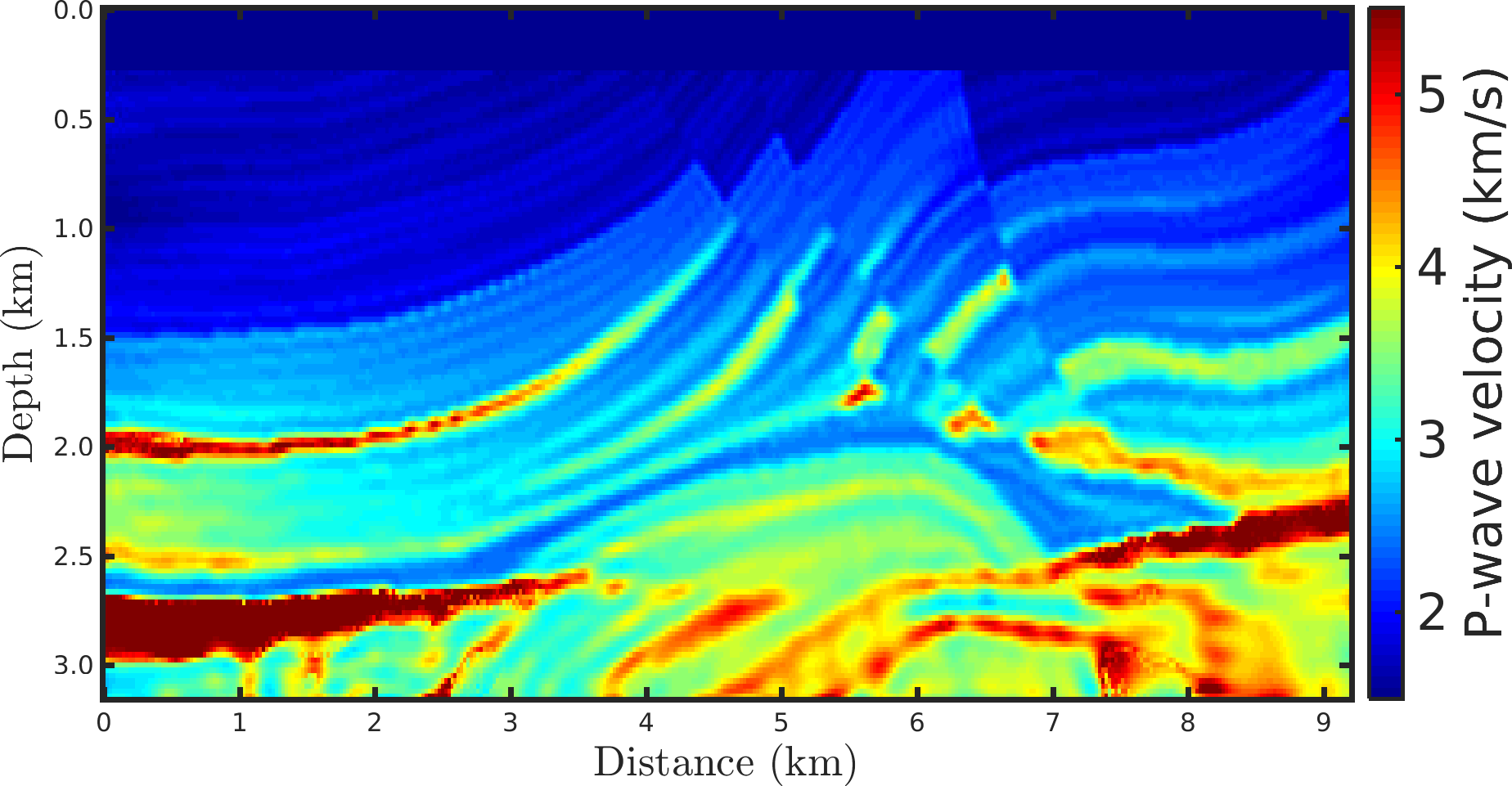}  
  \label{fig:noiseless_data_aFWI_timedomain_a_0_55}
\end{subfigure}
\begin{subfigure}{.5\textwidth}
  \centering
  \caption{$\alpha$-FWI  with $\alpha = 0.45$}
  \includegraphics[width=\linewidth]{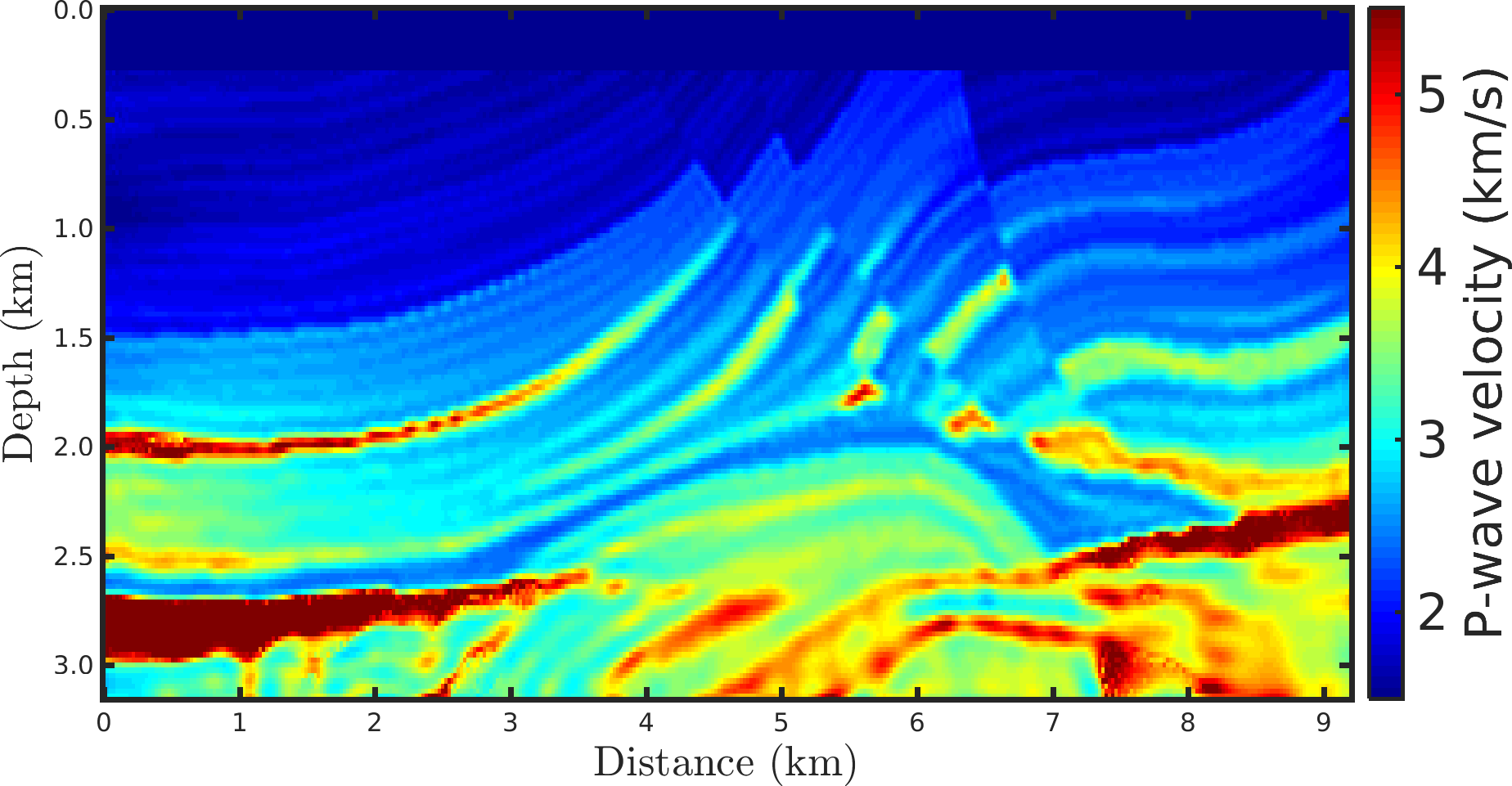}
  \label{fig:noiseless_data_aFWI_timedomain_a_0_45}
\end{subfigure}
\begin{subfigure}{.5\textwidth}
  \centering
  \caption{$\alpha$-FWI  with $\alpha = 0.35$}
  \includegraphics[width=\linewidth]{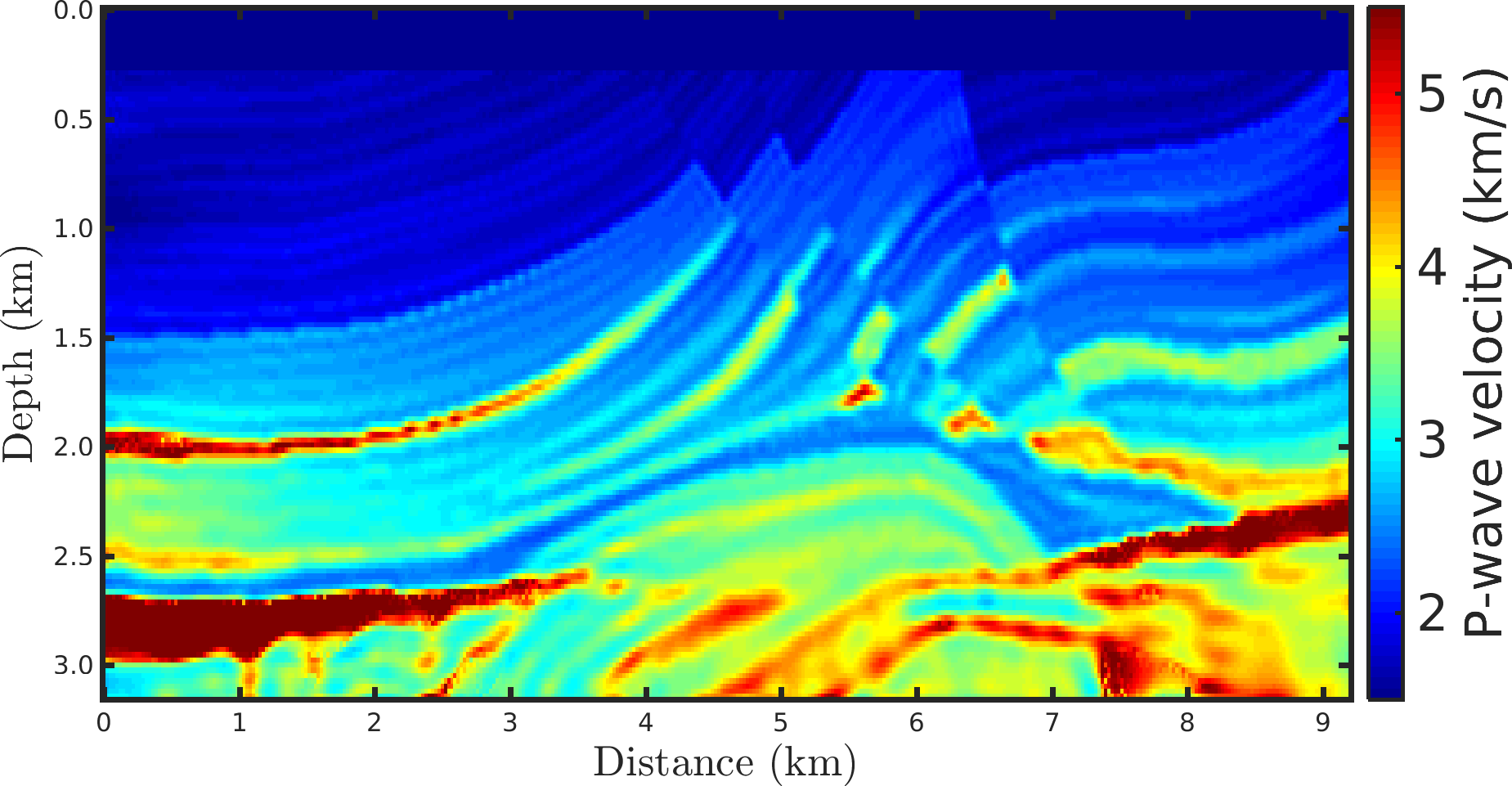}  
  \label{fig:noiseless_data_aFWI_timedomain_a_0_35}
\end{subfigure}
\caption{Reconstructed P-wave velocity models in the noiseless data case (first scenario) for the (a) classical FWI ($\alpha \rightarrow 1$), and $\alpha$-FWI with (b) $\alpha = 0.95$, (c) $\alpha = 0.85$, (d) $\alpha = 0.75$, (e) $\alpha = 0.65$, (f) $\alpha = 0.55$, (g) $\alpha = 0.45$, and (h) $\alpha = 0.35$.}
\label{fig:noiseless_data_aFWI_timedomain}
\end{figure}

Figure~\ref{fig:noiseless_data_aFWI_timedomain} shows the FWI results for the first scenario, in which is remarkable that the results are satisfactory, since the reconstructed velocity models are close to the true model (Fig.~\ref{fig:true_model_marmousi_test}) regardless of the $\alpha$-value. Such results were already expected once the observed data is not corrupted. We include this scenario just to demonstrate that our algorithms are in fine working order. To quantitatively compare the reconstructed models with the true model, we consider two statistical measures: (i) the Pearson product-moment correlation coefficient (or Pearson's \textit{R}, for short); and (ii) the normalized root-mean-square (\textit{NRMS}). The Pearson's \textit{R} measure the linear correlation between the reconstructed model and the true model, in which it varies between $-1$ and $1$, inclusive. In this regard, $R = 1$ means a strong correlation between the two models and, in our case, \textit{R}-values far from 1 mean low correlation. The \texttt{NRMS} varies from 0 (perfect model) to  $\infty$ (bad model) and is defined as:
\begin{equation}
    \texttt{NRMS} = \Bigg[\frac{\sum_i \big(c_i^{true}-c_i^{FWI}\big)^2}{\sum_i \big(c_i^{true}\big)^2}\Bigg]^{1/2}
\end{equation}
where $c^{true}$ is true model and $c^{FWI}$ corresponds to the  FWI result. The statistical measures for the noiseless data case (first scenario) are summarized in Table~\ref{tab:noiseless_timeFWI}. Indeed, all reconstructed models are strongly correlated with the true model ($R \geq 0.8$, according to the strength-scale suggested by Ref.~\cite{evans1996_PearsonClassifications}), although the cases $\alpha = 0.55$ and $\alpha = 0.65$ have the highest Pearson's \textit{R} and the lowest \textit{NRMS} value.

\begin{table}[]
\centering
\caption{Statistical measures between the true model and the reconstructed models in the noiseless data case (first scenario). {\footnotesize The Pearson's \textit{R} measures the linear correlations between the models, and the \textit{NRMS} measures the misfit between the true model and the reconstructed models.}}
\vspace{0.05cm}
\begin{tabular}{lccc}
\hline            
Strategy & $\alpha$-value & R & NRMS \\
\hline                               
classical FWI & $\alpha \rightarrow 1.0$ & $0.9129$ & $0.0203$ \\
\hline                               
& $\alpha = 0.95$ & $0.9164$ & $0.0193$ \\
& $\alpha = 0.85$ & $0.9154$ & $0.0195$ \\
& $\alpha = 0.75$ & $0.9163$ & $0.0193$ \\
$\alpha$-FWI & $\alpha = 0.65$ & $0.9176$ & $0.0190$ \\
& $\alpha = 0.55$ & $0.9177$ & $0.0190$ \\
& $\alpha = 0.45$ & $0.9133$ & $0.0201$ \\
& $\alpha = 0.35$ & $0.9124$ & $0.0203$ \\
\hline
\hline
\end{tabular}
\label{tab:noiseless_timeFWI}
\end{table}

\begin{figure}[]
\begin{subfigure}{.5\textwidth}
  \centering
  \caption{classical FWI ($\alpha \rightarrow 1$)}
  \includegraphics[width=\linewidth]{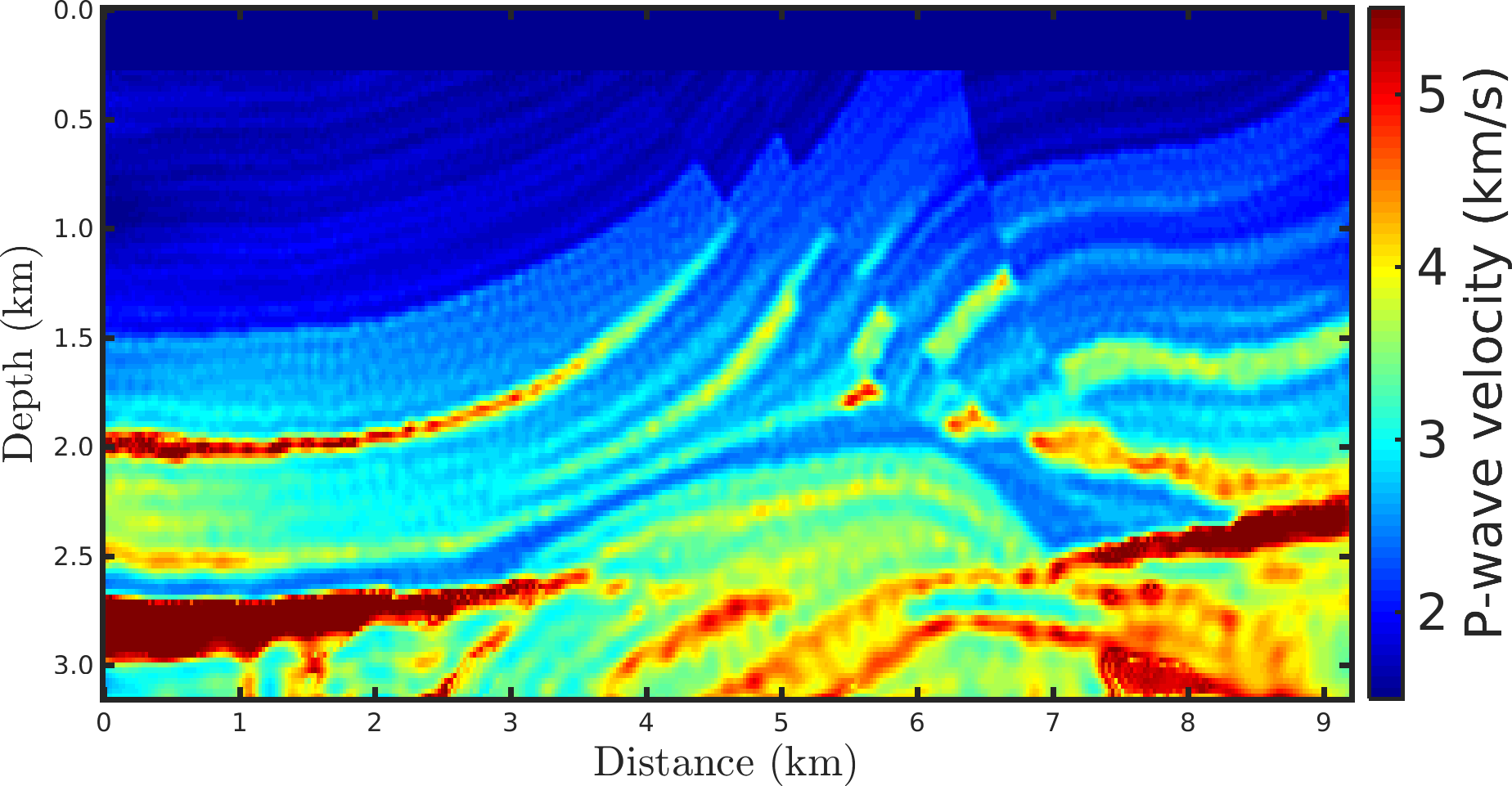}
  \label{fig:gauss_noise_70_data_classicalFWI_timedomain}
\end{subfigure}
\begin{subfigure}{.5\textwidth}
  \centering
  \caption{$\alpha$-FWI  with $\alpha = 0.95$}
  \includegraphics[width=\linewidth]{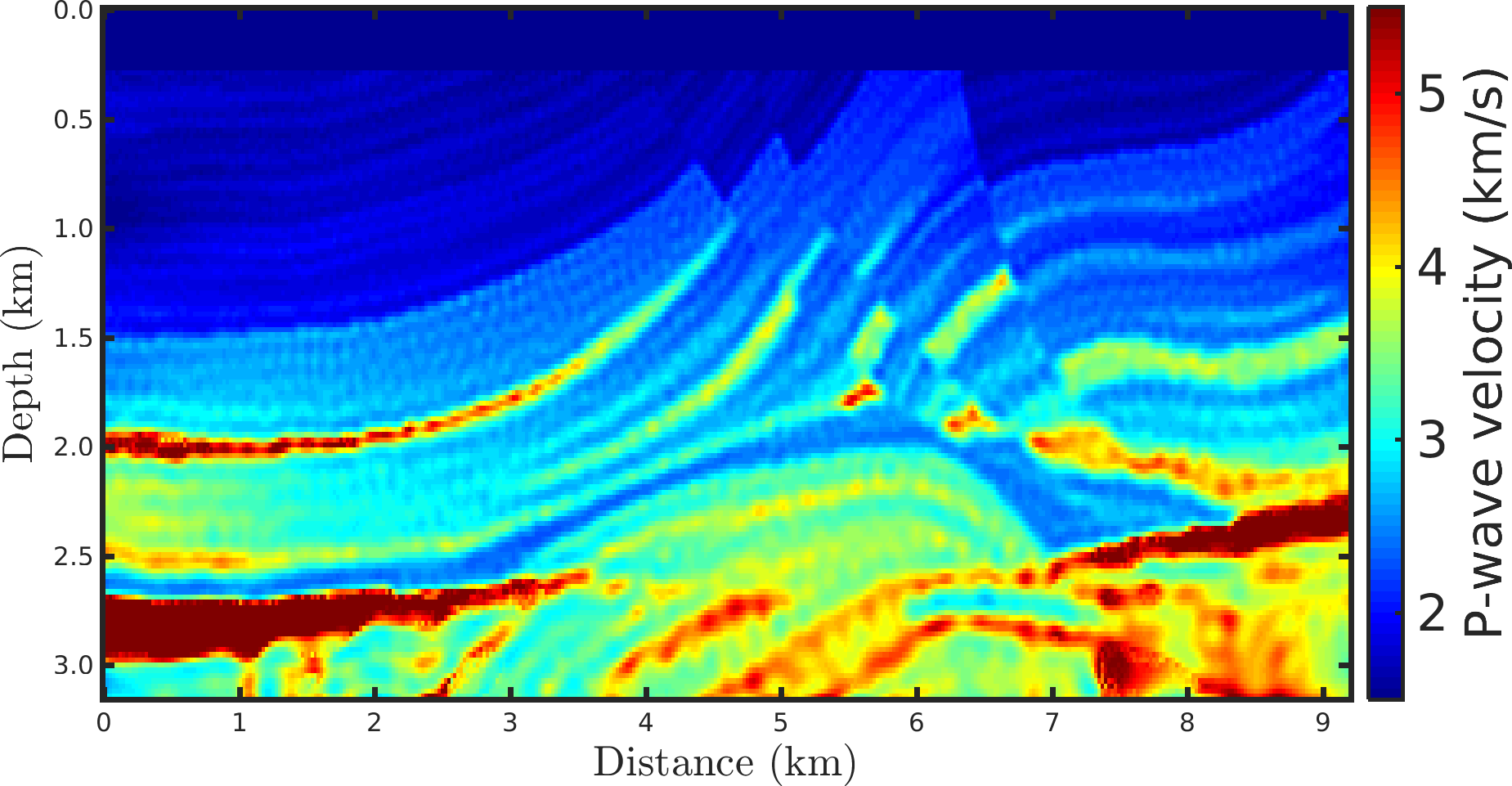}  
  \label{fig:gauss_noise_70_data_aFWI_timedomain_a_0_95}
\end{subfigure}
\begin{subfigure}{.5\textwidth}
  \centering
  \caption{$\alpha$-FWI  with $\alpha = 0.85$}
  \includegraphics[width=\linewidth]{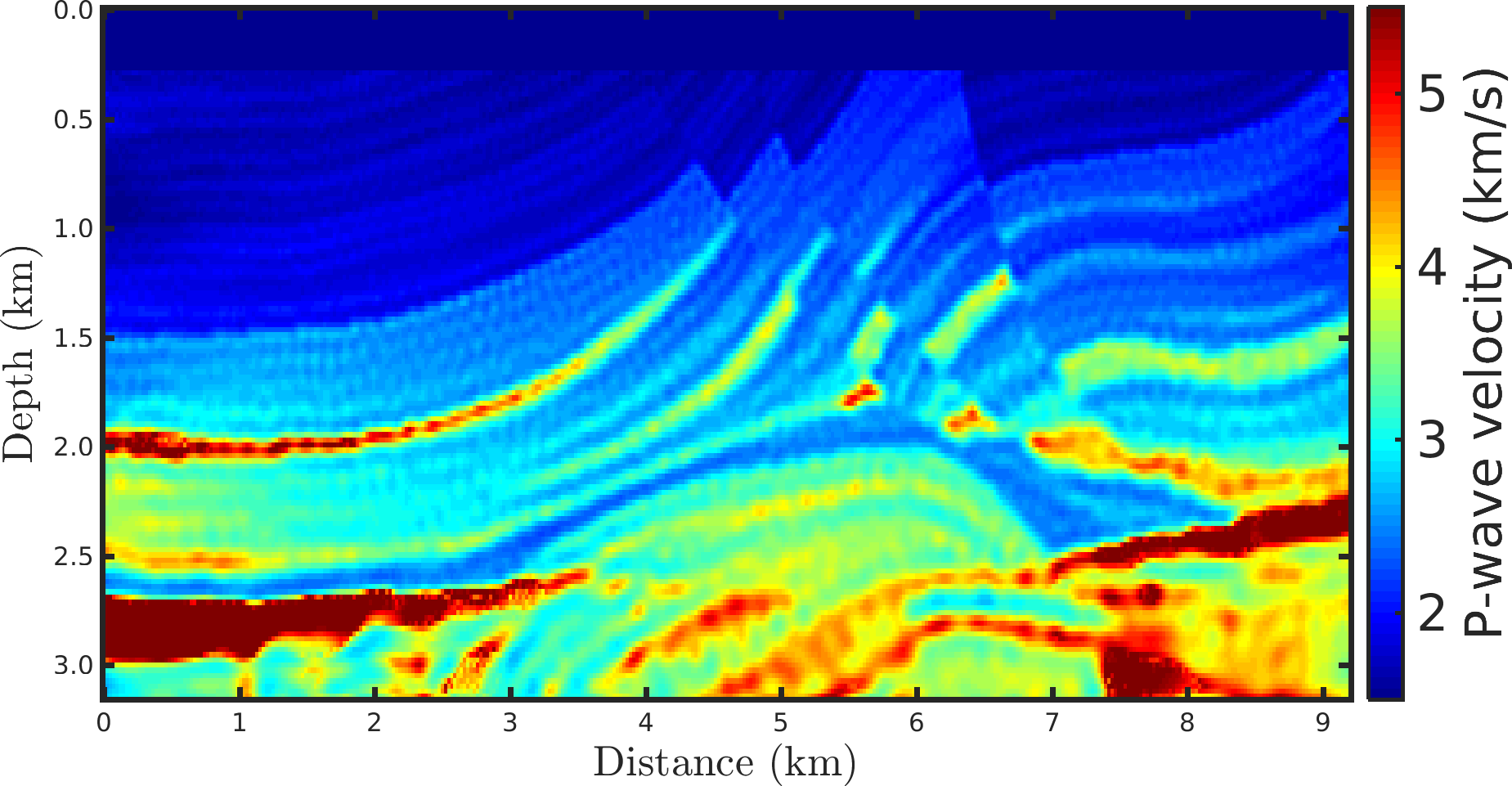}
  \label{fig:gauss_noise_70_data_aFWI_timedomain_a_0_85}
\end{subfigure}
\begin{subfigure}{.5\textwidth}
  \centering
  \caption{$\alpha$-FWI  with $\alpha = 0.75$}
  \includegraphics[width=\linewidth]{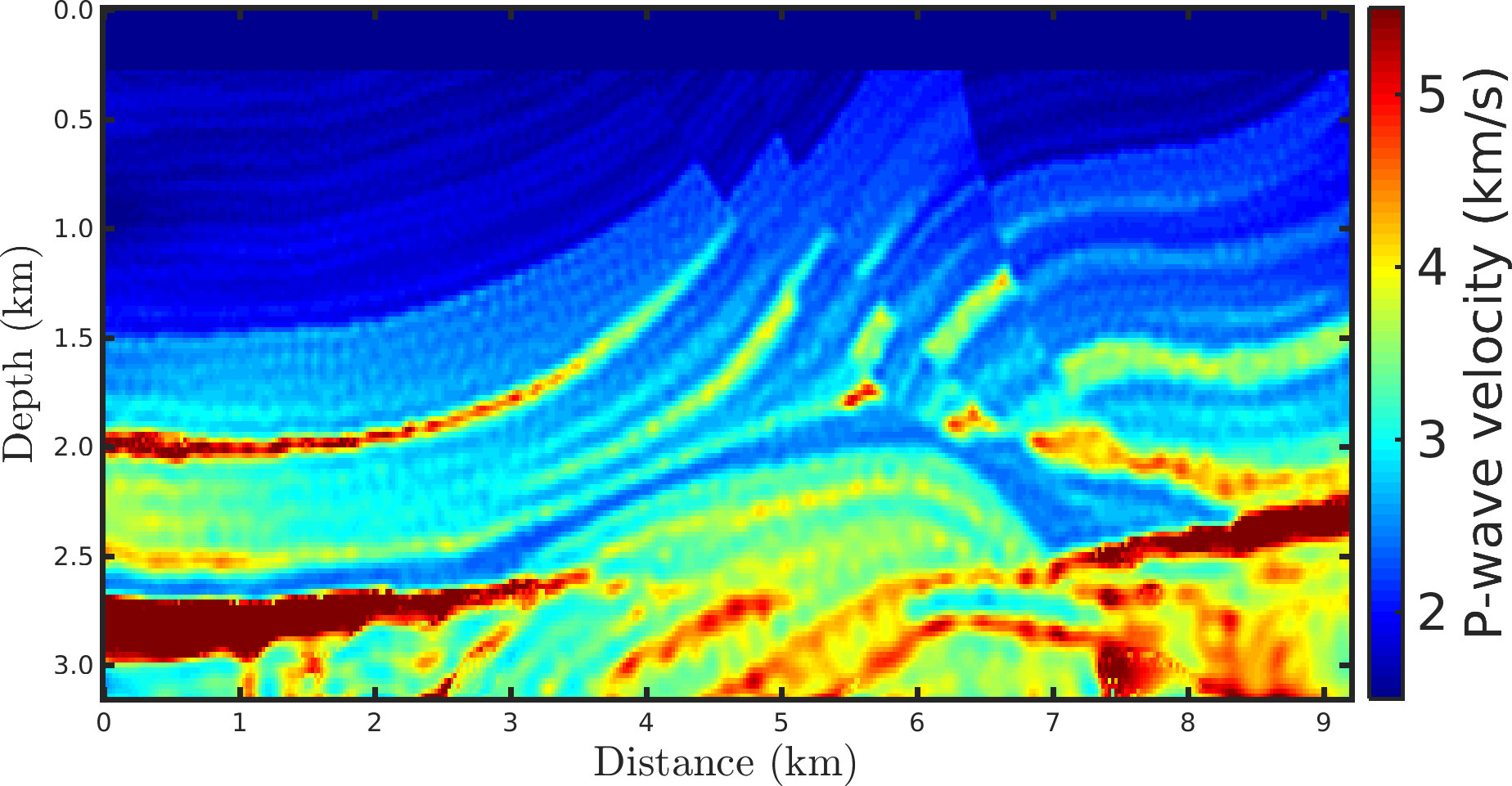}  
  \label{fig:gauss_noise_70_data_aFWI_timedomain_a_0_75}
\end{subfigure}
\begin{subfigure}{.5\textwidth}
  \centering
  \caption{$\alpha$-FWI  with $\alpha = 0.65$}
  \includegraphics[width=\linewidth]{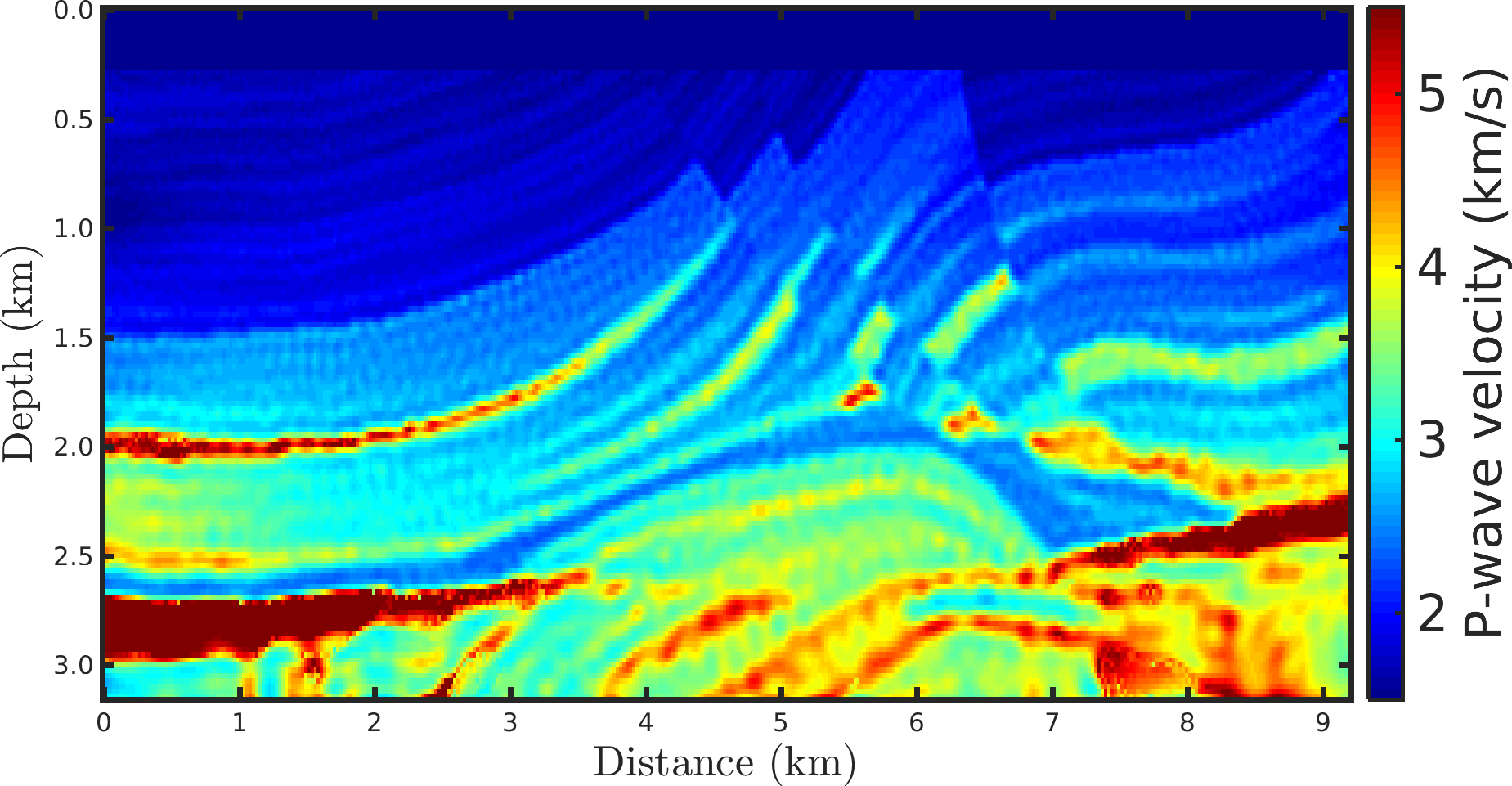}
  \label{fig:gauss_noise_70_data_aFWI_timedomain_a_0_65}
\end{subfigure}
\begin{subfigure}{.5\textwidth}
  \centering
  \caption{$\alpha$-FWI  with $\alpha = 0.55$}
  \includegraphics[width=\linewidth]{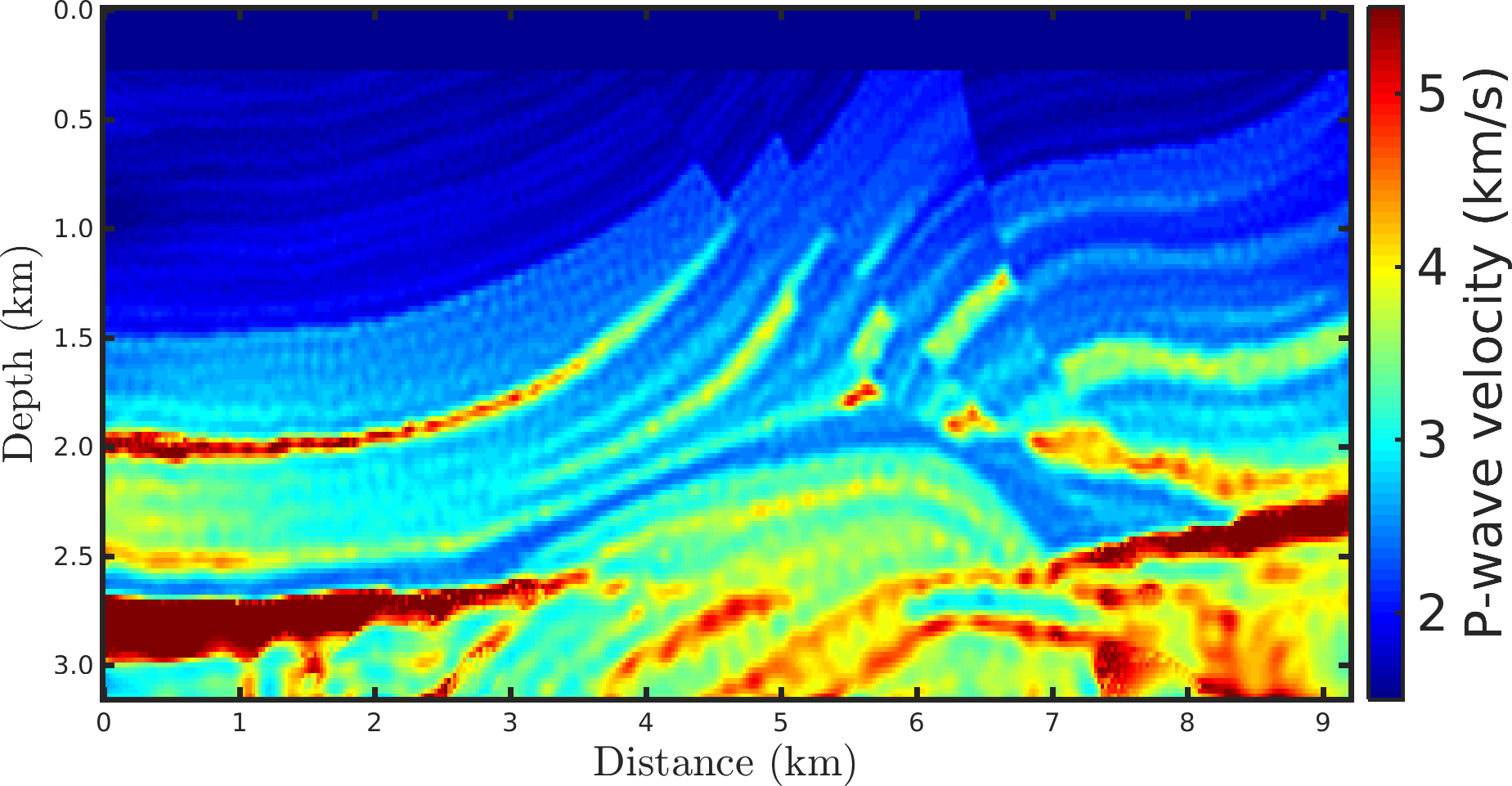}  
  \label{fig:gauss_noise_70_data_aFWI_timedomain_a_0_55}
\end{subfigure}
\begin{subfigure}{.5\textwidth}
  \centering
  \caption{$\alpha$-FWI  with $\alpha = 0.45$}
  \includegraphics[width=\linewidth]{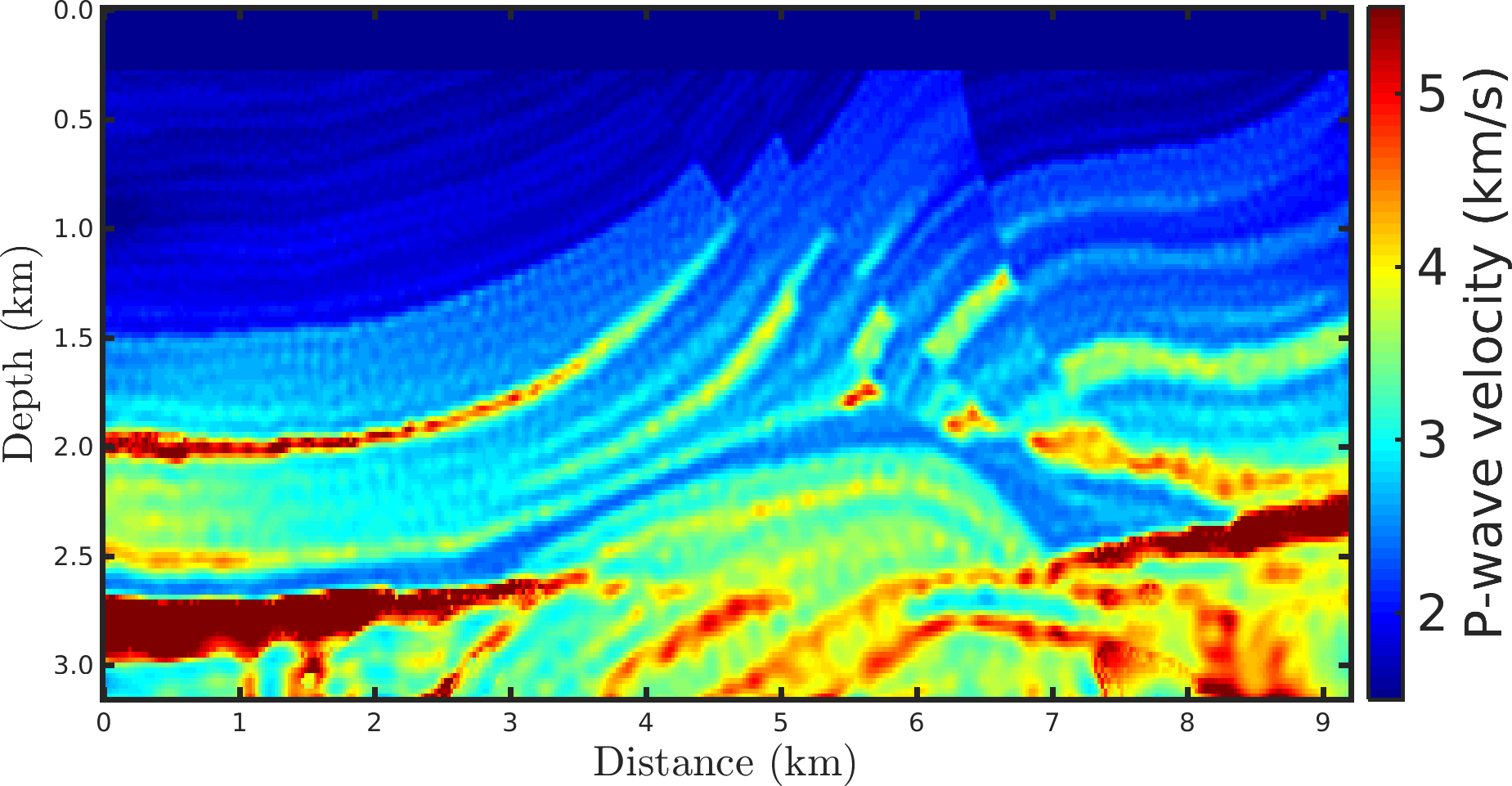}
  \label{fig:gauss_noise_70_data_aFWI_timedomain_a_0_45}
\end{subfigure}
\begin{subfigure}{.5\textwidth}
  \centering
  \caption{$\alpha$-FWI  with $\alpha = 0.35$}
  \includegraphics[width=\linewidth]{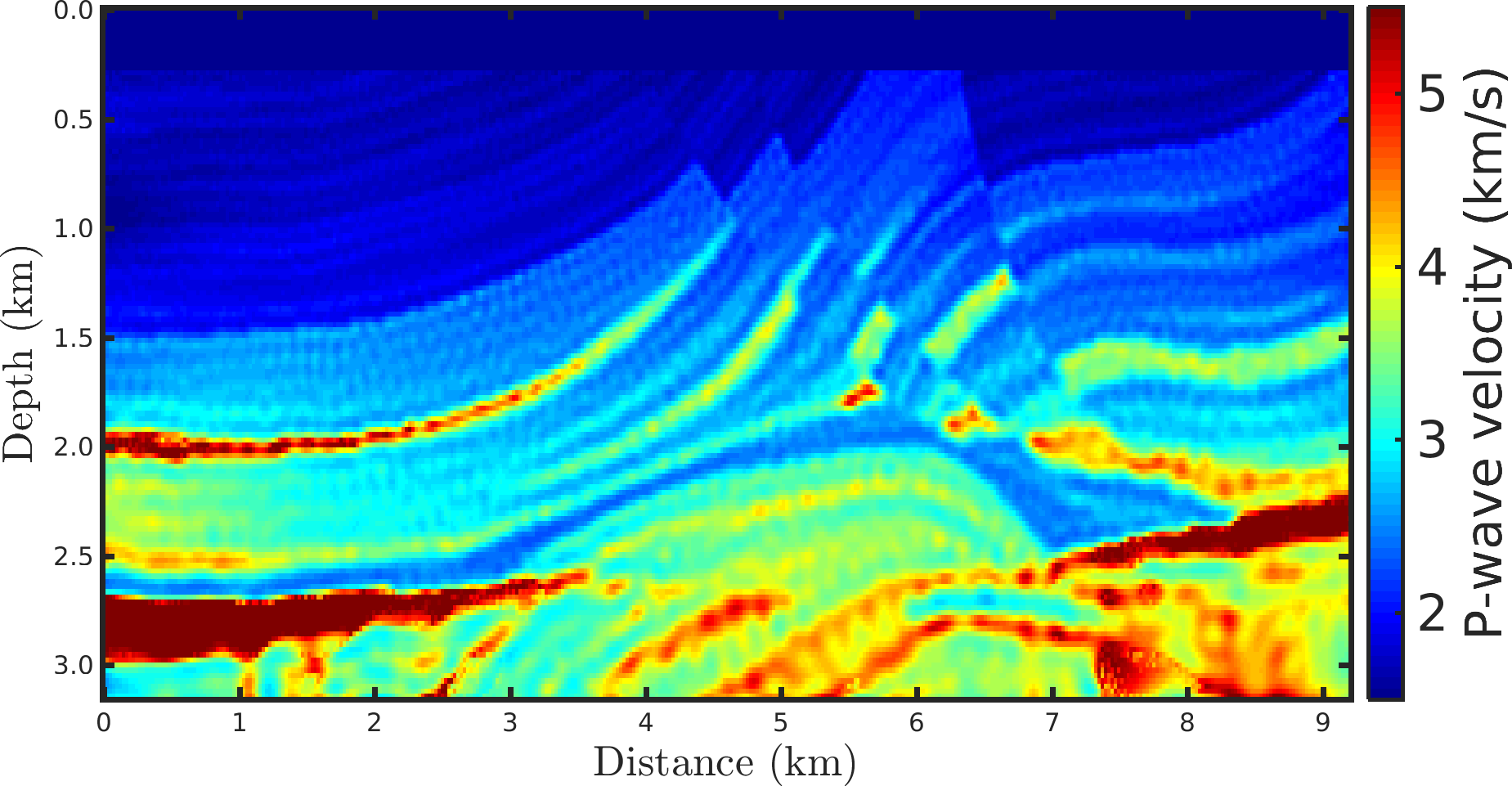}  
  \label{fig:gauss_noise_70_data_aFWI_timedomain_a_0_35}
\end{subfigure}
\caption{Reconstructed P-wave velocity models in the Gaussian noise case with $SNR = 70dB$ (second scenario) for the (a) classical FWI ($\alpha \rightarrow 1$), and $\alpha$-FWI with (b) $\alpha = 0.95$, (c) $\alpha = 0.85$, (d) $\alpha = 0.75$, (e) $\alpha = 0.65$, (f) $\alpha = 0.55$, (g) $\alpha = 0.45$, and (h) $\alpha = 0.35$.}
\label{fig:gauss_noise_70_data_aFWI_timedomain}
\end{figure}

\begin{figure}[]
\begin{subfigure}{.5\textwidth}
  \centering
  \caption{classical FWI ($\alpha \rightarrow 1$)}
  \includegraphics[width=\linewidth]{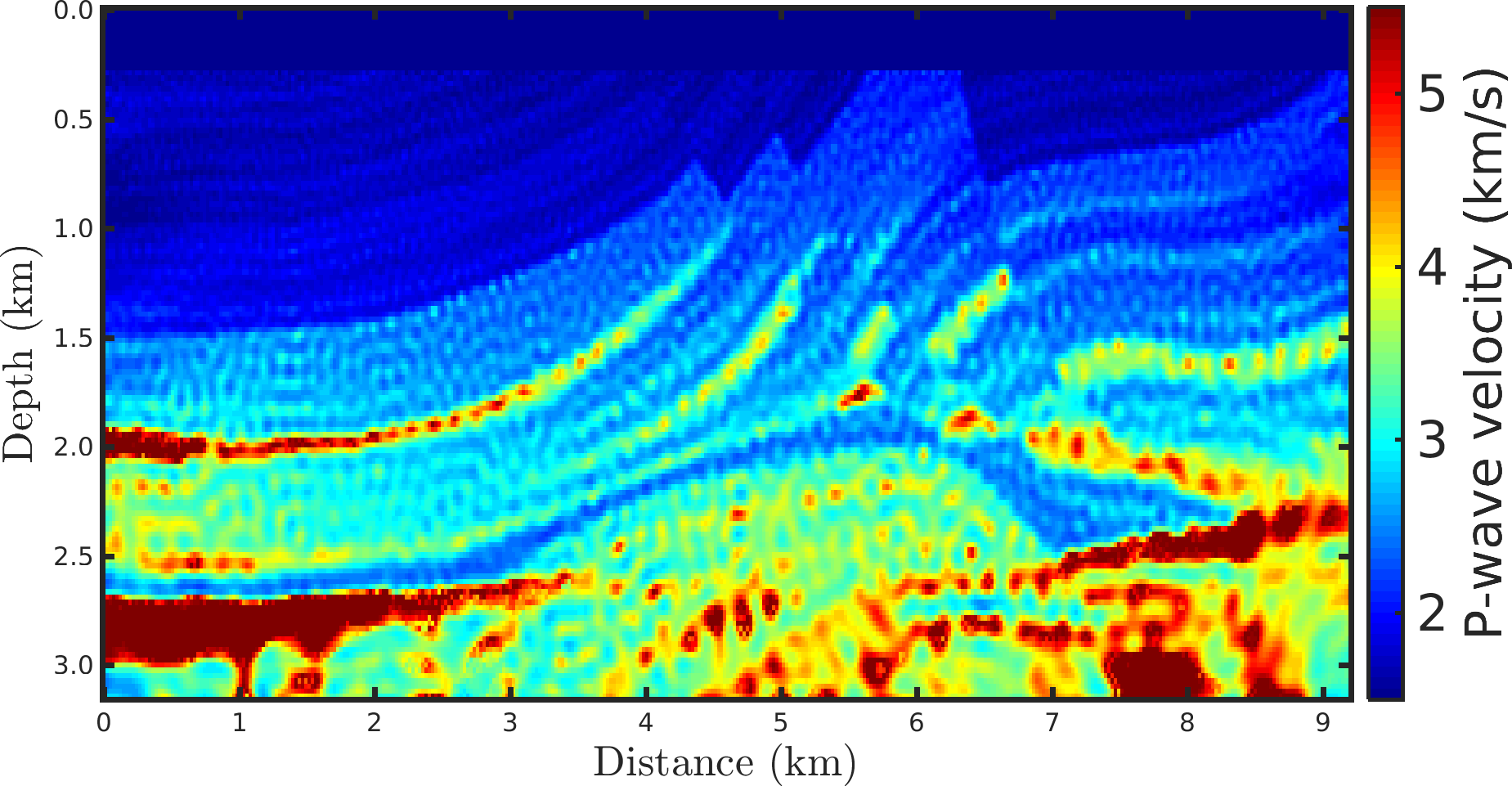}
  \label{fig:gauss_noise_60_data_classicalFWI_timedomain}
\end{subfigure}
\begin{subfigure}{.5\textwidth}
  \centering
  \caption{$\alpha$-FWI  with $\alpha = 0.95$}
  \includegraphics[width=\linewidth]{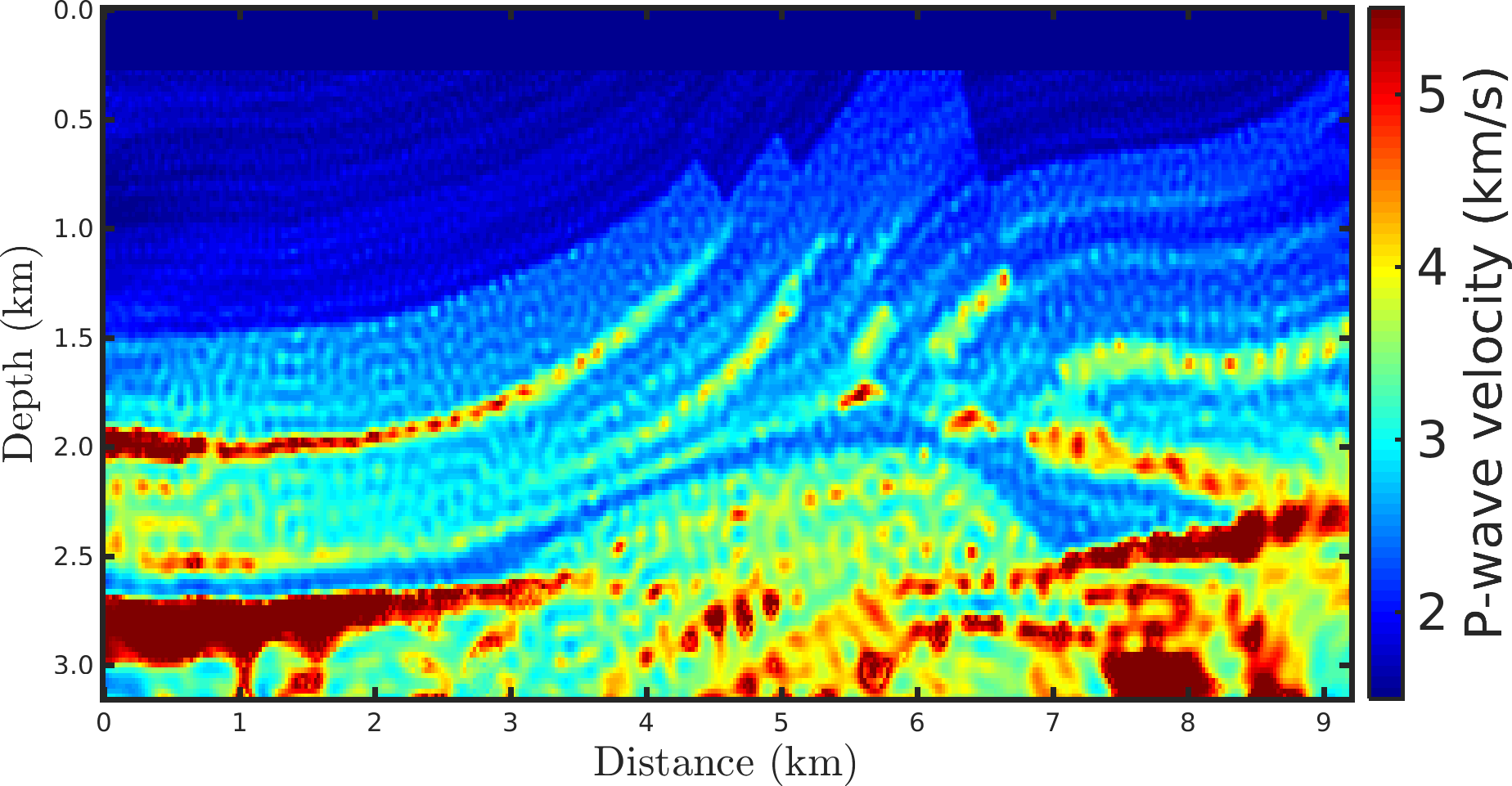}  
  \label{fig:gauss_noise_60_data_aFWI_timedomain_a_0_95}
\end{subfigure}
\begin{subfigure}{.5\textwidth}
  \centering
  \caption{$\alpha$-FWI  with $\alpha = 0.85$}
  \includegraphics[width=\linewidth]{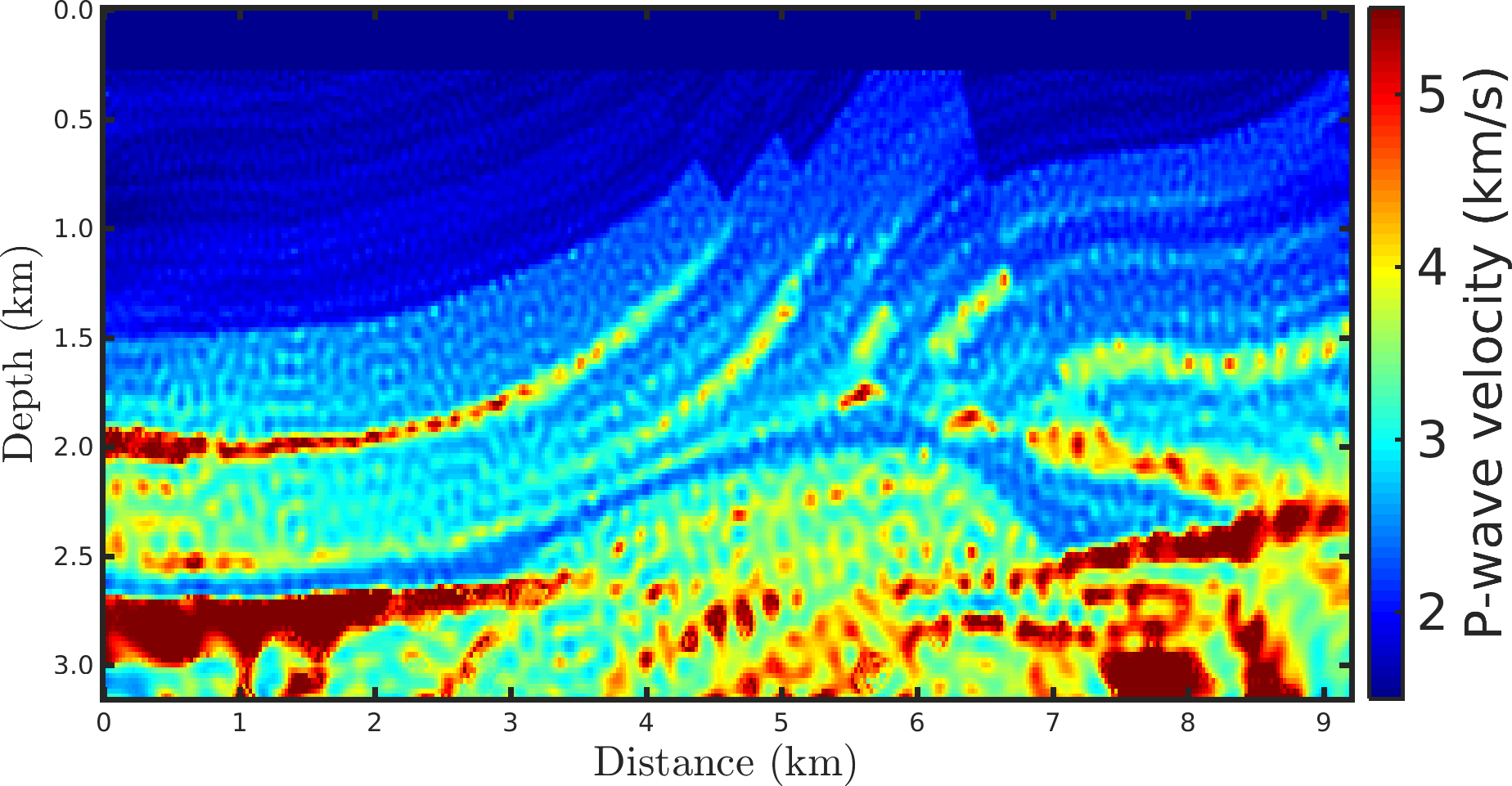}
  \label{fig:gauss_noise_60_data_aFWI_timedomain_a_0_85}
\end{subfigure}
\begin{subfigure}{.5\textwidth}
  \centering
  \caption{$\alpha$-FWI  with $\alpha = 0.75$}
  \includegraphics[width=\linewidth]{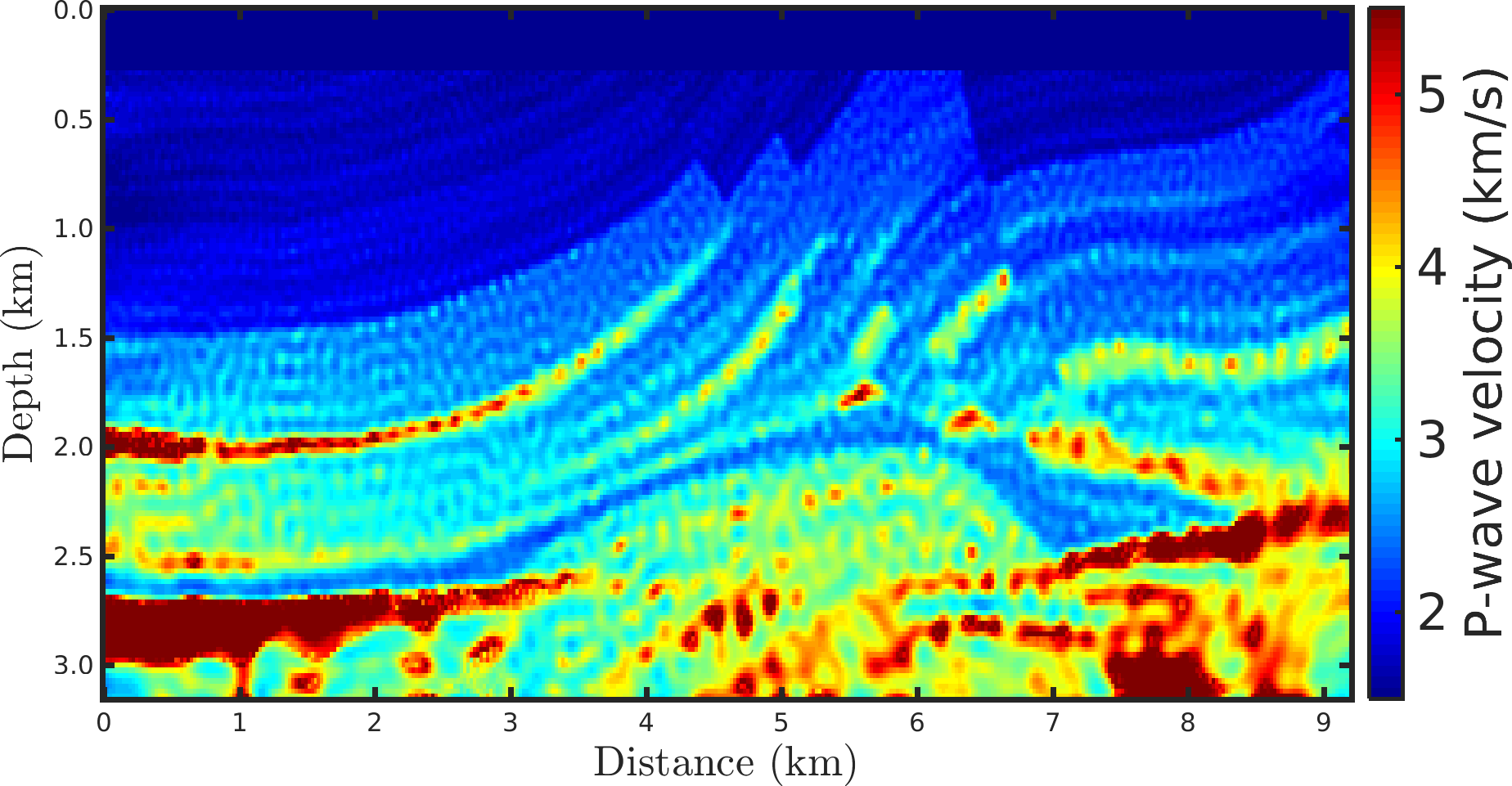}  
  \label{fig:gauss_noise_60_data_aFWI_timedomain_a_0_75}
\end{subfigure}
\begin{subfigure}{.5\textwidth}
  \centering
  \caption{$\alpha$-FWI  with $\alpha = 0.65$}
  \includegraphics[width=\linewidth]{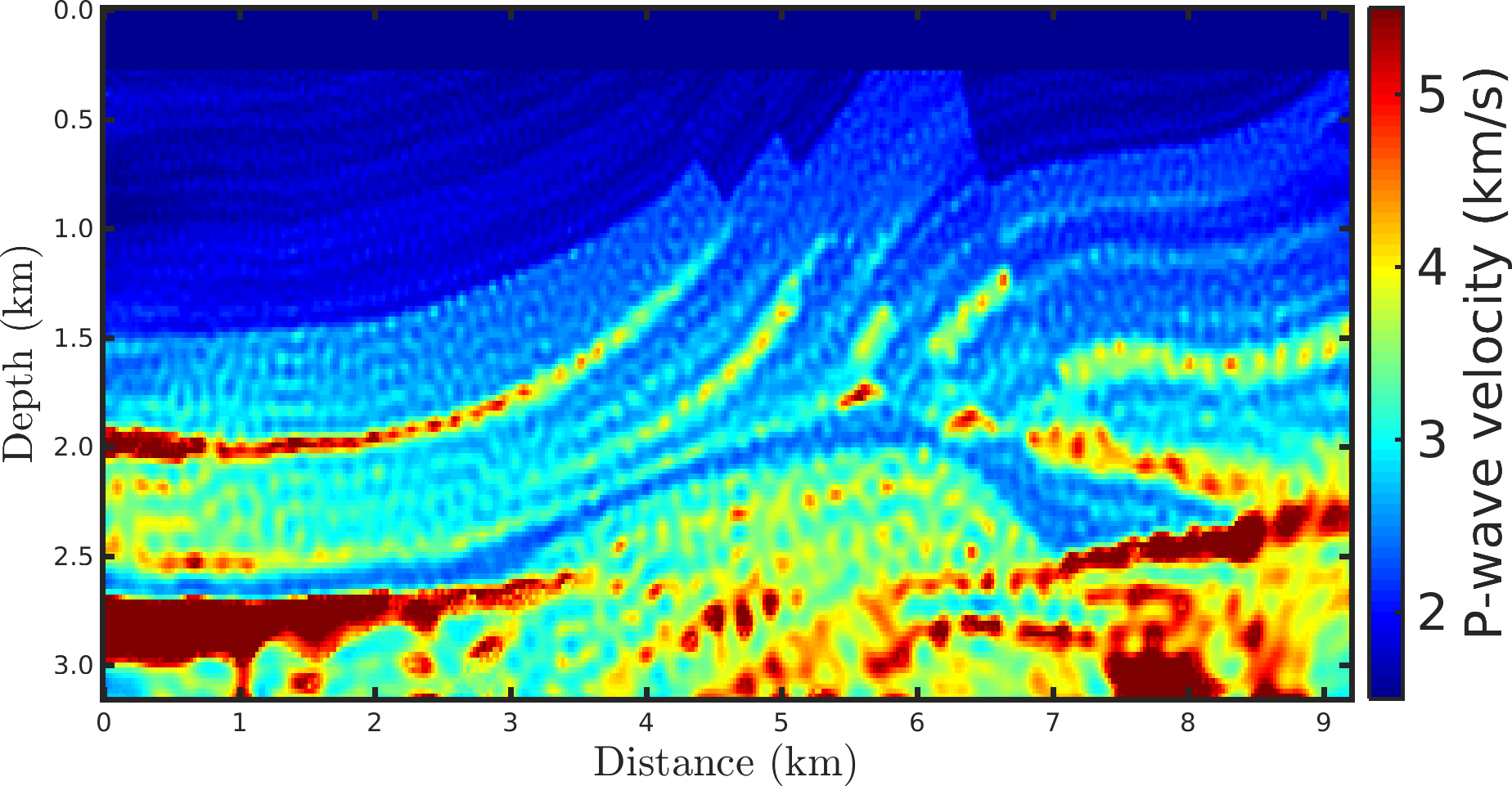}
  \label{fig:gauss_noise_60_data_aFWI_timedomain_a_0_65}
\end{subfigure}
\begin{subfigure}{.5\textwidth}
  \centering
  \caption{$\alpha$-FWI  with $\alpha = 0.55$}
  \includegraphics[width=\linewidth]{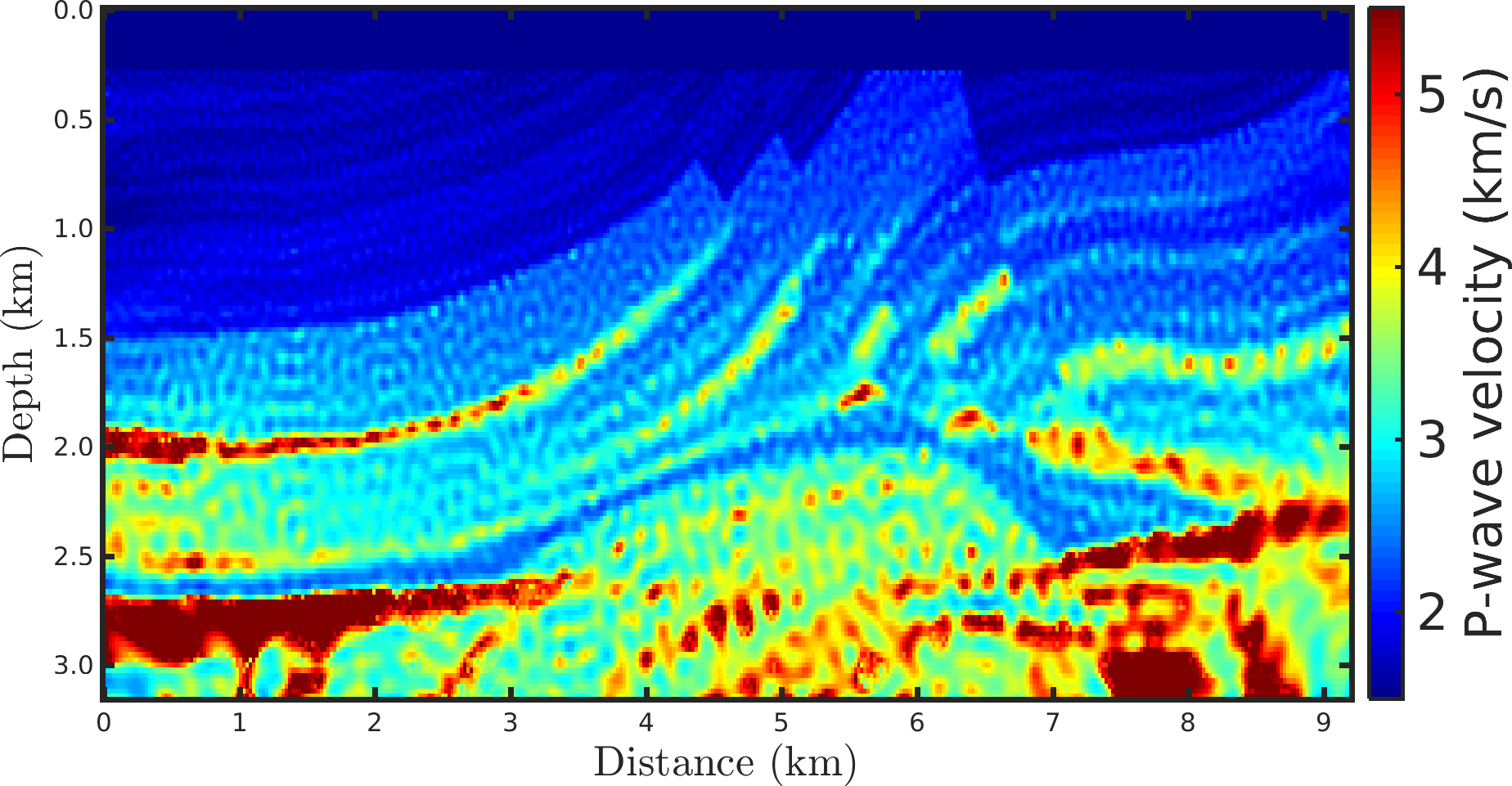}  
  \label{fig:gauss_noise_60_data_aFWI_timedomain_a_0_55}
\end{subfigure}
\begin{subfigure}{.5\textwidth}
  \centering
  \caption{$\alpha$-FWI  with $\alpha = 0.45$}
  \includegraphics[width=\linewidth]{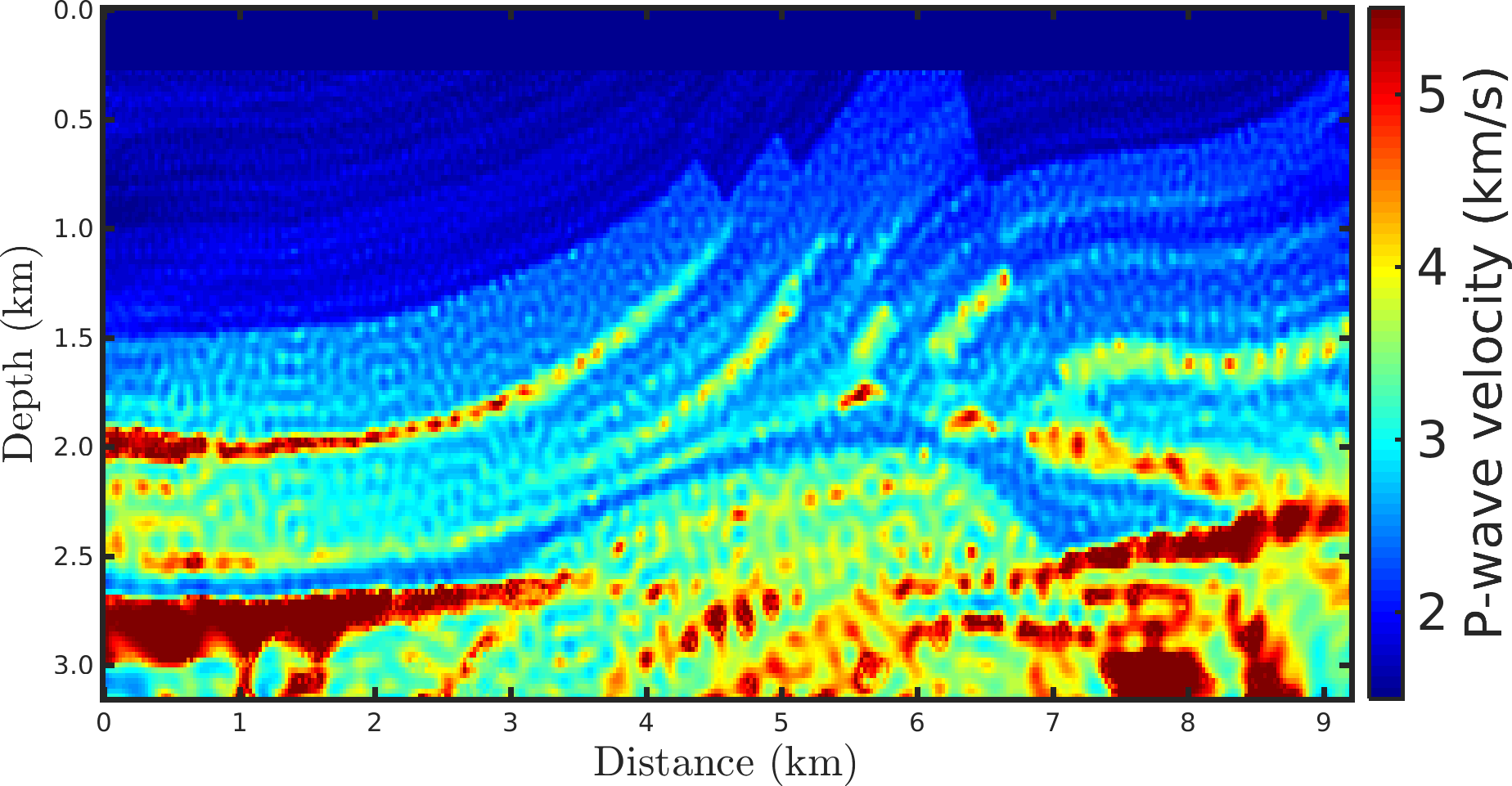}
  \label{fig:gauss_noise_60_data_aFWI_timedomain_a_0_45}
\end{subfigure}
\begin{subfigure}{.5\textwidth}
  \centering
  \caption{$\alpha$-FWI  with $\alpha = 0.35$}
  \includegraphics[width=\linewidth]{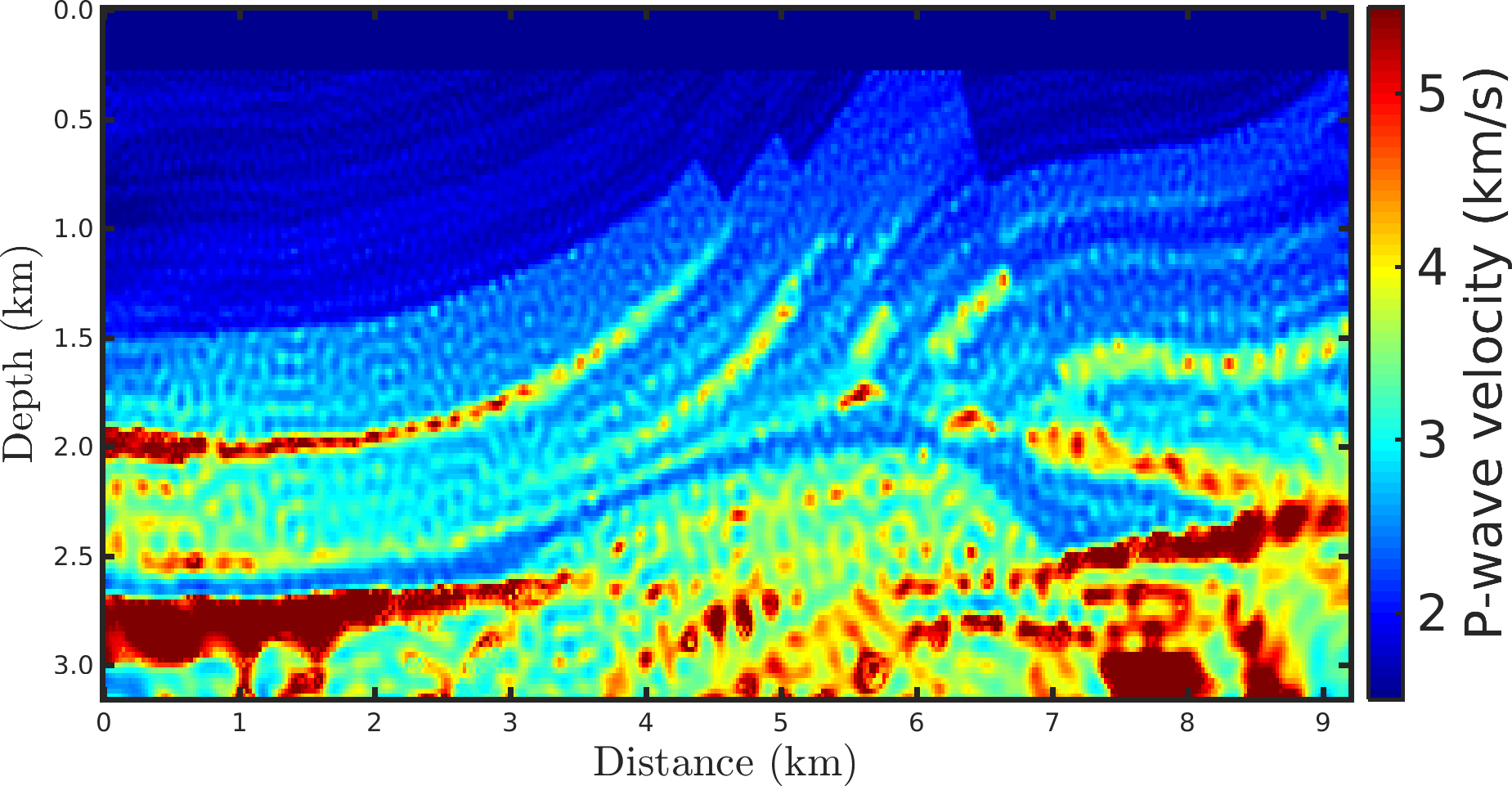}  
  \label{fig:gauss_noise_60_data_aFWI_timedomain_a_0_35}
\end{subfigure}
\caption{Reconstructed P-wave velocity models in the Gaussian noise case with $SNR = 60dB$ (third scenario) for the (a) classical FWI ($\alpha \rightarrow 1$), and $\alpha$-FWI with (b) $\alpha = 0.95$, (c) $\alpha = 0.85$, (d) $\alpha = 0.75$, (e) $\alpha = 0.65$, (f) $\alpha = 0.55$, (g) $\alpha = 0.45$, and (h) $\alpha = 0.35$.}
\label{fig:gauss_noise_60_data_aFWI_timedomain}
\end{figure}

Figures~\ref{fig:gauss_noise_70_data_aFWI_timedomain} and \ref{fig:gauss_noise_60_data_aFWI_timedomain} show the reconstructed models considering the second and third scenarios, respectively, in which the data is contaminated by Gaussian noise. From a visual inspection, we notice that just like the first scenario, the $\alpha$-FWI results are very satisfactory as the resulting models are close to the true model. In fact, regardless of the $\alpha$-value, the reconstructed P-wave models present the main structures of the Marmousi model despite the imprint of the noise. Again, case $\alpha = 0.55$ has a higher correlation and lower \textit{NRMS} error compared to the true model, as summarized in Table~\ref{tab:gauss_noise_timeFWI}.

\begin{table}[]
\centering
\caption{Statistical measures between the true model and the reconstructed models in the Gaussian noise cases with $SNR = 70dB$ (second scenario) and $SNR = 60dB$ (third scenario). {\footnotesize The Pearson's \textit{R} measures the linear correlations between the models, and the \textit{NRMS} measures the misfit between the true model and the reconstructed models.}}
\vspace{0.05cm}
\begin{tabular}{lc|cc|cc}
\hline            
\hline            
         &  & \hspace{.3cm} Second & Scenario ($70dB$)   & \hspace{.3cm} Third & Scenario ($60dB$) \\
%\hline            
Strategy & $\alpha$ & R & NRMS  &  R & NRMS \\
\hline                               
classical FWI & $\alpha \rightarrow 1.0$ & $0.9919$ & $0.0207$ &$0.9692$ & $0.0273$\\
\hline                               
& $\alpha = 0.95$ & $0.9907$ & $0.0209$ & $0.9698$ & $0.0272$ \\
& $\alpha = 0.85$ & $0.9835$ & $0.0229$ &$0.9699$ & $0.0271$ \\
& $\alpha = 0.75$ & $0.9906$ & $0.0210$ &$0.9681$ & $0.0275$ \\
$\alpha$-FWI & $\alpha = 0.65$ & $0.9920$ & $0.0206$ &$0.9693$ & $0.0271$ \\
& $\alpha = 0.55$ & $0.9922$ & $0.0205$ &$0.9701$ & $0.0270$ \\
& $\alpha = 0.45$ & $0.9936$ & $0.0202$ & $0.9698$ & $0.0271$ \\
& $\alpha = 0.35$ & $0.9911$ & $0.0208$ &$0.9691$ & $0.0275$ \\
\hline
\hline
\end{tabular}
\label{tab:gauss_noise_timeFWI}
\end{table}

%\begin{table}[]
%\centering
%\caption{Statistical measures between the true model and the reconstructed models in the Gaussian noise case with $SNR = 60dB$ (third scenario). {\footnotesize The Pearson's \textit{R} measures the linear correlations between the models and the \textit{NRMS} measures the misfit between the true model and the reconstructed models.}}
%\vspace{0.05cm}
%\begin{tabular}{lccc}
%\hline            
%Strategy & $\alpha$ & R  & NRMS \\
%\hline                               
%classical FWI & $\alpha \rightarrow 1.0$ & $0.9692$ & $0.0273$ \\
%\hline                               
%& $\alpha = 0.95$ & $0.9698$ & $0.0272$ \\
%& $\alpha = 0.85$ & $0.9699$ & $0.0271$ \\
%& $\alpha = 0.75$ & $0.9681$ & $0.0275$ \\
%$\alpha$-FWI & $\alpha = 0.65$ & $0.9693$ & $0.0271$ \\
%& $\alpha = 0.55$ & $0.9701$ & $0.0270$ \\
%& $\alpha = 0.45$ & $0.9698$ & $0.0271$ \\
%& $\alpha = 0.35$ & $0.9691$ & $0.0275$ \\
%\hline
%\hline
%\end{tabular}
%\label{tab:gauss_noise_snr60_timeFWI}
%\end{table}

\begin{figure}[]
\begin{subfigure}{.5\textwidth}
  \centering
  \caption{classical FWI ($\alpha \rightarrow 1$)}
  \includegraphics[width=\linewidth]{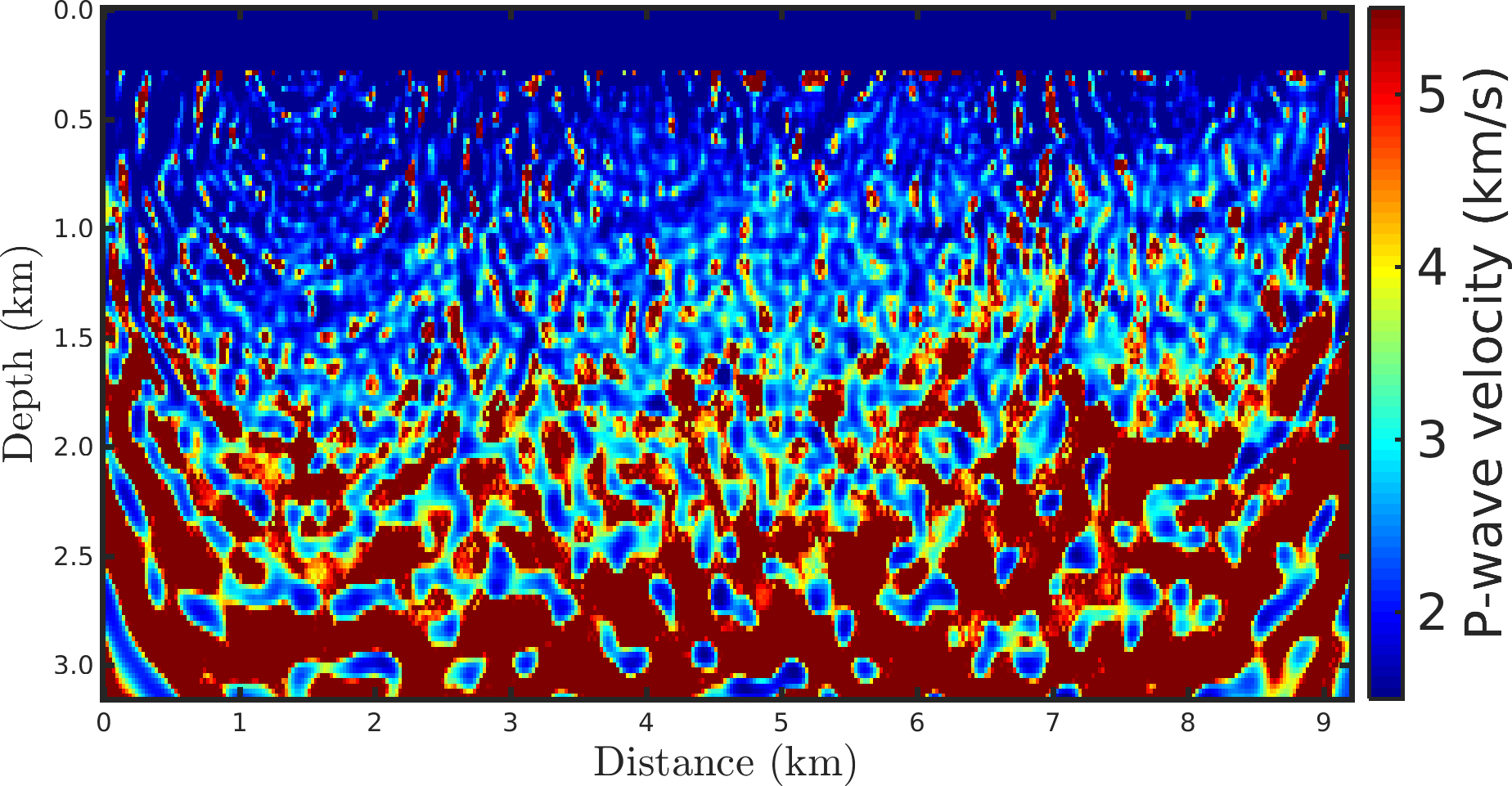}
  \label{fig:spike_0_5_gauss_noise_70_data_classicalFWI_timedomain}
\end{subfigure}
\begin{subfigure}{.5\textwidth}
  \centering
  \caption{$\alpha$-FWI  with $\alpha = 0.95$}
  \includegraphics[width=\linewidth]{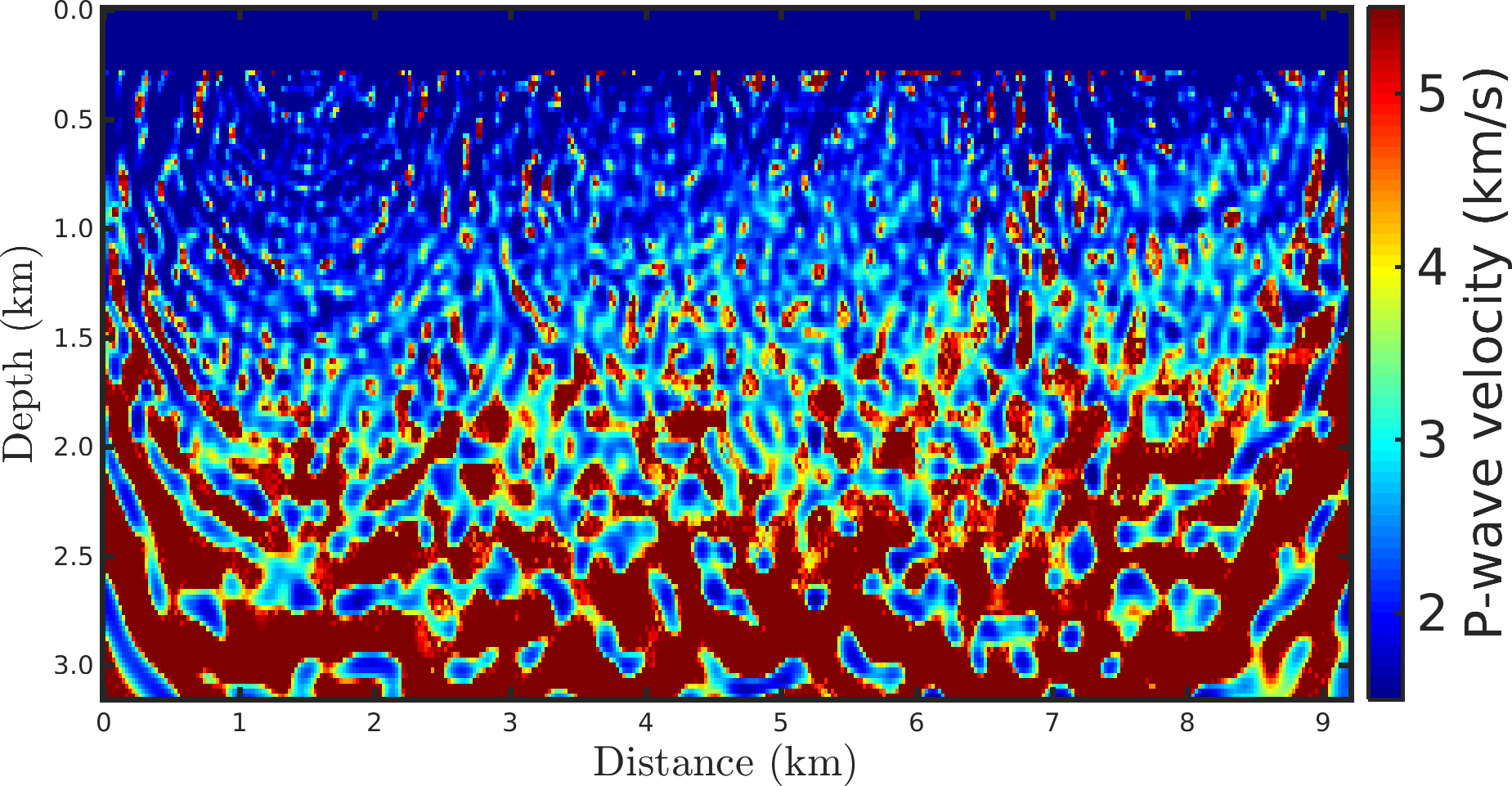}  
  \label{fig:spike_0_5_gauss_noise_70_data_aFWI_timedomain_a_0_95}
\end{subfigure}
\begin{subfigure}{.5\textwidth}
  \centering
  \caption{$\alpha$-FWI  with $\alpha = 0.85$}
  \includegraphics[width=\linewidth]{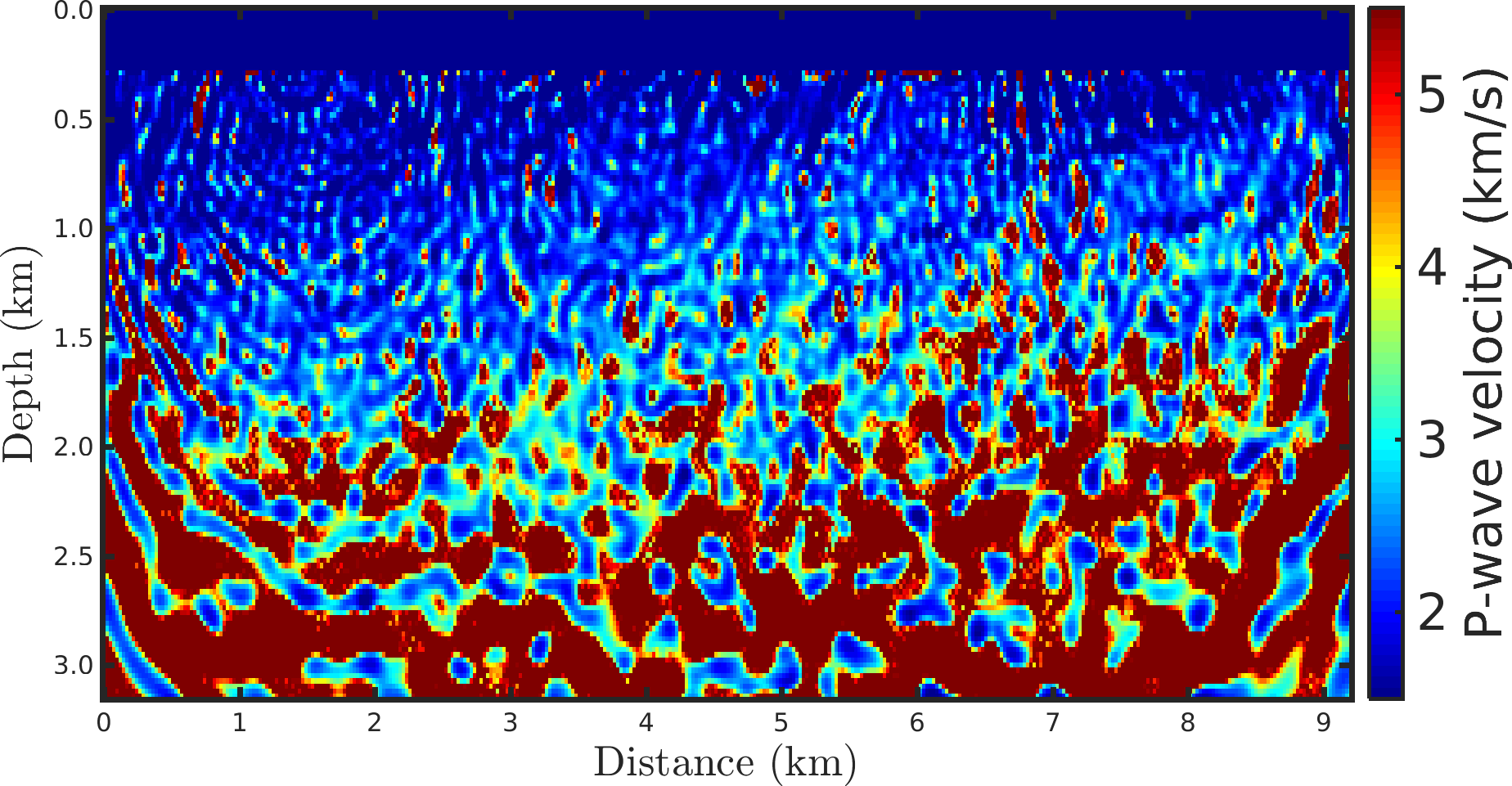}
  \label{fig:spike_0_5_gauss_noise_70_data_aFWI_timedomain_a_0_85}
\end{subfigure}
\begin{subfigure}{.5\textwidth}
  \centering
  \caption{$\alpha$-FWI  with $\alpha = 0.75$}
  \includegraphics[width=\linewidth]{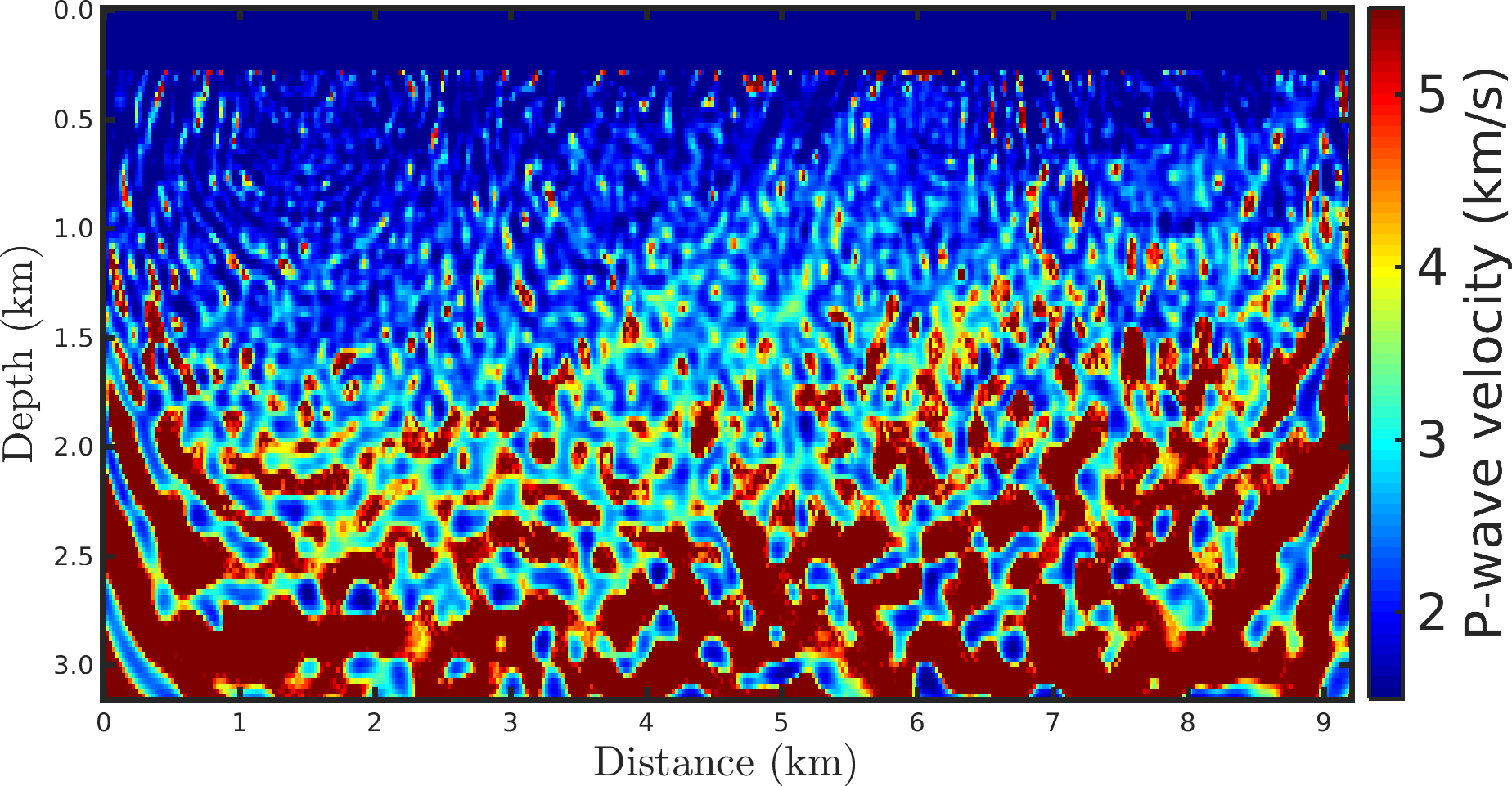}  
  \label{fig:spike_0_5_gauss_noise_70_data_aFWI_timedomain_a_0_75}
\end{subfigure}
\begin{subfigure}{.5\textwidth}
  \centering
  \caption{$\alpha$-FWI  with $\alpha = 0.65$}
  \includegraphics[width=\linewidth]{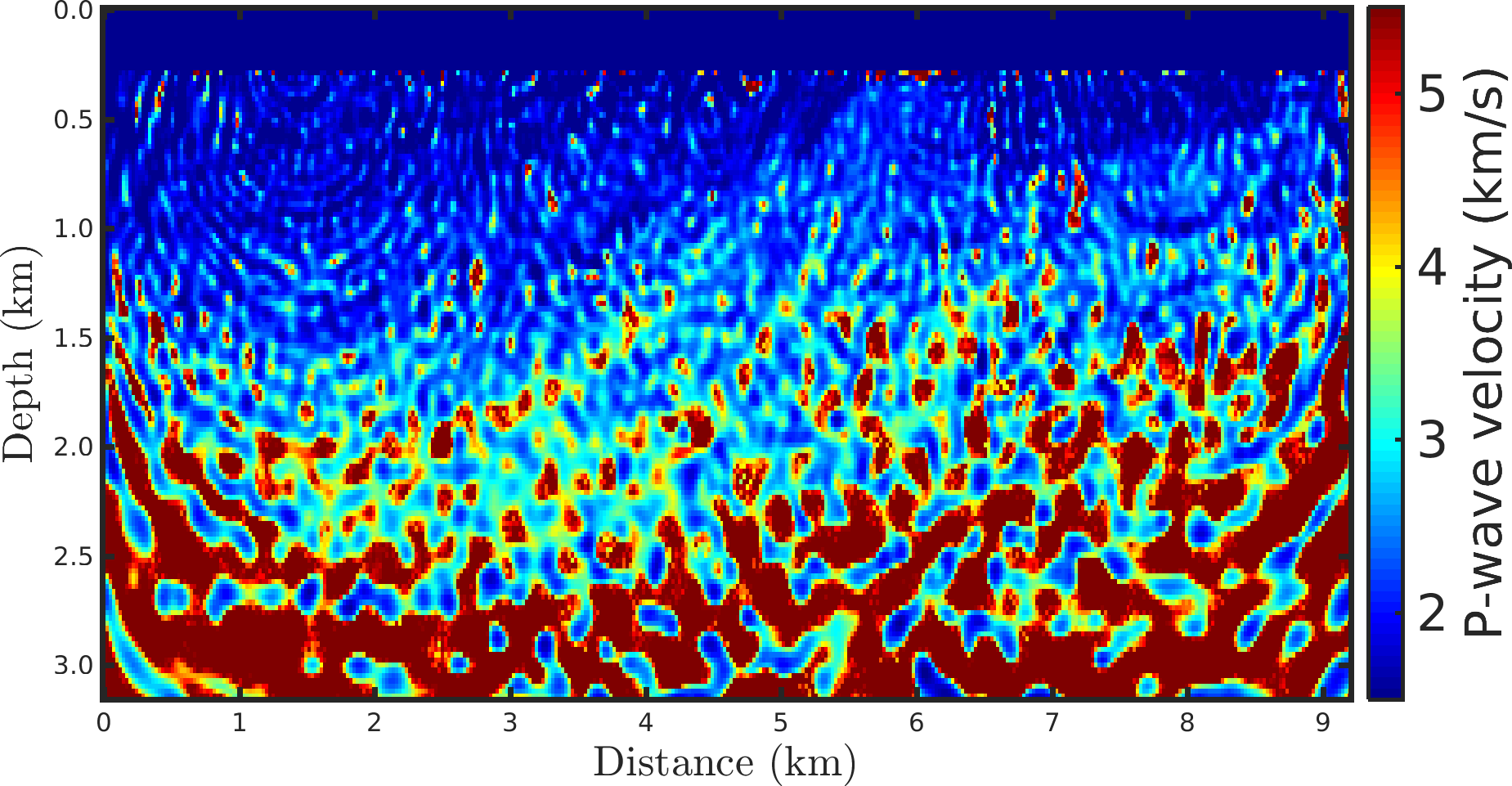}
  \label{fig:spike_0_5_gauss_noise_70_data_aFWI_timedomain_a_0_65}
\end{subfigure}
\begin{subfigure}{.5\textwidth}
  \centering
  \caption{$\alpha$-FWI  with $\alpha = 0.55$}
  \includegraphics[width=\linewidth]{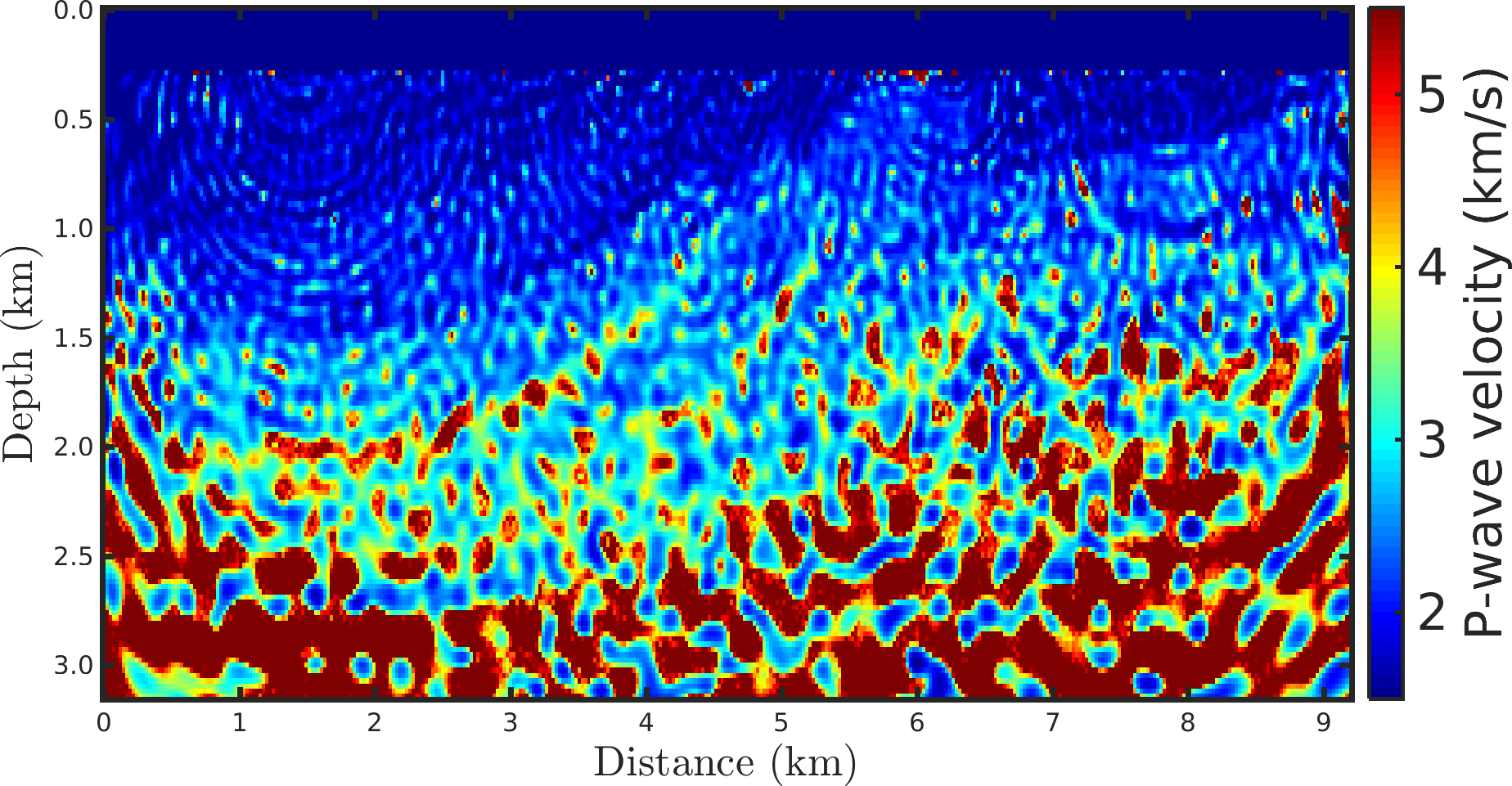}  
  \label{fig:spike_0_5_gauss_noise_70_data_aFWI_timedomain_a_0_55}
\end{subfigure}
\begin{subfigure}{.5\textwidth}
  \centering
  \caption{$\alpha$-FWI  with $\alpha = 0.45$}
  \includegraphics[width=\linewidth]{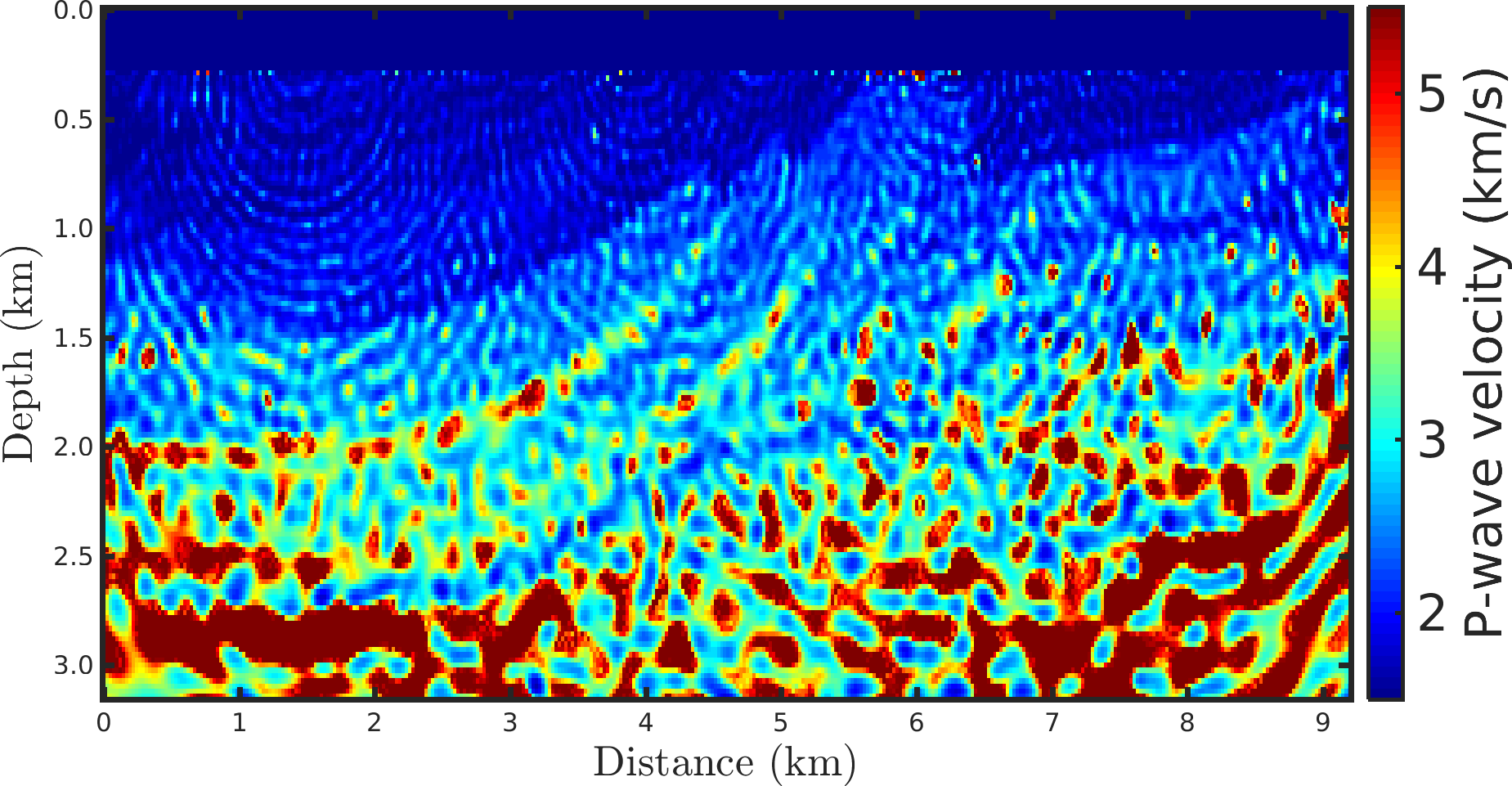}
  \label{fig:spike_0_5_gauss_noise_70_data_aFWI_timedomain_a_0_45}
\end{subfigure}
\begin{subfigure}{.5\textwidth}
  \centering
  \caption{$\alpha$-FWI  with $\alpha = 0.35$}
  \includegraphics[width=\linewidth]{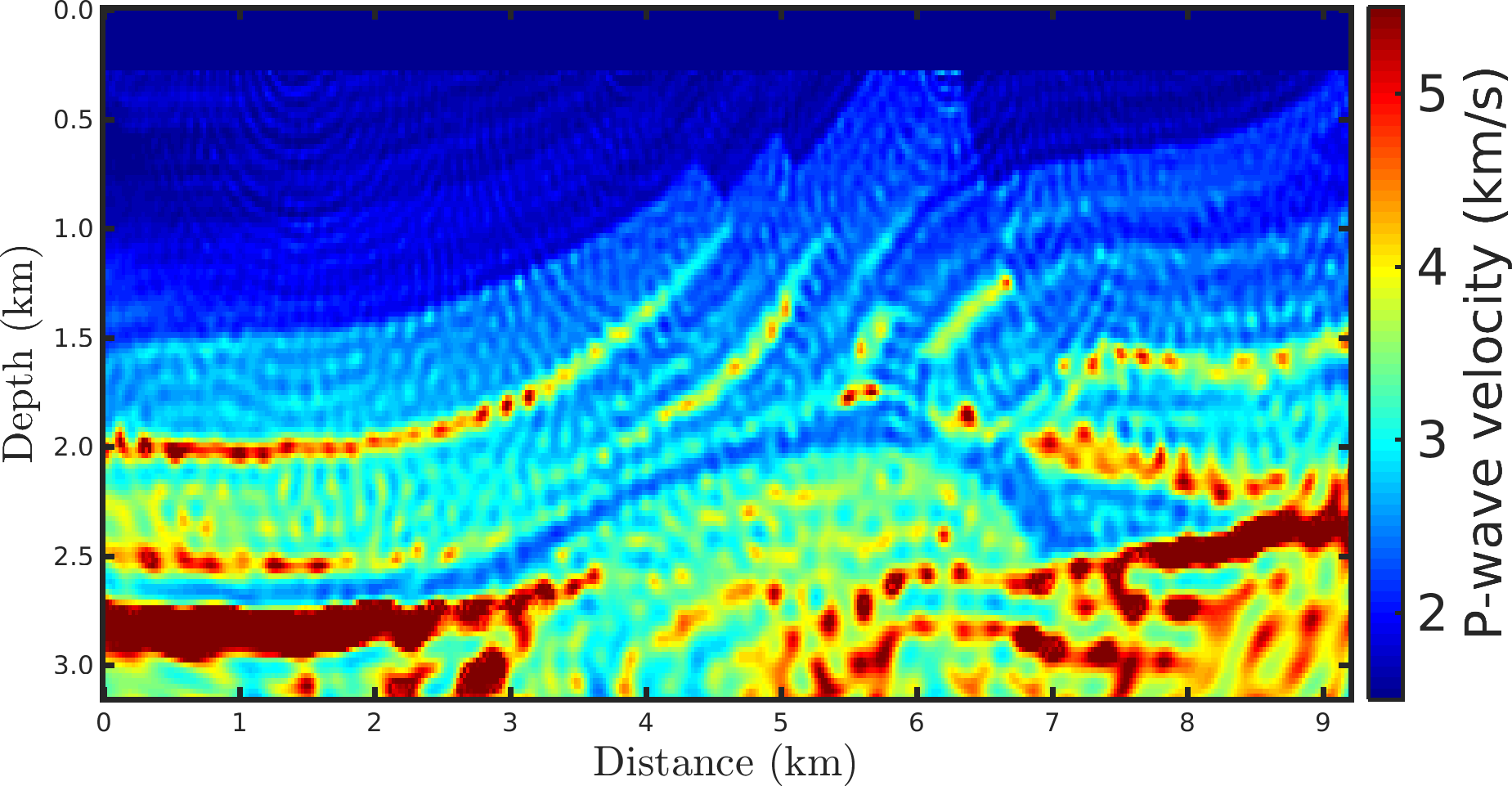}  
  \label{fig:spike_0_5_gauss_noise_70_data_aFWI_timedomain_a_0_35}
\end{subfigure}
\caption{Reconstructed P-wave velocity models in the Gaussian noise case with $SNR = 70dB$ and outliers (fourth scenario) for the (a) classical FWI ($\alpha \rightarrow 1$), and $\alpha$-FWI with (b) $\alpha = 0.95$, (c) $\alpha = 0.85$, (d) $\alpha = 0.75$, (e) $\alpha = 0.65$, (f) $\alpha = 0.55$, (g) $\alpha = 0.45$, and (h) $\alpha = 0.35$.}
\label{fig:spike_0_5_gauss_noise_70_data_aFWI_timedomain}
\end{figure}

In the fourth scenario, in which non-Gaussian noise is considered, the classical approach ($\alpha \rightarrow 1$) completely fails to reconstruct a P-wave velocity model, as depicted in Fig.~\ref{fig:spike_0_5_gauss_noise_70_data_classicalFWI_timedomain}. Such failure of the classical approach is associated with the assumption that the errors are Gaussian, when in practice this is not always true. We notice that as the $\alpha$-value decreases, which means a greater deviation from the Gaussian statistics, the resulting model is more suitable (see Fig.~\ref{fig:spike_0_5_gauss_noise_70_data_aFWI_timedomain}), especially in the case $\alpha = 0.35$ where the P-wave model is comparable with the reconstructed models in the previous scenarios. In fact, only the $\alpha$-FWI with $\alpha = 0.35$ is strongly correlated with the true model, in addition to having a low error which is in the same order as in the previous scenarios (see Table~\ref{tab:spike_0_5_gauss_noise_snr70_timeFWI}).

\begin{table}[]
\centering
\caption{Statistical measures between the true model and the reconstructed models in the Gaussian noise case with $SNR = 70dB$ and outliers (fourth scenario). {\footnotesize The Pearson's \textit{R} measures the linear correlations between the models, and the \textit{NRMS} measures the misfit between the true model and the reconstructed models.}}
\vspace{0.05cm}
\begin{tabular}{lccc}
\hline            
Strategy & $\alpha$ & R  & NRMS \\
\hline                               
classical FWI & $\alpha \rightarrow 1.0$ & $0.5789$ & $0.2372$ \\
\hline                               
& $\alpha = 0.95$ & $0.5942$ & $0.2189$ \\
& $\alpha = 0.85$ & $0.6021$ & $0.2136$ \\
& $\alpha = 0.75$ & $0.6360$ & $0.1831$ \\
$\alpha$-FWI & $\alpha = 0.65$ & $0.6500$ & $0.1591$ \\
& $\alpha = 0.55$ & $0.6804$ & $0.1328$ \\
& $\alpha = 0.45$ & $0.7343$ & $0.0889$ \\
& $\alpha = 0.35$ & $0.8432$ & $0.0380$ \\
\hline
\hline
\end{tabular}
\label{tab:spike_0_5_gauss_noise_snr70_timeFWI}
\end{table}

%%%%%%%%%%%%%%%%%%%%%%%%%%%%%%%%%%%%%%%%%%%%%%%%%%%%%%%%%%%%%%%%%%%%%%%%%
%%%%%%%%%%%%%%%%%%%%%%%%%%%%%%%%%%%%%%%%%%%%%%%%%%%%%%%%%%%%%%%%%%%%%%%%%
\subsection*{Brazilian pre-salt case study}
%%%%%%%%%%%%%%%%%%%%%%%%%%%%%%%%%%%%%%%%%%%%%%%%%%%%%%%%%%%%%%%%%%%%%%%%%
%%%%%%%%%%%%%%%%%%%%%%%%%%%%%%%%%%%%%%%%%%%%%%%%%%%%%%%%%%%%%%%%%%%%%%%%%

\begin{figure}[]
\begin{subfigure}{.5\textwidth}
  \centering
  \caption{}
  \includegraphics[width=\linewidth]{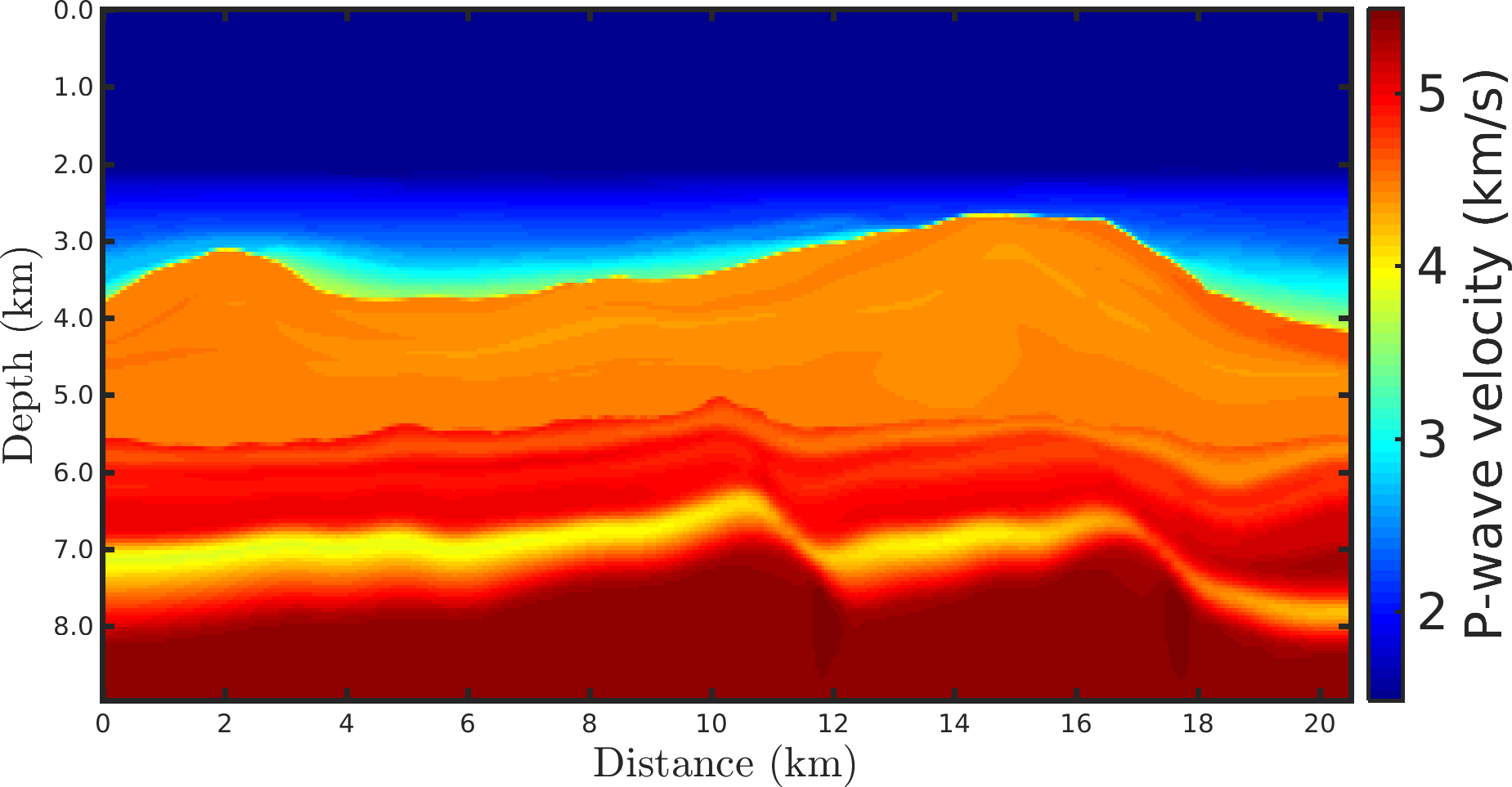}
  \label{fig:gdm_truemodel}
\end{subfigure}
\begin{subfigure}{.5\textwidth}
  \centering
  \caption{}
  \includegraphics[width=\linewidth]{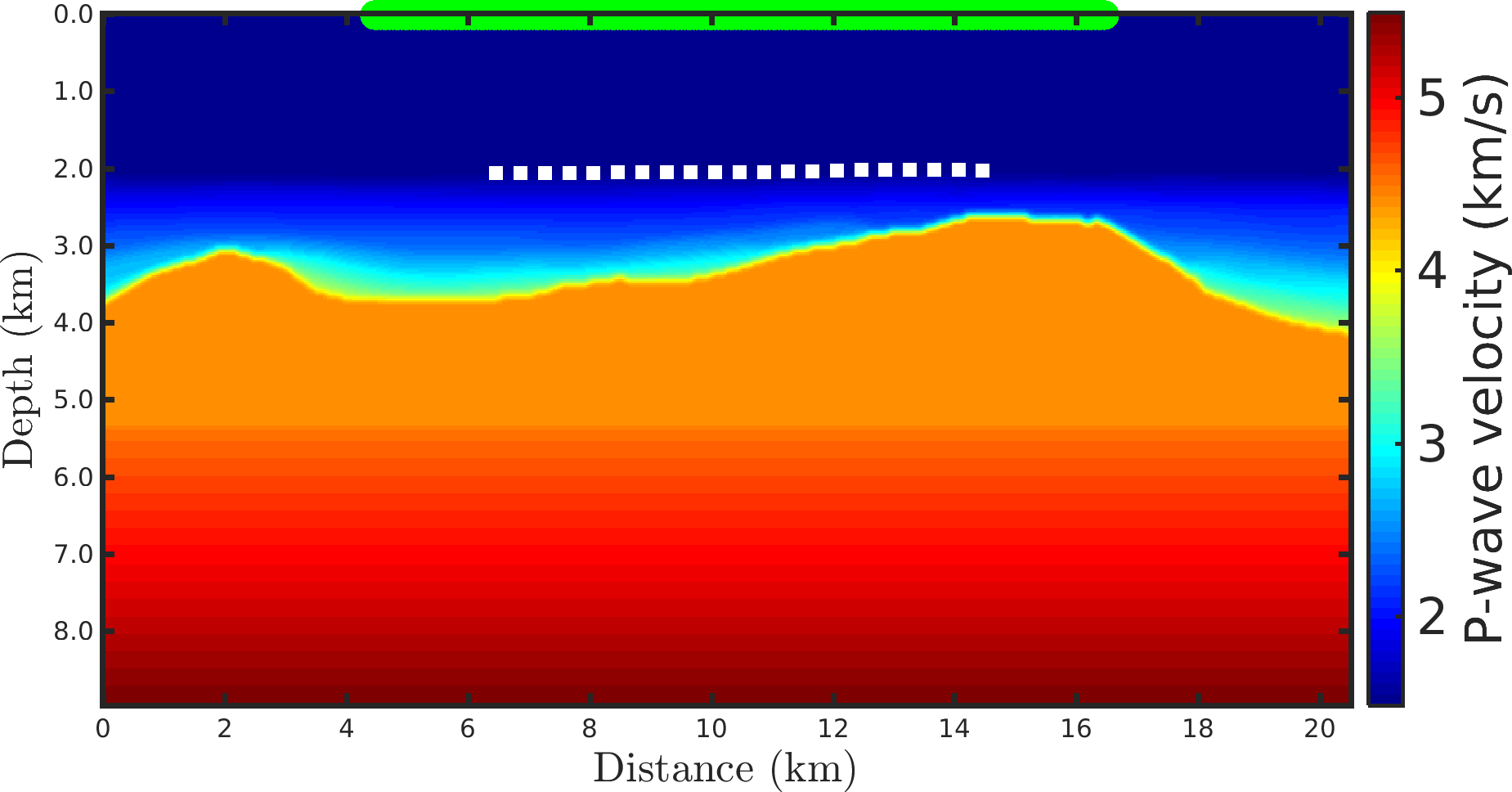}
  \label{fig:gdm_initialmodel}
\end{subfigure}
\caption{(a) Realistic P-wave velocity model based on the Brazilian pre-salt region, considered as the true model in this study. (b) Typical  model  of  Brazilian pre-salt field modified from \cite{daSilva_SEG2021_qFWI_2021}.
%(a) true model (b) initial model (salt velocity $4.4km/s$)... linear increasing from 5.4km in-depth from 4.4km/s to $~5.5km/s$ $v(z\geq 5400m) = 6.868 z + 3026$.... slowness smoothing $\sigma = 50m$.
}
\label{fig:fig:gdm_usedmodels}
\end{figure}

In this section, we explore the potential of the $\alpha$-PGF-FWI  for estimating P-wave velocities in a typical Brazilian pre-salt field. In this regard, we consider the realistic acoustic model depicted in Fig.~\ref{fig:gdm_truemodel} \cite{jorgeGdM_2020,daSilva_SEG2021_qFWI_2021} as the true model, which comprises a water layer in which the ocean floor is an average of $2$km in-depth, followed by post-salt sediments, a salt body with variable thickness and velocity, the pre-salt reservoir, and the bedrock as the model base. By using the true model, Fig.~\ref{fig:gdm_truemodel}, we generate a seismic dataset considering a OBN acquisition comprising $21$ nodes (receivers) equally spaced located every $400m$, from $6450m$ to $14450m$, deployed at the ocean floor (see the white squares in Fig.~\ref{fig:gdm_initialmodel}). In addition, we consider a line of seismic sources, spaced each one at $50$m, extending by $2$km beyond each one end of the node row (see the green line in Fig.~\ref{fig:gdm_initialmodel}). We call this acquisition geometry OBN classical. The acquisition time was $10s$, in which we employ a $5Hz$ Ricker wavelet as the seismic source. We perform several numerical experiments, which will be described next. In this section, in particular, we consider only the $\alpha \rightarrow 1$ (classical) and $\alpha = 0.35$ cases to compare our proposal with the classical approach. We notice that the water layer in all P-wave velocity models are assumed to be known and, therefore, are kept constant during the inversion process. 

%figure wavefiles frequency?

%For the same set of nodes, we consider two different arrangements regarding the seismic sources. First, we consider a line of seismic sources, spaced each one at $50$m, extending by $2$km beyond each one end of the node row (see the green line in Fig.~\ref{fig:gdm_initialmodel}). We call this acquisition geometry OBN classical. In the second one, we consider a so-called circular shot OBN acquisition geometry \cite{jorgeGdM_2020} which comprises only six seismic sources (see the red dots in Fig.~\ref{daSilva_SEG2021_qFWI_2021}) \cite{daSilva_SEG2021_qFWI_2021}. The circular shot OBN acquisition geometry allows recording of diving waves that are very important to maximize the illumination of the deepest regions of the velocity model, which, in our case, is the area of the pre-salt reservoir \cite{OBN_CGG_SEG2019,Costa_GdM_2020_UFF}. 

The first two experiment consists of imaging the pre-salt region, which remains a great challenge, using the $\alpha$-PGF-FWI. In the first one, we consider the noiseless data set generated by the acquisition geometry described previously. In the second one, we consider the same data set, but polluted by non-Gaussian noise. The non-Gaussian perturbations were generated through a Student's \textit{t}-distribution with three degrees of freedom. %Figure~\ref{} shows some examples of receiver-gathers. 
We solve the forward problem by solving the 2-D frequency-domain acoustic wave equation using the finite-difference method, in which we consider a 9-point stencil with a convolution-perfectly matched layer (C-PML) \cite{Roden_CPML_2000,Komatitsch_CPML_2007} in order to simulate an unlimited medium. 

We discretize the Helmholtz operator on a regular grid with a spacing of 25 m. In the inversion process, we consider a multiscale approach \cite{Kolb1986,multiscalebunks1995} in which we employ s
equentially three frequency groups: $\{2, ..., 3\}$,
$\{2, ..., 5\}$ and $\{2, ..., 7\}$Hz. In this framework, the data inversion is performed considering only the content of the data associated with the first group of frequencies, starting from the initial model shown in Fig.~\ref{fig:gdm_initialmodel}. Such a model comprises a water layer, followed by post-salt sediments (which are different from the true model), a salt body with variable thickness and constant velocity ($4.4km/s$), and a linear velocity increasing, from $5.4km$ in-depth, rang
the second frequencies group. Again, the resulting model is used as the initial model for the third group. For each frequencies group, we compute $20$ \textit{l}-BFGS iterations.

\begin{figure}[]
\begin{subfigure}{.5\textwidth}
  \centering
  \caption{classical PGF-FWI ($\alpha \rightarrow 1$)}
  \includegraphics[width=\linewidth]{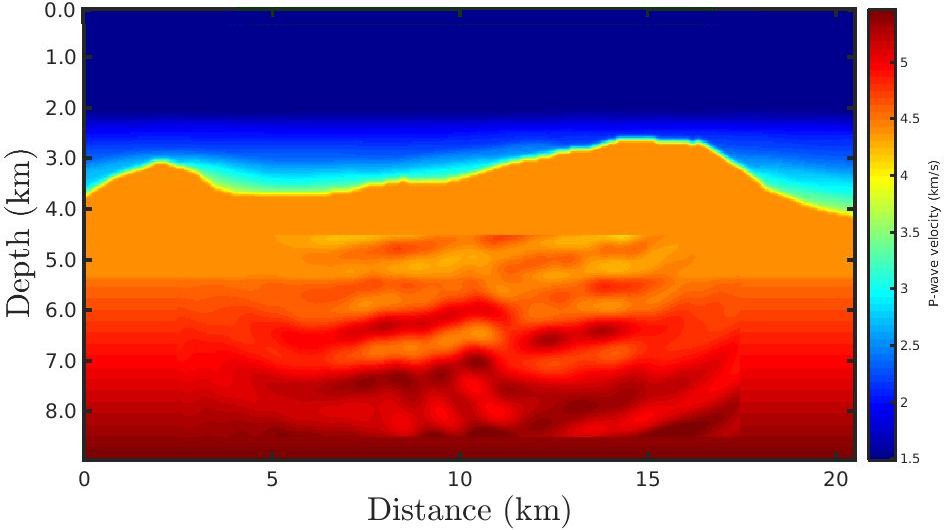}
  \label{fig:GdM_noiseless_PGF_classical_2_3Hz}
\end{subfigure}
\begin{subfigure}{.5\textwidth}
  \centering
  \caption{$\alpha$-PGF-FWI}
  \includegraphics[width=\linewidth]{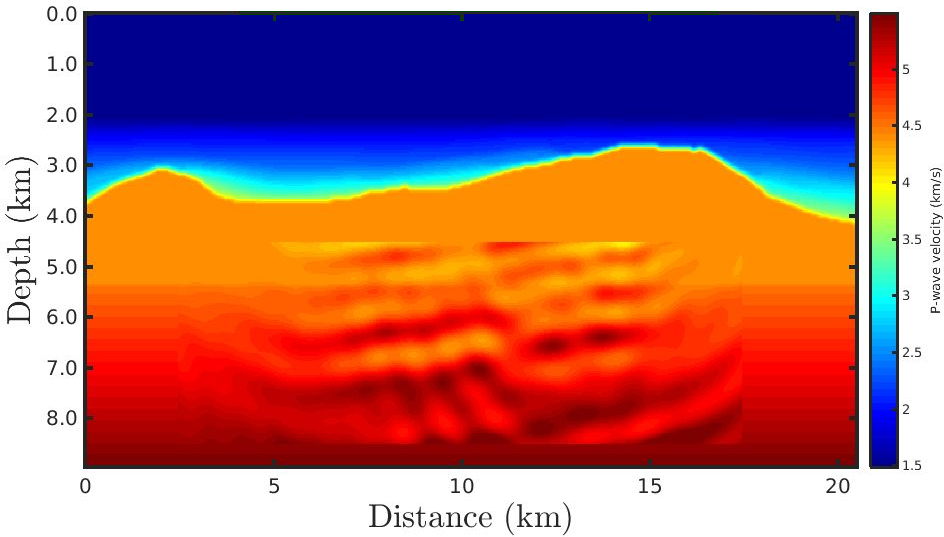}
  \label{fig:GdM_noiseless_PGF_proposal_2_3Hz}
\end{subfigure}
\begin{subfigure}{.5\textwidth}
  \centering
  \caption{classical PGF-FWI ($\alpha \rightarrow 1$)}
  \includegraphics[width=\linewidth]{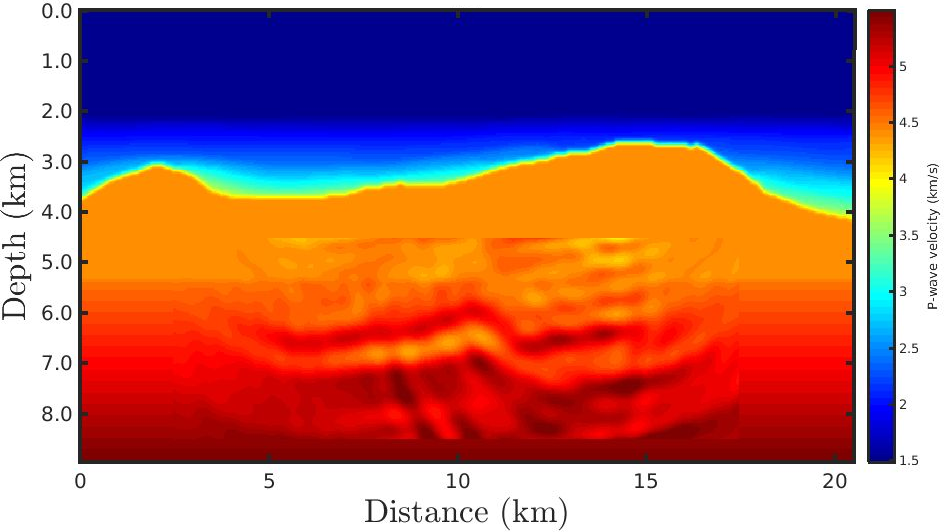}
  \label{fig:GdM_noiseless_PGF_classical_2_5Hz}
\end{subfigure}
\begin{subfigure}{.5\textwidth}
  \centering
  \caption{$\alpha$-PGF-FWI}
  \includegraphics[width=\linewidth]{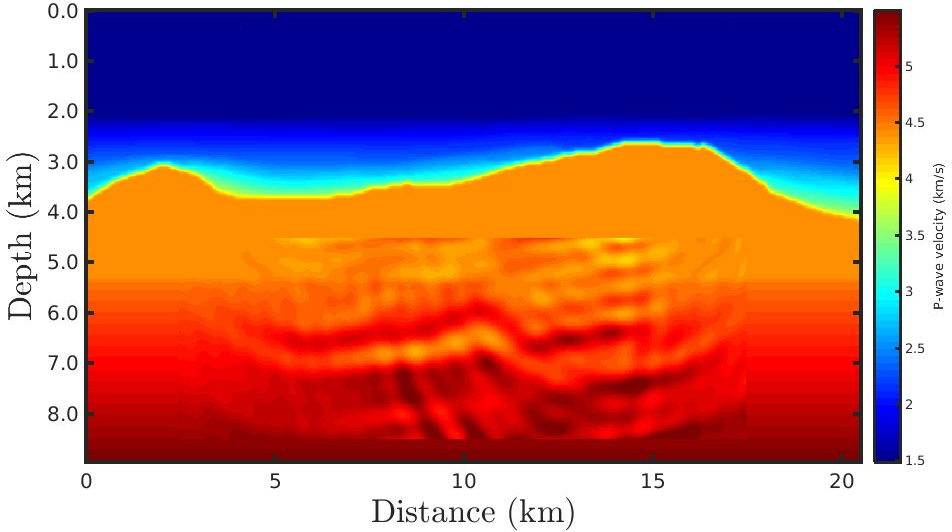}
  \label{fig:GdM_noiseless_PGF_proposal_2_5Hz}
\end{subfigure}
\begin{subfigure}{.5\textwidth}
  \centering
  \caption{classical PGF-FWI ($\alpha \rightarrow 1$)}
  \includegraphics[width=\linewidth]{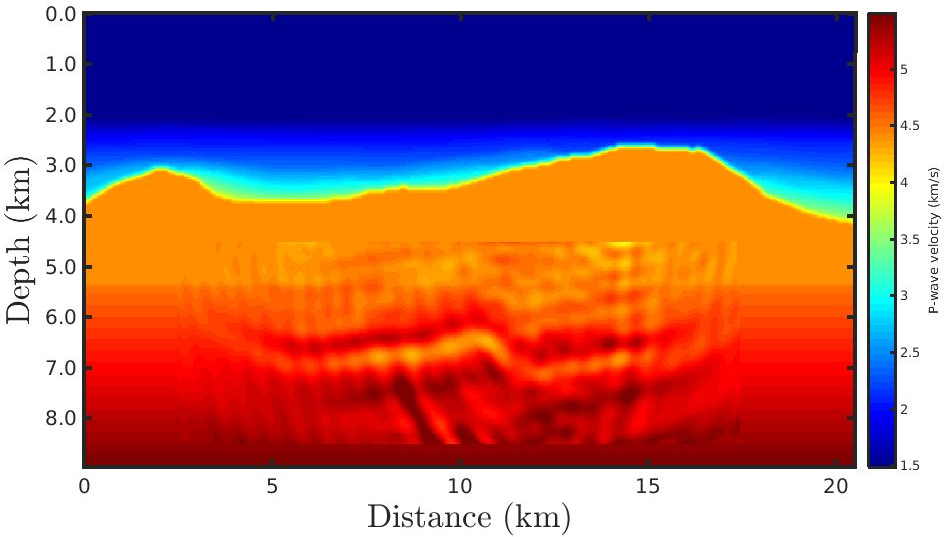}
  \label{fig:GdM_noiseless_PGF_classical_2_7Hz}
\end{subfigure}
\begin{subfigure}{.5\textwidth}
  \centering
  \caption{$\alpha$-PGF-FWI}
  \includegraphics[width=\linewidth]{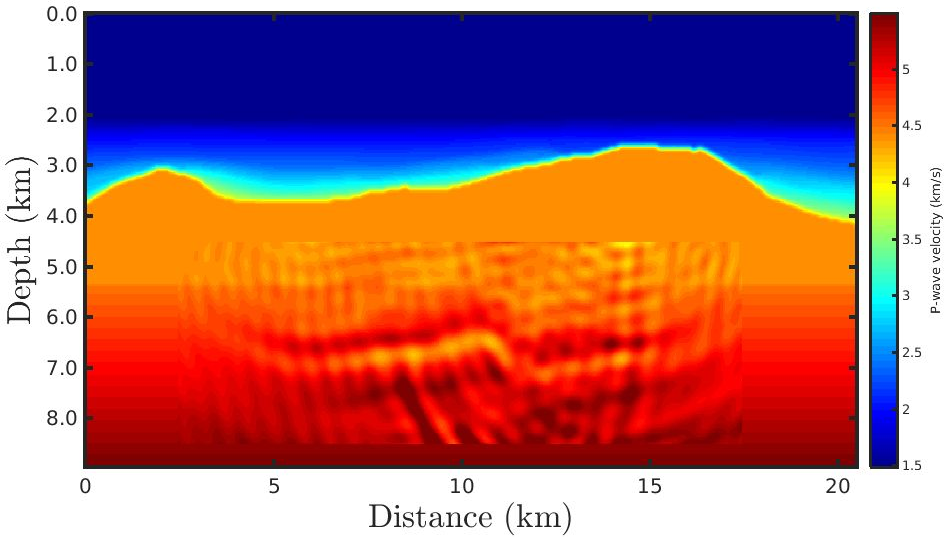}
  \label{fig:GdM_noiseless_PGF_proposal_2_7Hz}
\end{subfigure}
\caption{P-wave velocity models reconstructed, for the noiseless case, using $\alpha$-PGF-FWI for the (a)-(b) first, the (c)-(d) second, and the (e)-(f) third frequencies group, in which the left column refers to the classical approach ($\alpha \rightarrow 1$) and the right column refers to our proposal with $\alpha = 0.35$.}
\label{fig:GdM_noiseless_PGF}
\end{figure}

\begin{figure}[!htb]
\begin{subfigure}{.5\textwidth}
  \centering
  \caption{classical PGF-FWI ($\alpha \rightarrow 1$)}
  \label{fig:GdM_nonGaussianNoise_PGF_classical_2_3Hz}
  \includegraphics[width=\linewidth]{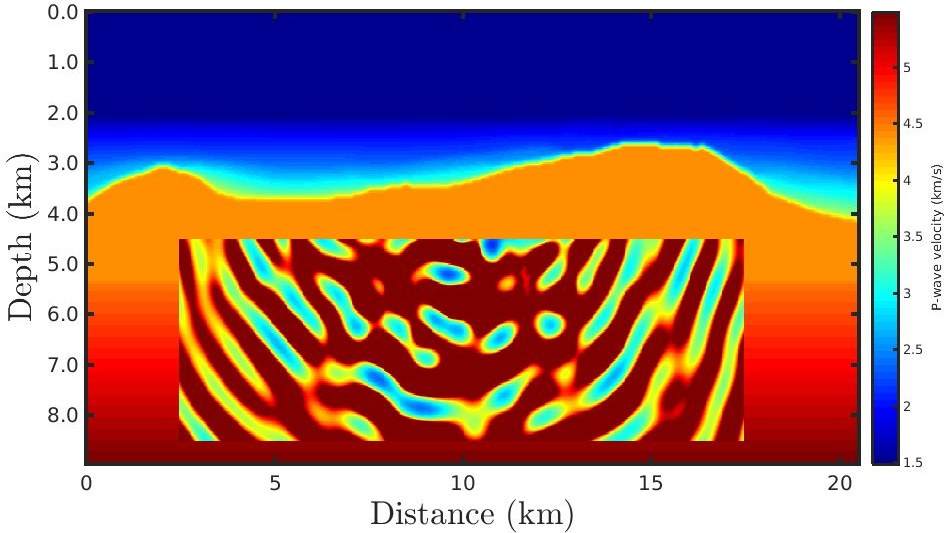}
  \label{fig:GdM_nonGaussianNoise_PGF_proposal_2_3Hz}
\end{subfigure}
\begin{subfigure}{.5\textwidth}
  \centering
  \caption{$\alpha$-PGF-FWI}
  \includegraphics[width=\linewidth]{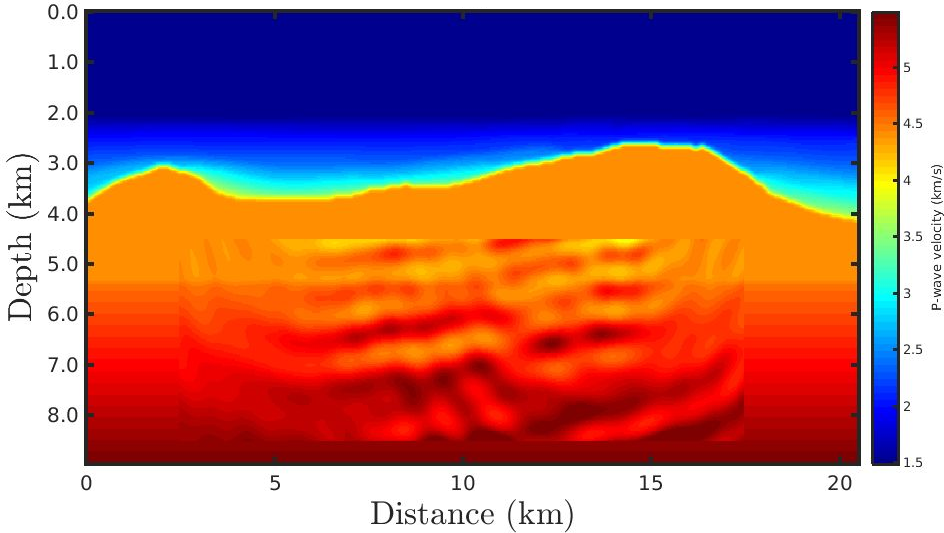}
\end{subfigure}
\begin{subfigure}{.5\textwidth}
  \centering
  \caption{classical PGF-FWI ($\alpha \rightarrow 1$)}
\includegraphics[width=\linewidth]{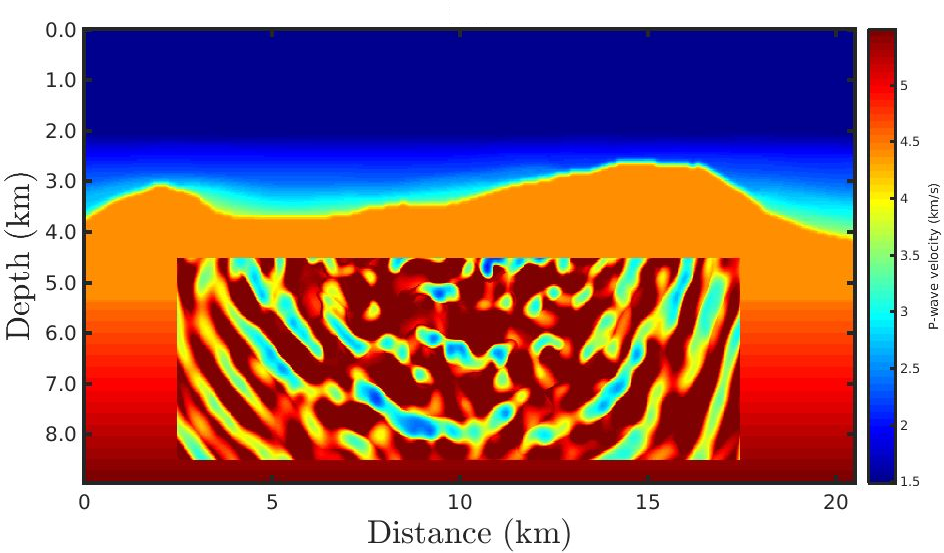}
  \label{fig:GdM_nonGaussianNoise_PGF_proposal_2_5Hz}
\end{subfigure}
\begin{subfigure}{.5\textwidth}
  \centering
  \caption{$\alpha$-PGF-FWI}
    \includegraphics[width=\linewidth]{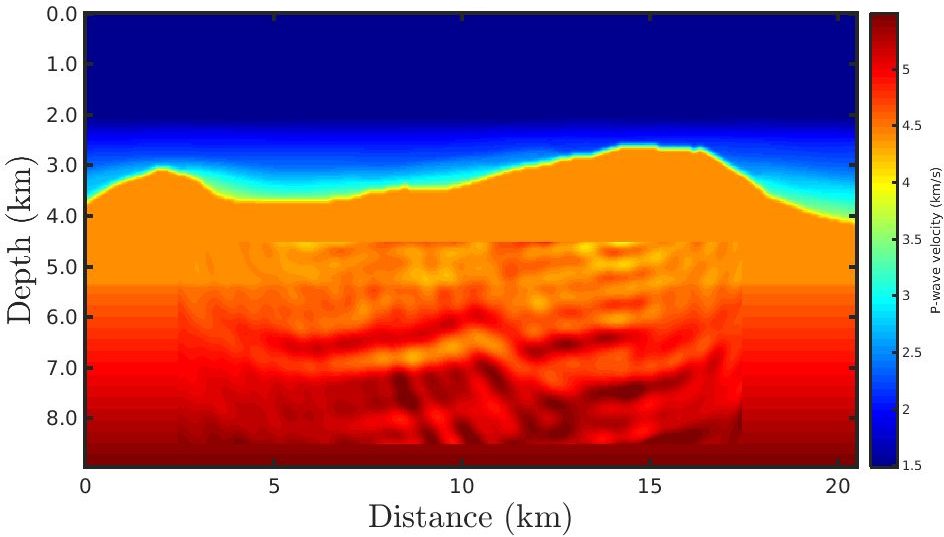}
  \label{fig:GdM_nonGaussianNoise_PGF_classical_2_5Hz}
\end{subfigure}
\begin{subfigure}{.5\textwidth}
  \centering
  \caption{classical PGF-FWI ($\alpha \rightarrow 1$)}
    \includegraphics[width=\linewidth]{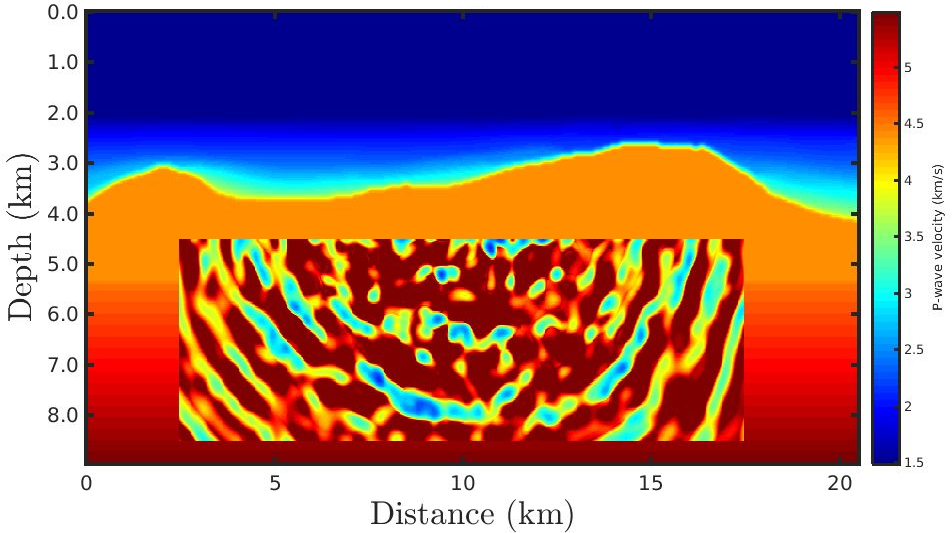}
  \label{fig:GdM_nonGaussianNoise_PGF_proposal_2_7Hz}
\end{subfigure}
\begin{subfigure}{.5\textwidth}
  \centering
  \caption{$\alpha$-PGF-FWI}
  \includegraphics[width=\linewidth]{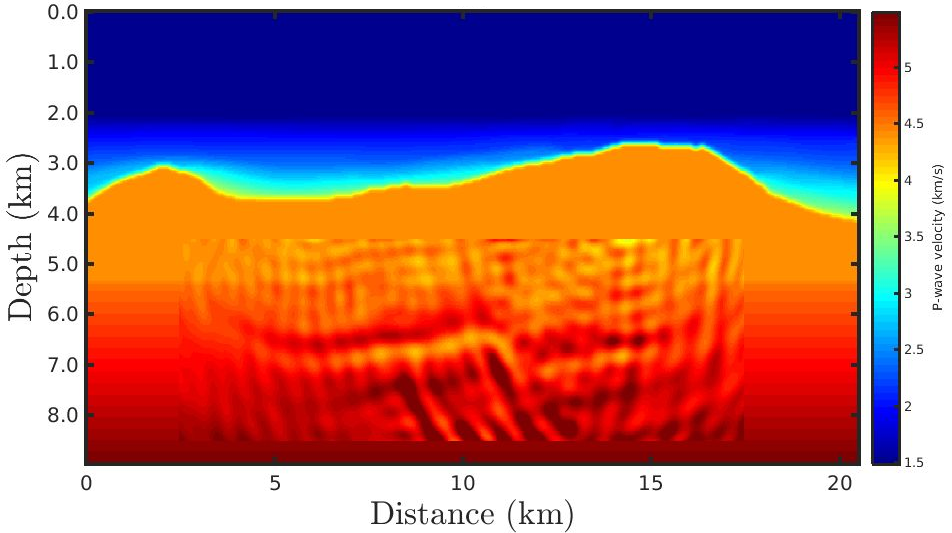}
  \label{fig:GdM_nonGaussianNoise_PGF_classical_2_7Hz}
\end{subfigure}
\caption{P-wave velocity models reconstructed, for the non-Gaussian noise case, using $\alpha$-PGF-FWI for the (a)-(b) first, the (c)-(d) second, and the (e)-(f) third frequencies group, in which the left column refers to the classical approach ($\alpha \rightarrow 1$) and the right column refers to our proposal with $\alpha = 0.35$.}
\label{fig:GdM_nonGaussianNoise_PGF}
\end{figure}

Figures~\ref{fig:GdM_noiseless_PGF} and \ref{fig:GdM_nonGaussianNoise_PGF} show the $\alpha$-PGF-FWI results for the noiseless and non-Gaussian noise cases, respectively, in which the left column refers to the classical approach ($\alpha \rightarrow 1$) and the right column refers to our proposal with $\alpha = 0.35$. We notice in Fig.~\ref{fig:GdM_noiseless_PGF} that if the data is noise-free, the resulting models are very similar regardless of the applied approach or data-frequency content. On the other hand, when non-Gaussian noise is considered the classical approach fails completely, in which the reconstructed model is biased and therefore very far from the true model. In contrast, our proposal is insensitive to non-Gaussian noise which generates a satisfactory reconstructed model (see the right column of Fig.~\ref{fig:GdM_nonGaussianNoise_PGF}) that is comparable to the case where the data is noise-free (Fig.~\ref{fig:GdM_noiseless_PGF}).

\begin{figure}[!htb]
\begin{subfigure}{.32\textwidth}
  \centering
  \caption{}
  \includegraphics[width=\linewidth]{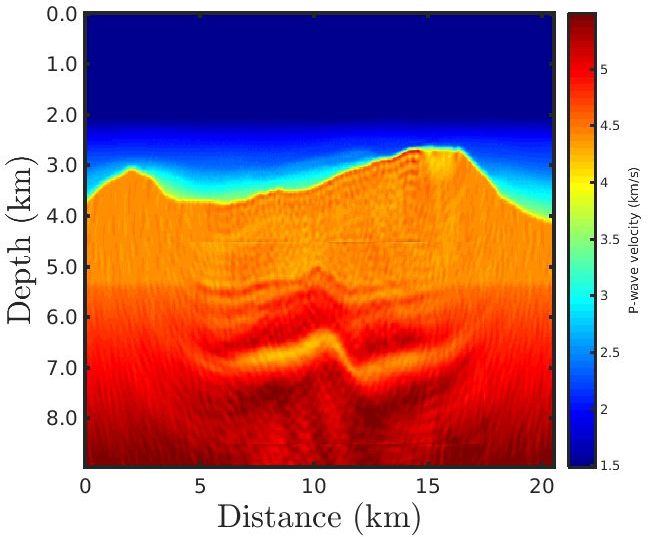}
  \label{fig:GdM_nonGaussianNoise_FWI_ourproposal_2_3Hz}
\end{subfigure}
\begin{subfigure}{.32\textwidth}
  \centering
  \caption{}
  \includegraphics[width=\linewidth]{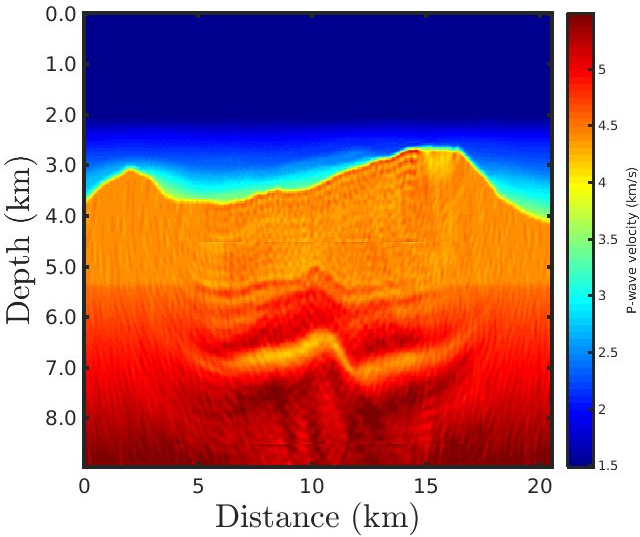}
  \label{fig:GdM_nonGaussianNoise_FWI_ourproposal_2_5Hz}
\end{subfigure}
\begin{subfigure}{.32\textwidth}
  \centering
  \caption{}
  \includegraphics[width=\linewidth]{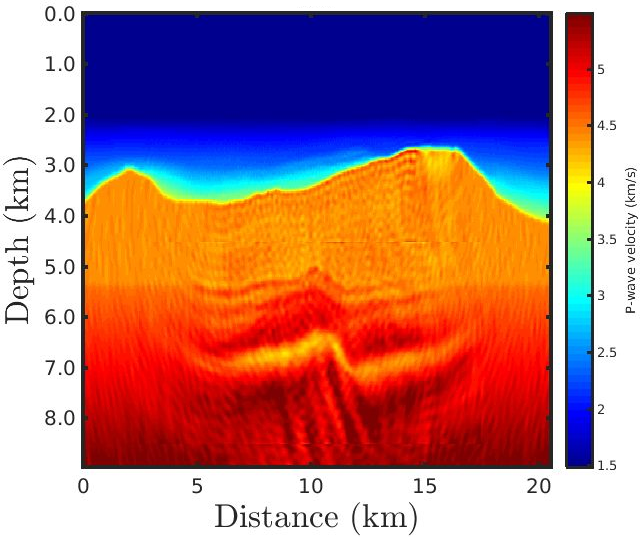}
  \label{fig:GdM_nonGaussianNoise_FWI_ourproposal_2_7Hz}
\end{subfigure}
\begin{subfigure}{.32\textwidth}
  \centering
  \caption{}
  \includegraphics[width=\linewidth]{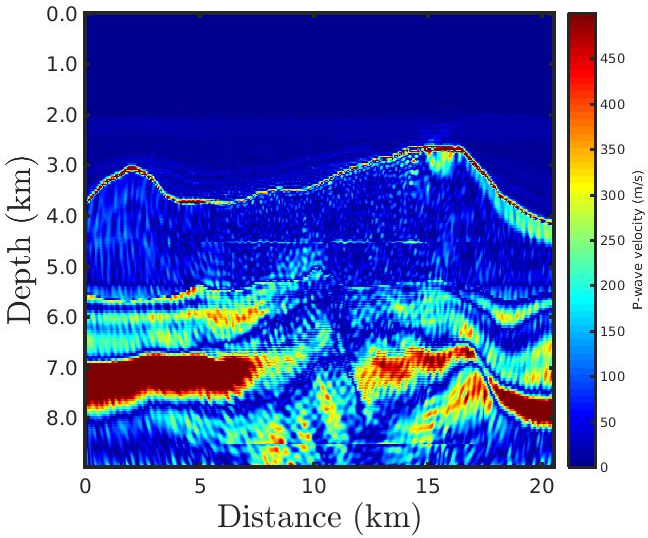}
  \label{fig:GdM_nonGaussianNoise_FWI_ourproposal_2_3Hz_error}
\end{subfigure}
\begin{subfigure}{.32\textwidth}
  \centering
  \caption{}
  \includegraphics[width=\linewidth]{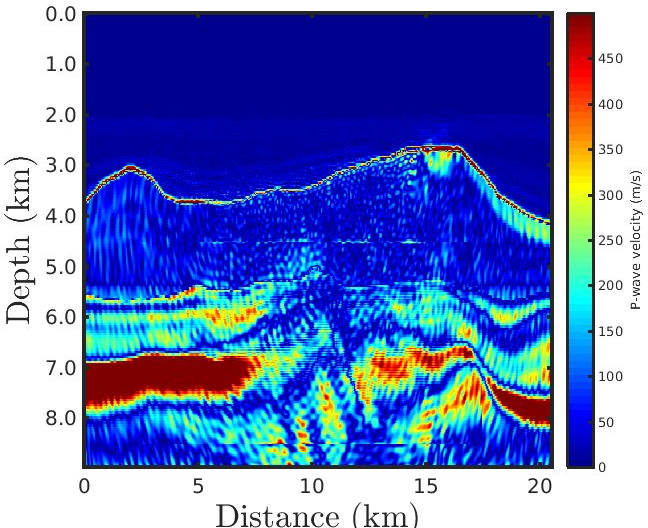}
  \label{fig:GdM_nonGaussianNoise_FWI_ourproposal_2_5Hz_error}
\end{subfigure}
\begin{subfigure}{.32\textwidth}
  \centering
  \caption{}
  \includegraphics[width=\linewidth]{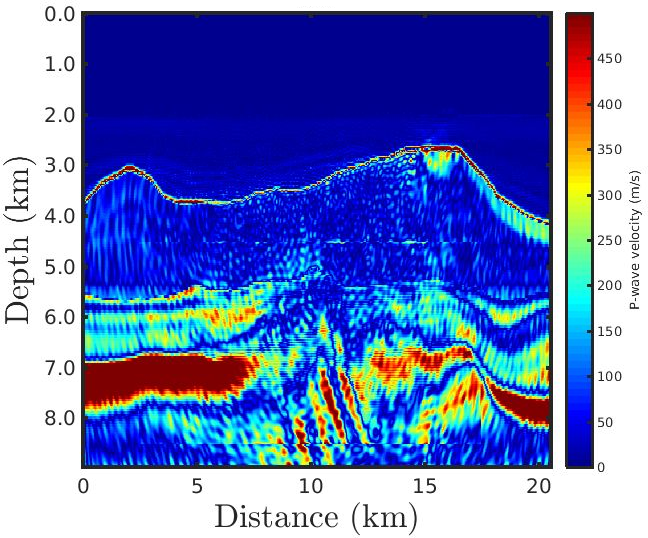}
  \label{fig:GdM_nonGaussianNoise_FWI_ourproposal_2_7Hz_error}
\end{subfigure}
\caption{Reconstructed P-wave models from the model in (a) Fig.~\ref{fig:GdM_nonGaussianNoise_PGF_classical_2_3Hz}, (b) Fig.~\ref{fig:GdM_nonGaussianNoise_PGF_classical_2_5Hz}, and Fig.~~\ref{fig:GdM_nonGaussianNoise_PGF_classical_2_7Hz} as initial model, using the time-domain FWI. Panels (d)-(f) show the absolute difference between the reconstructed models shown in panels (a)-(c) and the true model.}
\label{fig:GdM_nonGaussianNoise_FWI}
\end{figure}

Still considering the seismic data contaminated by non-Gaussian noise, we perform three inversions, in time-domain, using $\alpha$-FWI in which the difference between each simulation is just the initial model. In this regard, the P-wave models shown in Figs.~\ref{fig:GdM_nonGaussianNoise_PGF_classical_2_3Hz}, \ref{fig:GdM_nonGaussianNoise_PGF_classical_2_5Hz} and \ref{fig:GdM_nonGaussianNoise_PGF_classical_2_7Hz} are considered as initial model in three different data inversion processes. The resulting models are depicted in Fig.~\ref{fig:GdM_nonGaussianNoise_FWI}. In this figure, panels (d)-(f) show the absolute difference between the reconstructed models shown in panels (a)-(c) and the true model. As expected, areas of the P-wave reconstructed models outside the target region have similar results. However,  in the pre-salt reservoir region, the initial models presented in Figs.~\ref{fig:GdM_nonGaussianNoise_PGF_classical_2_3Hz} and \ref{fig:GdM_nonGaussianNoise_PGF_classical_2_5Hz} exhibit less error than the model presented in Fig.~\ref{fig:GdM_nonGaussianNoise_PGF_classical_2_7Hz}, as can be seen by the red fringes in the target region of Fig.~\ref{fig:GdM_nonGaussianNoise_FWI_ourproposal_2_7Hz_error}.

%%%%%%%%%%%%%%%%%%%%%%%%%%%%%%%%%%%%%%%%%%%%%%%%%%
\section{Conclusion \label{sec:conclusion}}
%%%%%%%%%%%%%%%%%%%%%%%%%%%%%%%%%%%%%%%%%%%%%%%%%%

To mitigate the effects of non-Gaussian noise and to reduce the computational cost in the reconstruction of P-wave velocity models using the FWI methodology, we proposed a new misfit function based on the Rényi $\alpha$-Gaussian distribution using the PGF method. We call our proposal by the abbreviation $\alpha$-PGF-FWI. The numerical studies with high-resolution complex models demonstrate the effectiveness of our proposal inversion methodology. Using the Marmousi model, we demonstrate the robustness of the Rényi statistics for noiseless circumstances and Gaussian and non-Gaussian noise scenarios. In addition, we have demonstrated that the combination of FWI based on the maximization of Rényi entropy together with the PGF technique is promising for monitoring target regions, such as Brazilian pre-salt reservoirs. 

It is worth mentioning that if the observed data does not contain noise or is polluted by Gaussian noise, $\alpha$-PGF-FWI and classical PGF-FWI have similar performances. However, the $\alpha$-PGF-FWI is robust to non-Gaussian noise, while the classical PGF-FWI estimates biased models in these circumstances. The reason is that large errors are better attenuated in the adjoint-source of our proposal compared to the classical approach, leading to a better weighting of waveforms crossing the less illuminated regions of the model. Indeed, the classical misfit function treats all residuals equally, while the proposed misfit function weights residual data according to their amplitude, giving less importance to large errors in the dataset.

The numerical results show that the $\alpha$-PGF-FWI is a powerful methodology to deal with non-Gaussian errors, which may become a valuable tool in geophysical imaging problems, especially when the data is polluted by large errors. It is also worth noting that our proposal can be applied to any problem of estimating physical parameters linked to partial differential equations similar to the Helmholtz equation, such as in biomedical imaging issues \cite{Guasch_BrainAWI_2020}.

\section{Data and Resources}

Plots and numerical simulations were done, respectively, with MATLAB R2016b and Julia Language,  on a computer hosting a quad-core (Intel Xeon E5-1620 v3) processor at 3.50GHz and 256GB RAM memories. All data supporting the results of this study are freely available as indicated in the manuscript.

\section{Acknowledgements} 
W.A. Barbosa and J.M. de Araújo gratefully acknowledge support from \textit{Shell Brasil} through the \textit{New Methods for Full Waveform Inversion} project at \textit{Universidade Federal do Rio Grande do Norte} and the strategic importance of the support given by ANP through the R\&D levy regulation. J.M. de Araújo thanks \textit{Conselho Nacional de Desenvolvimento Científico e Tecnológico} (CNPq) for his productivity fellowship (grant no. 313431/2018-3). Authors would like to thank Jorge Lopez from \textit{Shell} for the careful reading and comments that have improved the manuscript.

%Bibliography
%\bibliographystyle{unsrt}  
%\bibliography{references}  

\begin{thebibliography}{10}

\bibitem{VirieuxOperto_2009_overview}
Virieux J, Operto S.
\newblock {An overview of full-waveform inversion in exploration geophysics}.
\newblock Geophy. 2009 Nov;74(6):WCC1--WCC26.

\bibitem{FichtnerBook}
Fichtner A.
\newblock {Full Seismic Waveform Modelling and Inversion}.
\newblock Springer-Verlag Berlin Heidelberg.; 2011.

\bibitem{Tarantola_1984_FWI_Origem}
Tarantola A.
\newblock {Inversion of seismic reflection data in the acoustic approximation}.
\newblock Geophy. 1984 Aug;49(8):1259--1266.

\bibitem{RealDataApplicationsFWI2018}
Kurzmann A, Ga{\ss}ner L, Shigapov R, Thiel N, Athanasopoulos N, Bohlen T, Steinweg T.
\newblock {High Performance Computing in Science and Engineering. Real Data Applications of Seismic Full Waveform Inversion}.
\newblock Springer International Publishing.; 2018.

\bibitem{OptimalTransportRealData2021}
Górszczyk A, Brossier R, Métivier L.
\newblock {Graph-space optimal transport concept for time-domain full-waveform inversion of ocean-bottom seismometer data: Nankai Trough velocity structure reconstructed from a 1D model}.
\newblock J Geophys Res Solid Earth. 2021 May;126(5):e2020JB021504.

\bibitem{Bernard_FWIultrasonic_2017}
Bernard S, Monteiller V, Komatitsch D, Lasaygues P.
\newblock {Ultrasonic computed tomography based on full-waveform inversion for bone quantitative imaging}.
\newblock Phys Med Biol. 2017 Aug;62(17):7011--7035.

\bibitem{Guasch_BrainAWI_2020}
Guasch L, Calderón Agudo O, Tang M-X, Nachev P, Warner M.
\newblock {Full-waveform inversion imaging of the human brain}.
\newblock  npj Digit Med. 2020 Mar;3:28.

\bibitem{Hanasoge_FWIAstrofisica_2014}
Hanasoge SM.
\newblock {Full waveform inversion of solar interior flows}.
\newblock Astrophy J. 2014 Nov;797(23):1--9.

\bibitem{Hanasoge_FWIAstrofisica0_2014}
Hanasoge SM, Tromp J.
\newblock {Full waveform inversion for time-distance helioseismology}.
\newblock Astrophy J. 2014 Mar;784(1):69.

\bibitem{MENKE198479}
Menke W.
\newblock {Geophysical Data Analysis: Discrete Inverse Theory}.
\newblock Academic Press.; 1984.

\bibitem{tarantola2005book}
Tarantola A.
\newblock {Inverse problem theory - and methods for model parameter estimation}.
\newblock SIAM.; 2005.

\bibitem{Constable_1988_geophysicalJI}
Constable CG.
\newblock {Parameter estimation in non-Gaussian noise}.
\newblock Geophys J Int. 1988 Jul;94(1):131--142.

\bibitem{residualnormbrossier}
Brossier R, Operto S, Virieux J.
\newblock {Which data residual norm for robust elastic frequency-domain full waveform inversion?}.
\newblock Geophys. 2010 May;75(3):R37--R46.

\bibitem{EPJ_PLUS_PSI_Laplace_2021}
da Silva SLEF, dos Santos Lima GZ, Volpe EV, de Araújo JM, Corso G.
\newblock {Robust approaches for inverse problems based on Tsallis and Kaniadakis generalised statistics}.
\newblock Eur Phys J Plus. 2021 May;136:518.

\bibitem{q_GaussianPhysicaA}
da Silva SLEF, da Costa CAN, Carvalho PTC, de Araújo JM, Lucena LdS, Corso G.
\newblock {Robust full-waveform inversion using q-statistics}.
\newblock Phys A Stat Mech Appl. 2020 Jun;548(6):124473.

\bibitem{Suyari_2005_IEEE_TsallisErrorLaw}
Suyari H, Tsukada M.
\newblock {Law of error in Tsallis statistics}.
\newblock IEEE Trans Inf Theory. 2005 Feb;51(2):753--757.

\bibitem{studentT_1_1989}
Lange KL, Little RJA, Taylor JMG.
\newblock {Robust statistical modeling using the t distribution}.
\newblock J Am Stat Assoc. 1989 Dec;84(408):881--896.

\bibitem{studentT_tristan_2012}
Aravkin AY, Friedlander MP, Herrmann FJ, van Leeuwen T.
\newblock {Robust inversion, dimensionality reduction and randomized sampling}.
\newblock Math Program. 2012 Jun;134:101--125.

\bibitem{Ubaidillah_2017_studentT}
Ubaidillah A, Notodiputro KA, Kurnia A, Fitrianto A, Mangku IW.
\newblock {A robustness study of student-t distributions in regression models with application to infant birth weight data in Indonesia}.
\newblock IOP Conf Ser: Earth Environ Sci. 2017 Jan;58:012013.

\bibitem{GeneralizedGaussian_SEG2015}
Li Z, Liu Z, Song C, Hu G, Zhang J.
\newblock {Generalized Gaussian distribution based adaptive mixed-norm inversion for non-Gaussian noise}.
\newblock SEG Techn Progr Exp Abstr. 2015 Aug;2015(1):3926-3930.

\bibitem{kappa_GaussianPRE}
da Silva SLEF, Carvalho PTCC, de Araújo JM, Corso G.
\newblock {Full-waveform inversion based on Kaniadakis statistics}.
\newblock Phys. Rev. E. 2020 May;101(5):053311.

\bibitem{Sacchi_SEG_generalized_2020}
Carozzi F, Sacchi MD.
\newblock {Making seismic reconstruction more robust via a generalized loss function}.
\newblock SEG Techn Progr Exp Abstr. 2020 Aug;2020(1):3149-3153.

\bibitem{PSI_Jackson_2021}
Silva SA, da Silva SLEF,  de Souza RF, Marinho AA, de Araújo JM, Bezerra CG.
\newblock {Improving Seismic Inversion Robustness via Deformed Jackson Gaussian}.
\newblock Entropy. 2021 Jul;23(8):1081.

\bibitem{daSilva_et_al_NewKappaGaussian}
da Silva SLEF, Silva R, dos Santos Lima GZ, de Araújo JM, Corso G.
\newblock {An outlier-resistant $\kappa$-generalized approach for robust physical parameter estimation}.
\newblock arXiv 2021 Nov; arXiv:2111.09921.

\bibitem{HASEGAWA20093399}
Hasegawa Y, Arita M.
\newblock {Properties of the maximum q-likelihood estimator for independent random variables}.
\newblock Phys A Stat Mech Appl. 2009 Sep;388(17):3399--3412.

\bibitem{Ferrari_2010_AnnalsStat_lqLikelihood}
Ferrari D, Yang Y.
\newblock {Maximum Lq-likelihood estimation}.
\newblock Ann Statist. 2010 Apr;38(2):753--783.

\bibitem{PhysRevE.104.024107}
da Silva SLEF, Kaniadakis G.
\newblock {Robust parameter estimation based on the generalized log-likelihood in the context of Sharma-Taneja-Mittal measure}.
\newblock Phys Rev E. 2021 Aug;104(2):024107.

\bibitem{Carvalho_2021_Geophys_J_Int}
Carvalho PT, da Silva SLEF, Duarte EF, Brossier R, Corso G., de Araújo JM.
\newblock {Full waveform inversion based on the non-parametric estimate of the probability distribution of the residuals}.
\newblock Geophys J Int. 2022 Apr;229(1):35--55.

\bibitem{extensiveANDnonextensive_physicaA}
da Silva SLEF, dos Santos Lima GZ, de Araújo JM, Corso G.
\newblock {Extensive and nonextensive statistics in seismic inversion}.
\newblock Phys A Stat Mech Appl. 2021 Feb;563(1):125496.

\bibitem{Renyi1965informationTheory}
Rényi A.
\newblock {On the Foundations of Information Theory}.
\newblock Rev Inst Int Stat. 1965 Jan;33(1):1--14.

\bibitem{renyi1961measures}
Rényi A.
\newblock {On Measures of Entropy and Information}.
\newblock Proceedings of the Fourth Berkeley Symposium on Mathematical Statistics and Probability, Volume 1: Contributions to the Theory of Statistics. 1961.

\bibitem{lenzi_et_al_2000_physicaA_renyi}
Lenzi EK, Mendes RS, da Silva LR.
\newblock {Statistical mechanics based on Renyi entropy}.
\newblock Phys A Stat Mech Appl. 2000 Jun;280(3--4):337--345.

\bibitem{RenyiApp_Ecology_2015}
Vranken I, Baudry J, Aubinet M, Visser M, Bogaert J.
\newblock {A review on the use of entropy in landscape ecology: heterogeneity, unpredictability, scale dependence and their links with thermodynamics}.
\newblock Landscape Ecol. 2014 Oct;30(1):51--65.

\bibitem{XU2021107668}
Xu L, Bai L, Jiang X, Tan M, Zhang D, Luo B.
\newblock {Deep Rényi entropy graph kernel}.
\newblock Pattern Recognit. 2021 Mar;111(1):107668.

\bibitem{MachineLearningAppRenyi_2019}
Li XY, Zhu QS, Zhu MZ, Huang YM, Wu H, Wu SY.
\newblock {Machine learning study of the relationship between the geometric and entropy discord}.
\newblock EPL. 2019 Mar;127(2):20009.

\bibitem{e22020186}
Yu KS, Kim SH, Lim DW, Kim YS.
\newblock {A Multiple Rényi Entropy Based Intrusion Detection System for Connected Vehicles}.
\newblock Entropy. 2020 Feb;22(2):186.

\bibitem{QuantumRenyiapp_2015}
Lewkowycz A, Perlmutter E.
\newblock {Universality in the geometric dependence of Rényi entropy}.
\newblock J High Energ Phys. 2015 Jan;2015(1):80.

\bibitem{science.aau4963_2019}
Brydges T, Elben A, Jurcevic P, Vermersch B,  Maier C, Lanyon BP, Zoller P, Blatt R, Roos CF.
\newblock {Probing Rényi entanglement entropy via randomized measurements}.
\newblock Science. 2019 Apr;364(6437):260--263.

\bibitem{PhysRevB.90.115408}
Power SR, Jauho AP.
\newblock {Electronic transport in disordered graphene antidot lattice devices}.
\newblock Phys Rev B. 2014 Sep;90(11):115408.

\bibitem{PhysRevB.91.125408}
Settnes M, Power SR, Lin J, Petersen DH, Jauho AP.
\newblock {Patched Green's function techniques for two-dimensional systems: Electronic behavior of bubbles and perforations in graphene}.
\newblock Phys Rev B. 2015 Mar;91(12):125408.

\bibitem{FERREIRA2002355}
Ferreira MA, Bauer GEW, Wapenaar CPA.
\newblock {Recursive Green functions technique applied to the propagation of elastic waves in layered media}.
\newblock Ultrasonics. 2002 May;40(1):355--359.

\bibitem{Moura2020}
Moura FA, Barbosa WA, Duarte EF, Silva DP, Ferreira MS, Lucena LS, de Araújo JM.
\newblock {Patched Green's function method applied to acoustic wave propagation in disordered media: an interdisciplinary approach}.
\newblock J Geophys Eng. 2020 Sep;17(5):914--922.

\bibitem{almeida2020}
Barbosa WA, Duarte EF, Moura FA, Ferreira MS, de Araújo JM.
\newblock {Target-oriented wave propagator using the Patched Green Function method}.
\newblock EAGE Conf Proceed. 2020 Jun;2020(1):1--5.

\bibitem{FWIPGF2021}
da Silva D, Duarte EF, Almeida W, Ferreira M,  Moura FA,  de Araújo JM.
\newblock {Target-oriented inversion using the Patched Green's function method}.
\newblock Geophys. 2021 Nov;86(6):R811--R823.

\bibitem{Marfurt_1984}
Marfurt KJ.
\newblock {Accuracy of finite-difference and finite-element modeling of the scalar and elastic wave equations}.
\newblock Geophys. 1984 May; 49(5):533--549.

\bibitem{Pratt_1999}
Pratt RG.
\newblock {Seismic waveform inversion in the frequency domain, Part 1: Theory and verification in a physical scale model}.
\newblock Geophys. 1999 May; 64(3):888--901.

\bibitem{SEISCOPE_optimization_2016}
Métivier L, Brossier R.
\newblock {The SEISCOPE optimization toolbox: A large-scale nonlinear optimization library based on reverse communication}.
\newblock Geophys. 2016 Mar; 81(2):F1--F15.

\bibitem{Haber_2000}
Haber E, Ascher UM, Oldenburg D.
\newblock {On optimization techniques for solving nonlinear inverse problems}.
\newblock Inv Probl. 2000 Oct; 16(5):1263--1280.

\bibitem{Plessix_AdjointReview_GJI_2006}
Plessix R-E.
\newblock {A review of the adjoint-state method for computing the gradient of a functional with geophysical applications}.
\newblock Geophys J Int. 2006 Nov; 167(2):495--503.

\bibitem{ClaerboutRobustErraticData1973}
Claerbout JF, Muir F.
\newblock {Robust modeling with erratic data}.
\newblock Geophys. 1973 Oct; 38(1):826--844.

\bibitem{PhysRevLett.93.130601}
Bashkirov AG.
\newblock {Maximum Renyi Entropy Principle for Systems with Power-Law Hamiltonians}.
\newblock Phys Rev Lett. 2004 Sep; 93(13):130601.

\bibitem{LEONENKO20101981}
Leonenko N, Seleznjev O.
\newblock {Statistical inference for the $\epsilon$-entropy and the quadratic Rényi entropy}.
\newblock J Multivar Anal. 2010 Oct; 101(9):1981--1994.

\bibitem{t_Student2003}
Costa J, Hero A, Vignat C.
\newblock {On Solutions to Multivariate Maximum $\alpha$-Entropy Problems. In: Energy Minimization Methods in Computer Vision and Pattern Recognition}.
\newblock Springer, Berlin, Heidelberg. 2003.

\bibitem{t_Student_JOHNSON2007}
Johnson O, Vignat C.
\newblock {Some results concerning maximum Rényi entropy distributions}.
\newblock Ann inst Henri Poincare (B) Probab Stat. 2007 May; 43(3):339--351.

\bibitem{RenyMaximum2019}
Tanaka HA, Nakagawa M, Oohama Y.
\newblock {A Direct Link between Rényi–Tsallis Entropy and Hölder’s Inequality—Yet Another Proof of Rényi–Tsallis Entropy Maximization}.
\newblock Entropy. 2019 May; 21(6):549.

\bibitem{doniach1974green}
Doniach S, Sondheimer EH.
\newblock {Green's functions for solid state physicists}.
\newblock Frontiers in physics. 1974.

\bibitem{economou2006}
Economou EN.
\newblock {Green's Functions in Quantum Physics}.
\newblock Springer Series in Solid-State Sciences. 2006.

\bibitem{sheng2006introduction}
Sheng P.
\newblock {Introduction to Wave Scattering, Localization and Mesoscopic Phenomena}.
\newblock Springer Series in Materials Science. 2006.

\bibitem{ByrdNocedalDetails}
Byrd RH, Lu P, Nocedal J, Zhu C.
\newblock {A Limited Memory Algorithm for Bound Constrained Optimization}.
\newblock J Sci Comput. 1995 May; 16(5):1190--1208.

\bibitem{wolfeOriginal}
Wolfe P.
\newblock {Convergence Conditions for Ascent Methods}.
\newblock SIAM Rev. 1969 Apr; 11(2):226--235.

\bibitem{rickerSource}
Ricker N.
\newblock {Further developments in the wavelet theory of seismogram structure}.
\newblock Bull Seism Soc Am. 1943 Jul;33(3):197--228.

\bibitem{marmousi}
Martin GS, Wiley R, Marfurt KJ.
\newblock {Marmousi2: An elastic upgrade for Marmousi}.
\newblock Lead Edge. 2006 Feb; 25(2):156--166.

\bibitem{Elboth_Geophysics_74_SweelNoise_2009}
Elboth T, Reif BA, Andreassen Ø.
\newblock {Flow and swell noise in marine seismic data}.
\newblock Geophys. 2009 Mar; 74(2):Q17--Q25.

\bibitem{evans1996_PearsonClassifications}
Evans JD.
\newblock {Straightforward Statistics for the Behavioral Sciences}.
\newblock Brooks/Cole Publishing Company. 1996.

\bibitem{jorgeGdM_2020}
Lopez J, Neto F, Cabrera M, Cooke S, Grandi S, Roehl D.
\newblock {Refraction seismic for pre-salt reservoir characterization  and  monitoring}.
\newblock SEG Techn Progr Exp Abstr. 2020 Aug;2020(1):2365--2369.

\bibitem{daSilva_SEG2021_qFWI_2021}
da Silva SLEF, Lopez JL, de Araújo JM, Corso G.
\newblock {Multiscale q-FWI applied to circular shot OBN acquisition for accurate presalt velocity estimates}.
\newblock In: First International Meeting for Applied Geoscience $\&$ Energy Expanded Abstracts. 2021 Sep;2021(1):712--716.

\bibitem{Roden_CPML_2000}
Roden JA, Gedney SD.
\newblock {Convolution PML (CPML): An efficient FDTD implementation of the CFS–PML for arbitrary media}.
\newblock Microw Opt Technol Lett. 2000 Oct;27(5):334--339.

\bibitem{Komatitsch_CPML_2007}
Komatitsch D, Martin R.
\newblock {An unsplit convolutional perfectly matched layer improved at grazing incidence for the seismic wave equation}.
\newblock Geophys. 2007 Sep;72(5):SM155--SM167.

\bibitem{Kolb1986}
Kolb P, Collino F, Lailly P.
\newblock {Pre-stack inversion of a 1-D medium}.
\newblock Proc IEEE. 1986 Mar; 74(3):498--505.

\bibitem{multiscalebunks1995}
Bunks C, Saleck FM, Zaleski S, Chavent G.
\newblock {Multiscale seismic waveform inversion}.
\newblock Geophys. 1995 Sep; 60(5):1457--1473.



\end{thebibliography}

\end{document}